\documentclass[aps,preprint,nofootinbib]{revtex4}%
\usepackage{amsfonts}
\usepackage{amsmath}
\usepackage{amssymb}
\usepackage{graphicx}%
\providecommand{\U}[1]{\protect\rule{.1in}{.1in}}

\begin{document}
\title{The Hamiltonian formulation of N-bein, Einstein-Cartan, gravity in any
dimension: the Progress Report \\(Extended version of a talk given on CAIMS-2009, June 11-14, London, Canada)}
\author{N. Kiriushcheva}
\affiliation{Department of Applied Mathematics, University of Western Ontario, London, Canada}
\email{nkiriush@uwo.ca}
\author{S.V. Kuzmin}
\affiliation{Faculty of Arts and Social Science, Huron University College and Department of
Applied Mathematics, University of Western Ontario, London, Canada}
\email{skuzmin@uwo.ca}

\begin{abstract}
The Hamiltonian formulation of N-bein, Einstein-Cartan, gravity, using its
first order form in any dimension higher than two, is analyzed. This
Hamiltonian formulation allows to explicitly show where peculiarities of three
dimensional case (\textit{A.M.Frolov et al, 0902.0856 [gr-qc]}) occur and make
a conjecture, based on presented in this report results, that there is one
general for \textit{all} dimensions characteristic of N-bein formulation of
gravity: after elimination of second class constraints the algebra of Poisson
brackets among remaining first class secondary constraints is the Poincar\'{e}
algebra and in all dimensions N-bein, Cartan-Einstein, gravity \textit{is the
Poincar\'{e} gauge theory}. The gauge symmetry corresponding to the algebra of
first class constraints has two parameters- rotational (Lorentz) and
translational. Translational invariance is common to all dimensions but some
terms in general expressions for gauge transformations of N-beins and
connections are zero in a particular, three dimensional, case.

The proof of our conjecture is outlined in detail. Some straightforward but
tedious calculations remain to be completed to call our conjecture - a theorem
and will be reported later.

\end{abstract}
\maketitle

\section{Introduction}

In this Report we continue the analysis of the Hamiltonian formulation of
Einstein-Cartan, N-bein gravity (tetrads in four dimensional case), using its
first-order form. This analysis was started in \cite{3D} where the complete
treatment of three dimensional case ($3D$, $D$ is the dimension of spacetime)
was discussed. In the considering Hamiltonian formulation we will compare
quite often higher dimensional cases with \cite{3D} because Dirac's method of
constraint dynamics \cite{Diracbook} (and his more technical initial articles
\cite{Diracarticles} based on the course of lectures given at Canadian
Mathematical Seminar, Vancouver, August-September 1949) is perfectly suitable
for such a task as this is the general method not related to a particular
dimension. To be able to use advantages of this method, we do not specialize
our formulation to any particular representation of variables which is valid
only in a particular dimension and work with variables in which the Lagrangian
of N-bein gravity is originally formulated. Note that this is a Progress
Report, not a regular article or review, and our main goal here is to provide
details of calculations (still in progress). Some related references can be
found in \cite{3D}.

The Lagrangian of Einstein-Cartan, N-bein, gravity written in its first order
form is%

\begin{equation}
L\left(  e_{\mu\left(  \alpha\right)  },\omega_{\mu\left(  \alpha\beta\right)
}\right)  =-e\left(  e^{\mu\left(  \alpha\right)  }e^{\nu\left(  \beta\right)
}-e^{\nu\left(  \alpha\right)  }e^{\mu\left(  \beta\right)  }\right)  \left(
\omega_{\nu\left(  \alpha\beta\right)  ,\mu}+\omega_{\mu\left(  \alpha
\gamma\right)  }\omega_{\nu~~\beta)}^{~(\gamma}\right)  , \label{eqnPR1}%
\end{equation}

where the covariant N-beins $e_{\gamma\left(  \rho\right)  }$ and the
connections $\omega_{\nu\left(  \alpha\beta\right)  }$ ($\omega_{\nu\left(
\alpha\beta\right)  }=-\omega_{\nu\left(  \beta\alpha\right)  }$) are treated
as independent fields, and $e=\det\left(  {e_{\gamma\left(  \rho\right)  }%
}\right)  $.\footnote{Usually variables $e_{\gamma\left(  \rho\right)  }$ and
$\omega_{\nu\left(  \alpha\beta\right)  }$ are named tetrads and spin
connection, but such names are specialized for $D=4$. As we consider the
Hamiltonian formulation in any dimension ($D>2$), we will call $e_{\gamma
\left(  \rho\right)  }$ and $\omega_{\nu\left(  \alpha\beta\right)  }$ N-beins
and connections, respectively.} Greek letters indicate covariant indices
$\alpha=0,1,2,...\left(  D-1\right)  $. Indices in brackets $(..)$ denote the
internal (\textquotedblleft Lorentz\textquotedblright) indices, whereas
indices without brackets are external or \textquotedblleft
world\textquotedblright\ indices. Internal and external indices are raised and
lowered by the Minkowski tensor $\tilde{\eta}_{\alpha\beta}=\left(
-,+,+,...\right)  $ and the metric tensor $g_{\mu\nu}=e_{\mu\left(
\alpha\right)  }e_{\nu}^{\left(  \alpha\right)  }$, respectively (here and
below we will use tilde for any combination with only internal indices and do
not use brackets in these cases, except for antisymmetric indices). We assume
that N-beins are invertible and $e^{\mu\left(  \alpha\right)  }e_{\mu\left(
\beta\right)  }=\tilde{\delta}_{\beta}^{\alpha}$, $e^{\mu\left(
\alpha\right)  }e_{\nu\left(  \alpha\right)  }=\delta_{\nu}^{\mu}$. For the
Hamiltonian formulation where we have to separate space and time
\textit{indices }(not a spacetime itself on space and time) we are using $0$
for an external time index ($\left(  0\right)  $ for an internal
\textquotedblleft time\textquotedblright\ index) and Latin letters for spatial
external indices $k=1,2,...\left(  D-1\right)  $ ($\left(  k\right)  $ for
internal \textquotedblleft spatial\textquotedblright\ indices).

For the N-bein gravity, equation (\ref{eqnPR1}) represents the first order
form because variation of this Lagrangian with respect to connections treated
as independent variables gives an equation of motion that can be solved for
connections and the solution of it gives exactly the definition of a connection%

\begin{equation}
\omega_{\sigma}^{~\left(  \alpha\beta\right)  }=\frac{1}{2}e_{\sigma\left(
\lambda\right)  }\left(  A^{\varepsilon\left(  \alpha\right)  \mu\left(
\lambda\right)  }e_{\varepsilon,\mu}^{\left(  \beta\right)  }+A^{\varepsilon
\left(  \lambda\right)  \mu\left(  \beta\right)  }e_{\varepsilon,\mu}^{\left(
\alpha\right)  }-A^{\varepsilon\left(  \beta\right)  \mu\left(  \alpha\right)
}e_{\varepsilon,\mu}^{\left(  \lambda\right)  }\right)  \label{eqnPR2}%
\end{equation}
where%

\begin{equation}
A^{\mu\left(  \alpha\right)  \nu\left(  \beta\right)  }\equiv e^{\mu\left(
\alpha\right)  }e^{\nu\left(  \beta\right)  }-e^{\mu\left(  \beta\right)
}e^{\nu\left(  \alpha\right)  }. \label{eqnPR3}%
\end{equation}

The standard, second order form, is (\ref{eqnPR1}) with connections as short
notation for the combination (\ref{eqnPR2}), not independent variables.

The proof that connections can be treated as independent fields, i.e. solving
a corresponding variational equation for $\omega_{\nu\left(  \alpha
\beta\right)  }$ in terms of $e_{\gamma\left(  \rho\right)  }$, closely
resembles a solution of similar problem in formulation of metric gravity when
possibility to treat affine connections as independent fields for the standard
Einstein-Hilbert Lagrangian is discussed. This solution was first given by
Einstein \cite{Einstein} and the similar problem for particular combinations
of affine connections was considered in \cite{AOP}. In case of N-bein
formulation such a proof, to the best of our knowledge, was not published
anywhere and for the pedagogical reason we provide it in next Section. This
proof also allows us to establish notation that will be useful in considering
the Hamiltonian formulation of N-bein gravity and to solve similar equation
that arises in higher than three dimensions of N-bein formulation during
elimination of second class constraints.

\section{The proof that connections can be treated as independent variables}

First of all, to make the analysis more transparent, using integration by
parts, we rewrite the Lagrangian (\ref{eqnPR1}) in the form%

\begin{equation}
L\left(  e_{\mu\left(  \alpha\right)  },\omega_{\mu\left(  \alpha\beta\right)
}\right)  =eB^{\gamma\left(  \rho\right)  \mu\left(  \alpha\right)  \nu\left(
\beta\right)  }e_{\gamma\left(  \rho\right)  ,\mu}\omega_{\nu\left(
\alpha\beta\right)  }-eA^{\mu\left(  \alpha\right)  \nu\left(  \beta\right)
}\omega_{\mu\left(  \alpha\gamma\right)  }\omega_{\nu~~\beta)}^{~(\gamma},
\label{eqnPR5}%
\end{equation}
where%

\begin{equation}
B^{\gamma\left(  \rho\right)  \mu\left(  \alpha\right)  \nu\left(
\beta\right)  }=e^{\gamma\left(  \rho\right)  }A^{\mu\left(  \alpha\right)
\nu\left(  \beta\right)  }+e^{\gamma\left(  \alpha\right)  }A^{\mu\left(
\beta\right)  \nu\left(  \rho\right)  }+e^{\gamma\left(  \beta\right)  }%
A^{\mu\left(  \rho\right)  \nu\left(  \alpha\right)  }. \label{eqnPR6}%
\end{equation}

Of course, equations of motion for (\ref{eqnPR1}) and (\ref{eqnPR5}) remain
the same and variation with respect to a connection is%

\begin{equation}
\frac{\delta L}{\delta\omega_{\sigma\left(  \lambda\tau\right)  }}%
=-eA^{\sigma\left(  \lambda\right)  \nu\left(  \beta\right)  }\omega
_{\nu..\beta)}^{..(\tau}+eA^{\sigma\left(  \tau\right)  \nu\left(
\beta\right)  }\omega_{\nu..\beta)}^{..(\lambda}+eB^{\gamma\left(
\rho\right)  \mu\left(  \lambda\right)  \sigma\left(  \tau\right)  }%
e_{\gamma\left(  \rho\right)  ,\mu}=0. \label{eqnPR6a}%
\end{equation}

In case of affine-metric gravity, to solve similar equations one has to have
all free indices in one position (covariant or contravariant). For the N-bein
Lagrangian it is more complicated because we have indices of different nature
(internal and external) and cannot use permutation of indices immediately,
like in Einstein's solution \cite{Einstein}. So, the first step is to consider
combinations with the same indices and try to solve for them. To accomplish
this we introduce the following combinations%

\begin{equation}
\tilde{\omega}_{.....\beta)}^{\lambda(\tau}\equiv e^{\nu\left(  \lambda
\right)  }\omega_{\nu...\beta)}^{..(\tau}. \label{eqnPR7}%
\end{equation}
Using (\ref{eqnPR7}) the variation (\ref{eqnPR6a}) can be rewritten as%

\begin{equation}
-e^{\sigma\left(  \lambda\right)  }\tilde{\omega}_{.....\beta)}^{\beta(\tau
}+e^{\sigma\left(  \beta\right)  }\tilde{\omega}_{.....\beta)}^{\lambda(\tau
}+e^{\sigma\left(  \tau\right)  }\tilde{\omega}_{.....\beta)}^{\beta(\lambda
}-e^{\sigma\left(  \beta\right)  }\tilde{\omega}_{....\beta)}^{\tau(\lambda
}=D^{\sigma\left(  \lambda\tau\right)  } \label{eqnPR8}%
\end{equation}
where%

\begin{equation}
D^{\sigma\left(  \lambda\tau\right)  }\equiv-B^{\gamma\left(  \rho\right)
\mu\left(  \lambda\right)  \sigma\left(  \tau\right)  }e_{\gamma\left(
\rho\right)  ,\mu}. \label{eqnPR9}%
\end{equation}

Now, contracting with $e_{\sigma}^{\left(  \varepsilon\right)  }$ to eliminate
only one external index that is left and to have all free indices of the same
nature (internal only, in our case), we obtain%

\begin{equation}
-\tilde{\eta}^{\varepsilon\lambda}\tilde{\omega}_{.....\beta)}^{\beta(\tau
}+\tilde{\omega}^{\lambda\left(  \tau\varepsilon\right)  }+\tilde{\eta
}^{\varepsilon\tau}\tilde{\omega}_{.....\beta)}^{\beta(\lambda}-\tilde{\omega
}^{\tau\left(  \lambda\varepsilon\right)  }=\tilde{D}^{\varepsilon\left(
\lambda\tau\right)  }, \label{eqnPR10}%
\end{equation}
where%

\begin{equation}
\tilde{D}^{\varepsilon\left(  \lambda\tau\right)  }\equiv e_{\sigma}^{\left(
\varepsilon\right)  }D^{\sigma\left(  \lambda\tau\right)  }. \label{eqnPR11}%
\end{equation}

If we have only the second and fourth terms in equation (\ref{eqnPR10}) then
it can be solved using Einstein's permutation \cite{Einstein} (see below),
however we have two terms of lower tensorial dimensions with only one free
index $\tilde{\omega}_{.....\beta)}^{\beta(\tau}$ (first and third terms). To
find these \textquotedblleft traces\textquotedblright\ we contract
(\ref{eqnPR10}) with $\tilde{\eta}_{\varepsilon\lambda}$%

\begin{equation}
\left(  D-2\right)  \tilde{\omega}_{.....\beta)}^{\beta(\tau}=\tilde{D}%
_{\beta}^{..\left(  \tau\beta\right)  }. \label{eqnPR12}%
\end{equation}

In two dimensions this equation cannot be solved and, as a result, connections
can be treated as independent fields only for $D>2$. In higher than two
dimensions we have%

\begin{equation}
\tilde{\omega}_{.....\beta)}^{\beta(\tau}=\frac{1}{D-2}\tilde{D}_{\beta
}^{..\left(  \tau\beta\right)  }. \label{eqnPR14}%
\end{equation}
Upon substitution of (\ref{eqnPR14}), equation (\ref{eqnPR10}) becomes%

\begin{equation}
+\tilde{\omega}^{\lambda\left(  \tau\varepsilon\right)  }-\tilde{\omega}%
^{\tau\left(  \lambda\varepsilon\right)  }=\tilde{D}^{\prime\varepsilon\left(
\lambda\tau\right)  } \label{eqnPR15}%
\end{equation}
where%

\begin{equation}
\tilde{D}^{\prime\varepsilon\left(  \lambda\tau\right)  }=\tilde
{D}^{\varepsilon\left(  \lambda\tau\right)  }+\tilde{\eta}^{\varepsilon
\lambda}\frac{1}{D-2}\tilde{D}_{\sigma}^{..\left(  \tau\sigma\right)  }%
-\tilde{\eta}^{\varepsilon\tau}\frac{1}{D-2}\tilde{D}_{\sigma}^{..\left(
\lambda\sigma\right)  }. \label{eqnPR16}%
\end{equation}

Now, performing permutation of indices in (\ref{eqnPR15}) $\left(  \lambda
\tau\varepsilon\right)  +$ $\left(  \varepsilon\lambda\tau\right)  -\left(
\tau\varepsilon\lambda\right)  $ (in the way as was done by Einstein in
\cite{Einstein}, we obtain the solution%

\begin{equation}
2\tilde{\omega}^{\lambda\left(  \tau\varepsilon\right)  }=\tilde{D}%
^{\prime\varepsilon\left(  \lambda\tau\right)  }+\tilde{D}^{\prime\tau\left(
\varepsilon\lambda\right)  }-\tilde{D}^{^{\prime}\lambda\left(  \varepsilon
\tau\right)  }. \label{eqnPR17}%
\end{equation}
To find an explicit form of this solution we, using properties (\ref{eqnPR6})
and definition (\ref{eqnPR11}), obtain%

\begin{equation}
\tilde{D}^{\varepsilon\left(  \lambda\tau\right)  }=-\tilde{\eta}%
^{\varepsilon\tau}A^{\mu\left(  \lambda\right)  \gamma\left(  \rho\right)
}e_{\gamma\left(  \rho\right)  ,\mu}-\tilde{\eta}^{\varepsilon\lambda}%
A^{\mu\left(  \rho\right)  \gamma\left(  \tau\right)  }e_{\gamma\left(
\rho\right)  ,\mu}-\tilde{\eta}^{\varepsilon\rho}A^{\mu\left(  \tau\right)
\gamma\left(  \lambda\right)  }e_{\gamma\left(  \rho\right)  ,\mu}.
\label{eqnPR18}%
\end{equation}
The \textquotedblleft trace\textquotedblright\ of this combination is%

\begin{equation}
\tilde{D}_{\varepsilon}^{..\left(  \lambda\varepsilon\right)  }=-\left(
D-2\right)  A^{\mu\left(  \lambda\right)  \gamma\left(  \rho\right)
}e_{\gamma\left(  \rho\right)  ,\mu} \label{eqnPR19}%
\end{equation}
and, using definition of $\tilde{D}^{\prime\varepsilon\left(  \lambda
\tau\right)  }$ (\ref{eqnPR16}), it immediately follows%

\begin{equation}
\tilde{D}^{\prime\varepsilon\left(  \lambda\tau\right)  }=-A^{\mu\left(
\tau\right)  \gamma\left(  \lambda\right)  }e_{\gamma,\mu}^{\left(
\varepsilon\right)  } \label{eqnPR20}%
\end{equation}
that upon substitution into (\ref{eqnPR17}) gives us the solution for
$\tilde{\omega}^{\lambda\left(  \tau\varepsilon\right)  },$ or contracting
with $e_{\nu\left(  \varepsilon\right)  }$ (see (\ref{eqnPR7}))%

\begin{equation}
e_{\nu\left(  \varepsilon\right)  }\tilde{\omega}^{\varepsilon\left(
\lambda\tau\right)  }=\omega_{\nu}^{~~\left(  \mu\gamma\right)  }=\frac{1}%
{2}e_{\nu\left(  \varepsilon\right)  }\left(  A^{\mu\left(  \tau\right)
\gamma\left(  \varepsilon\right)  }e_{\gamma,\mu}^{\left(  \lambda\right)
}-A^{\mu\left(  \lambda\right)  \gamma\left(  \varepsilon\right)  }%
e_{\gamma,\mu}^{\left(  \tau\right)  }-A^{\mu\left(  \lambda\right)
\gamma\left(  \tau\right)  }e_{\gamma,\mu}^{\left(  \varepsilon\right)
}\right)  \label{eqnPR21}%
\end{equation}
which is definition of a connection $\omega_{\nu}^{~~\left(  \mu\gamma\right)
}$. In all higher than two dimensions $\omega_{\nu\left(  \alpha\beta\right)
}$ can be treated as an independent variable and equivalence of two
Lagrangians, $L\left(  e_{\mu\left(  \alpha\right)  },\omega_{\mu\left(
\alpha\beta\right)  }\right)  $ and $L\left(  e_{\mu\left(  \alpha\right)
}\right)  $, is established.

\section{The N-bein Hamiltonian}

As for any first order formulation (at most linear in \textquotedblleft
velocities\textquotedblright\ - time derivatives of fields), the first step of
the Hamiltonian formulation is strikingly simple. Separating terms with
\textquotedblleft velocities\textquotedblright\ in our (\ref{eqnPR5})%

\begin{equation}
L=eB^{\gamma\left(  \rho\right)  0\left(  \alpha\right)  \nu\left(
\beta\right)  }e_{\gamma\left(  \rho\right)  ,0}\omega_{\nu\left(  \alpha
\beta\right)  }+eB^{\gamma\left(  \rho\right)  k\left(  \alpha\right)
\nu\left(  \beta\right)  }e_{\gamma\left(  \rho\right)  ,k}\omega_{\nu\left(
\alpha\beta\right)  }-eA^{\mu\left(  \alpha\right)  \nu\left(  \beta\right)
}\omega_{\mu\left(  \alpha\gamma\right)  }\omega_{\nu~~\beta)}^{~(\gamma}
\label{eqnPR30}%
\end{equation}
we can just read off the total Hamiltonian%

\[
H_{T}=\underset{\phi^{0\left(  \rho\right)  }}{\underbrace{\pi^{0\left(
\rho\right)  }}}\dot{e}_{0\left(  \rho\right)  }+\underset{\phi^{k\left(
\rho\right)  }}{\underbrace{\left(  \pi^{k\left(  \rho\right)  }-eB^{k\left(
\rho\right)  0\left(  \alpha\right)  m\left(  \beta\right)  }\omega_{m\left(
\alpha\beta\right)  }\right)  }}\dot{e}_{k\left(  \rho\right)  }%
+\underset{\Phi^{\mu\left(  \alpha\beta\right)  }}{\underbrace{\Pi^{\mu\left(
\alpha\beta\right)  }}}\dot{\omega}_{\mu\left(  \alpha\beta\right)  }-
\]

\begin{equation}
\underset{H_{c}=-L\left(  \text{part without \textquotedblleft
velocities\textquotedblright}\right)  }{\underbrace{-eB^{\gamma\left(
\rho\right)  k\left(  \alpha\right)  \nu\left(  \beta\right)  }e_{\gamma
\left(  \rho\right)  ,k}\omega_{\nu\left(  \alpha\beta\right)  }%
+eA^{\mu\left(  \alpha\right)  \nu\left(  \beta\right)  }\omega_{\mu\left(
\alpha\gamma\right)  }\omega_{\nu~~\beta)}^{~(\gamma}}} \label{eqnPR31}%
\end{equation}
where $\pi^{\mu\left(  \rho\right)  }$ and $\Pi^{\mu\left(  \alpha
\beta\right)  }$ are momenta conjugate to N-beins and connections. As in any
Hamiltonian formulation of a first order action, the number of primary
constraints ($\phi^{\mu\left(  \rho\right)  }$, $\Phi^{\mu\left(  \alpha
\beta\right)  }$) equals the number of independent variables, or canonical
variables, with the fundamental Poisson brackets (PB)%

\begin{equation}
\left\{  e_{\mu\left(  \alpha\right)  }\left(  \mathbf{x}\right)  ,\pi
^{\gamma\left(  \rho\right)  }\left(  \mathbf{y}\right)  \right\}
=\delta_{\mu}^{\gamma}\tilde{\delta}_{\alpha}^{\rho}\delta\left(
\mathbf{x}-\mathbf{y}\right)  ,\text{ \ \ }\left\{  \omega_{\lambda\left(
\alpha\beta\right)  }\left(  \mathbf{x}\right)  ,\Pi^{\rho\left(  \mu
\nu\right)  }\left(  \mathbf{y}\right)  \right\}  =\tilde{\Delta}_{\left(
\alpha\beta\right)  }^{\left(  \mu\nu\right)  }\delta_{\lambda}^{\rho}%
\delta\left(  \mathbf{x}-\mathbf{y}\right)  \label{eqnPR32}%
\end{equation}
where%

\begin{equation}
\tilde{\Delta}_{\left(  \alpha\beta\right)  }^{\left(  \mu\nu\right)  }%
\equiv\frac{1}{2}\left(  \tilde{\delta}_{\alpha}^{\mu}\tilde{\delta}_{\beta
}^{\nu}-\tilde{\delta}_{\alpha}^{\nu}\tilde{\delta}_{\beta}^{\mu}\right)  .
\label{eqnPR33}%
\end{equation}
The rest of PB are zero. (In the text we often write a PB without the factor
$\delta\left(  \mathbf{x}-\mathbf{y}\right)  $).

Note that properties of antisymmetry (for any pair of internal and external
indices) of coefficient functions $A,B$ are very helpful in calculations. For
example, the absence of term with connections in primary constraint
$\phi^{0\left(  \rho\right)  }$ is just a consequence of this property,
because in this case we have $B^{0\left(  \rho\right)  0\left(  \alpha\right)
m\left(  \beta\right)  }$ with two equal indices, and so it is zero.
Similarly, calculation of PBs among primary constraints $\phi^{\mu\left(
\rho\right)  }$ is almost obvious if similar properties of next generation of
coefficient functions is used. We have already introduced $B$ which is the
result of following variation%

\begin{equation}
\frac{\delta}{\delta e_{\gamma\left(  \rho\right)  }}\left(  eA^{\mu\left(
\alpha\right)  \nu\left(  \beta\right)  }\right)  =eB^{\gamma\left(
\rho\right)  \mu\left(  \alpha\right)  \nu\left(  \beta\right)  }
\label{eqnPR36}%
\end{equation}
(it can be expressed in terms of $A$ (see (\ref{eqnPR6}))). Next generation of
such functions is%

\begin{equation}
\frac{\delta}{\delta e_{\sigma\left(  \tau\right)  }}\left(  eB^{\gamma\left(
\rho\right)  \mu\left(  \alpha\right)  \nu\left(  \beta\right)  }\right)
=eC^{\sigma\left(  \tau\right)  \gamma\left(  \rho\right)  \mu\left(
\alpha\right)  \nu\left(  \beta\right)  } \label{eqnPR37}%
\end{equation}
where $C$ is again antisymmetric function in any pair of internal or external
indices and can be expressed in terms of $B$ (in the same way as $B$ is
expressed in terms of $A$ in (\ref{eqnPR6}))%

\[
C^{\sigma\left(  \tau\right)  \gamma\left(  \rho\right)  \mu\left(
\alpha\right)  \nu\left(  \beta\right)  }=
\]

\begin{equation}
e^{\sigma\left(  \tau\right)  }B^{\gamma\left(  \rho\right)  \mu\left(
\alpha\right)  \nu\left(  \beta\right)  }-e^{\sigma\left(  \rho\right)
}B^{\gamma\left(  \tau\right)  \mu\left(  \alpha\right)  \nu\left(
\beta\right)  }+e^{\sigma\left(  \alpha\right)  }B^{\gamma\left(
\beta\right)  \mu\left(  \tau\right)  \nu\left(  \rho\right)  }-e^{\sigma
\left(  \beta\right)  }B^{\gamma\left(  \alpha\right)  \mu\left(  \tau\right)
\nu\left(  \rho\right)  }. \label{eqnPR38}%
\end{equation}

These relations (\ref{eqnPR36}, \ref{eqnPR37}) and antisymmetry of these
functions will be very helpful for further calculations; we will call them
$ABC$ properties. For calculation of PBs among primary constraints $\phi
^{\mu\left(  \rho\right)  }$ the antisymmetry of $C$ is sufficient%

\begin{equation}
\left\{  \phi^{\mu\left(  \sigma\right)  },\phi^{\nu\left(  \rho\right)
}\right\}  =eC^{\mu\left(  \sigma\right)  \nu\left(  \rho\right)  0\left(
\alpha^{\prime\prime}\right)  m^{\prime\prime}\left(  \beta^{\prime\prime
}\right)  }\omega_{m^{\prime\prime}\left(  \alpha^{\prime\prime}\beta
^{\prime\prime}\right)  }-eC^{\nu\left(  \rho\right)  \mu\left(
\sigma\right)  0\left(  \alpha^{\prime}\right)  m^{\prime}\left(
\beta^{\prime}\right)  }\omega_{m^{\prime}\left(  \alpha^{\prime}\beta
^{\prime}\right)  }=0. \label{eqnPR39}%
\end{equation}
The rest of PBs among primary constraints, $\left\{  \Pi^{\mu\left(
\alpha\beta\right)  },\Pi^{\mu\left(  \alpha\beta\right)  }\right\}  =\left\{
\Pi^{\mu\left(  \alpha\beta\right)  },\pi^{0\left(  \rho\right)  }\right\}
=\left\{  \pi^{0\left(  \rho\right)  },\pi^{0\left(  \rho\right)  }\right\}
=0$, follows just from definition of the fundamental PB (\ref{eqnPR32}). The
only non-zero PB among primary constraints is%

\begin{equation}
\left\{  \phi^{k\left(  \rho\right)  },\Phi^{m\left(  \alpha\beta\right)
}\right\}  =-eB^{k\left(  \rho\right)  0\left(  \alpha\right)  m\left(
\beta\right)  }. \label{eqnPR40}%
\end{equation}

Based on this simple analysis, it is clear that only $\phi^{0\left(
\rho\right)  }$ and $\Phi^{0\left(  \alpha\beta\right)  }$ are candidates for
first class constraints in any dimension higher than two (this is the case
also in $3D$ \cite{3D}). The gauge invariance is derivable if algebra of PBs
among all first class constraints is known, an explicit form of
transformations depends on this algebra and form of constraints. However,
gauge parameters of transformations are defined by primary first class
constraints only. It is clear from the first and very simple step of the Dirac
procedure that for the primary first class constraints $\phi^{0\left(
\rho\right)  }$ and $\Phi^{0\left(  \alpha\beta\right)  }$ the only possible
gauge parameters are $t_{\left(  \rho\right)  }$ and $r_{\left(  \alpha
\beta\right)  }$. There are no primary first class constraints that allow to
have any parameter with an external index. As a result, diffeomorphism
invariance that needs such a parameter, $\xi_{\mu}$, cannot be the
\textit{gauge symmetry} of N-bein gravity in any dimension. In works claiming
that diffeomorphism is the gauge symmetry of N-bein gravity (even in a
particular dimension) by referring to \textquotedblleft
results\textquotedblright\ of the Hamiltonian formulation with the
\textquotedblleft diffeomorphism constraint\textquotedblright\ (spatial or
full, it does not matter) the non-canonical change of variables must be
performed. So, any connection with the Einstein-Cartan formulation would be
lost. Of course, such theories which differ from the original ones, despite
that they are obtained by abandoning mathematical rules of ordinary mechanics,
can be still considered as some toy models. We do not see any reason why the
original formulation of Einstein should be abandoned and, in general, we doubt
that any toy obtained by non-canonical transformations of a theory which has a
mathematical beauty and experimental conformations has even small hope to
produce any meaningful result. Contrary, a bad theory, in principle, can be
\textquotedblleft converted\textquotedblright\ into a good one by
non-canonical transformations, because equivalence with bad one is lost, but
such an approach is not scientific and chances to win such a \textquotedblleft
game\textquotedblright\ are infinitesimally small.

Our interest is to find the Hamiltonian formulation of the original
Einstein-Cartan theory and we continue with the next step of the Dirac procedure.

\section{Time development of primary constraints}

The next step of the Dirac procedure is time development of primary
constraints which is a PB of constraints with the total Hamiltonian
(\ref{eqnPR31})%

\begin{equation}
\dot{\phi}^{\mu\left(  \rho\right)  }=\left\{  \phi^{\mu\left(  \rho\right)
},H_{T}\right\}  , \label{eqnPR50}%
\end{equation}
and similarly for $\Phi^{\mu\left(  \alpha\beta\right)  }$. Note that PBs are
calculated with the total Hamiltonian which includes all primary constraints.
It is customary to call \textquotedblleft velocities\textquotedblright\ in
front of primary constraints as \textquotedblleft undetermined
multipliers\textquotedblright\ but they are rarely used. In case of N-bein
gravity the true \textquotedblleft multiplier\textquotedblright\ nature of
this coefficients (actually, they are the Lagrange multipliers for the second
class constraints) becomes important which we clarify in Section VI. So, to
make this more transparent, we rewrite the Hamiltonian (\ref{eqnPR31}) in the
following form%

\begin{equation}
H_{T}=H_{c}+\lambda_{\mu\left(  \rho\right)  }\phi^{\mu\left(  \rho\right)
}+\Lambda_{\mu\left(  \alpha\beta\right)  }\Phi^{\mu\left(  \alpha
\beta\right)  }. \label{eqnPR51}%
\end{equation}

Considering time development of primary constraints we classify them into
three distinct groups.

\textbf{First group} consist of $\phi^{k\left(  m\right)  }$ and
$\Phi^{p\left(  k0\right)  }$ (see (\ref{eqnPR31}))%

\begin{equation}
\phi^{k\left(  m\right)  }=\pi^{k\left(  m\right)  }-2eB^{k\left(  m\right)
0\left(  q\right)  p\left(  0\right)  }\omega_{p\left(  q0\right)  }%
\underset{=0\text{ (if }D=3\text{)}}{\underbrace{-eB^{k\left(  m\right)
0\left(  p\right)  n\left(  q\right)  }\omega_{n\left(  pq\right)  }}%
}=0,\text{ \ \ \ \ } \label{eqnPR52}%
\end{equation}

\begin{equation}
\Phi^{p\left(  k0\right)  }=\Pi^{p\left(  k0\right)  }=0 \label{eqnPR53}%
\end{equation}
and their time development%

\begin{equation}
\dot{\Pi}^{p\left(  m0\right)  }=-\frac{\delta H_{c}}{\delta\omega_{p\left(
m0\right)  }}+\lambda_{k\left(  q\right)  }eB^{k\left(  q\right)  0\left(
m\right)  p\left(  0\right)  }, \label{eqnPR54}%
\end{equation}

\begin{equation}
\dot{\phi}^{k\left(  m\right)  }=-\frac{\delta H_{c}}{\delta e_{k\left(
m\right)  }}-2\Lambda_{p\left(  q0\right)  }eB^{k\left(  m\right)  0\left(
q\right)  p\left(  0\right)  }\underset{=0\text{ (if }D=3\text{)}}%
{\underbrace{-\Lambda_{p\left(  nq\right)  }eB^{k\left(  m\right)  0\left(
n\right)  p\left(  q\right)  }}.} \label{eqnPR55}%
\end{equation}

In this group of equations there are extra terms compare with three
dimensional case \cite{3D}. However, for this group such terms neither affect
possibility to solve nor the way of solving these equations for $\Pi^{p\left(
k0\right)  }$, $\omega_{p\left(  q0\right)  }$, separately or together, with
the corresponding to them multipliers $\lambda_{k\left(  q\right)  }$,
$\Lambda_{p\left(  q0\right)  }$ (determined) in all dimensions. Of course, as
in many known cases we can use a short cut and just solve pair of equations
(\ref{eqnPR52}-\ref{eqnPR53}) for $\Pi^{p\left(  k0\right)  }$, $\omega
_{p\left(  q0\right)  }$, and substitute the solution into $H_{c}$ and the
rest of constraints. This is the Hamiltonian reduction - elimination of a pair
of canonical variables by solving a pair of second class constraints (this is
what we did in \cite{3D}). Of course, we have to be careful and calculate the
Dirac brackets which after such eliminations quite often coincide with Poisson
brackets. In the extended form (keeping multipliers and the corresponding
equations), constraints are second class if we can solve these equations for
the corresponding to them multipliers.

Note that coefficients in front of all fields that we want to find,
$\omega_{p\left(  q0\right)  }$, $\lambda_{k\left(  q\right)  },$ and
$\Lambda_{p\left(  q0\right)  }$ ($\Pi^{p\left(  k0\right)  }=0$ is trivial),
are $B-$functions of the same structure:%

\begin{equation}
B^{k\left(  m\right)  0\left(  q\right)  p\left(  0\right)  }. \label{eqnPR56}%
\end{equation}

To solve all above equations, we need the inverse for this particular
combination. It was found in \cite{3D} that there is a combination (not a new
variable), similar to those used by Dirac $\left(  \gamma^{km}=g^{km}%
-\frac{g^{k0}g^{m0}}{g^{00}}\text{, }\gamma^{km}g_{mn}=\delta_{n}^{k}\right)
$ when he considered Hamiltonian formulation of second order metric gravity
\cite{DiracP}. For N-bein gravity it is%

\begin{equation}
\gamma^{k\left(  m\right)  }\equiv e^{k\left(  m\right)  }-\frac{e^{k\left(
0\right)  }e^{0\left(  m\right)  }}{e^{0\left(  0\right)  }} \label{eqnPR57}%
\end{equation}
with properties%

\begin{equation}
\gamma^{m\left(  p\right)  }e_{m\left(  q\right)  }=\tilde{\delta}_{q}%
^{p},\text{ \ \ \ \ \ \ }\gamma^{n\left(  q\right)  }e_{m\left(  q\right)
}=\delta_{m}^{n}. \label{eqnPR58}%
\end{equation}
that allow to rewrite%

\begin{equation}
B^{0\left(  0\right)  k\left(  q\right)  p\left(  m\right)  }=e^{0\left(
0\right)  }E^{k\left(  q\right)  p\left(  m\right)  } \label{eqnPR59}%
\end{equation}
where%

\begin{equation}
E^{k\left(  m\right)  p\left(  q\right)  }\equiv\gamma^{k\left(  m\right)
}\gamma^{p\left(  q\right)  }-\gamma^{k\left(  q\right)  }\gamma^{p\left(
m\right)  }. \label{eqnPR60}%
\end{equation}

Note that all $B$-coefficient in (\ref{eqnPR52}), (\ref{eqnPR54}) and
(\ref{eqnPR55}) by permutations of indices can be converted in such a form.
For any dimension $D>2$ (which is consistent with restriction on possibility
to solve for connections (\ref{eqnPR14})) we can find the inverse of
$E^{k\left(  m\right)  p\left(  q\right)  }$%

\begin{equation}
I_{m\left(  q\right)  a\left(  b\right)  }\equiv\frac{1}{D-2}e_{m\left(
q\right)  }e_{a\left(  b\right)  }-e_{m\left(  b\right)  }e_{a\left(
q\right)  }, \label{eqnPR61}%
\end{equation}

\begin{equation}
I_{m\left(  q\right)  a\left(  b\right)  }E^{a\left(  b\right)  n\left(
p\right)  }=E^{n\left(  p\right)  a\left(  b\right)  }I_{a\left(  b\right)
m\left(  q\right)  }=\delta_{m}^{n}\tilde{\delta}_{q}^{p}. \label{eqnPR62}%
\end{equation}
So, for example, we have%

\begin{equation}
\omega_{k\left(  q0\right)  }=-\frac{1}{2ee^{0\left(  0\right)  }}I_{k\left(
q\right)  m\left(  p\right)  }\pi^{m\left(  p\right)  }+\frac{1}{2e^{0\left(
0\right)  }}I_{k\left(  q\right)  m\left(  p\right)  }B^{m\left(  p\right)
0\left(  a\right)  n\left(  b\right)  }\omega_{n\left(  ab\right)  }
\label{eqnPR63}%
\end{equation}
and%

\begin{equation}
\lambda_{a\left(  b\right)  }=-I_{a\left(  b\right)  p\left(  m\right)  }%
\frac{1}{ee^{0\left(  0\right)  }}\frac{\delta H_{c}}{\delta\omega_{p\left(
m0\right)  }}. \label{eqnPR64}%
\end{equation}
These solutions suggest to rewrite all $B$ with different combinations of
indices in such \textquotedblleft$\gamma-E$-form\textquotedblright. Using the
identity ((\ref{eqnPR6}), (\ref{eqnPR57}) and (\ref{eqnPR60}) has to be used
to prove it)%

\begin{equation}
B^{0\left(  p\right)  m\left(  a\right)  n\left(  b\right)  }=e^{0\left(
p\right)  }E^{m\left(  a\right)  n\left(  b\right)  }+e^{0\left(  a\right)
}E^{m\left(  b\right)  n\left(  p\right)  }+e^{0\left(  b\right)  }E^{m\left(
p\right)  n\left(  a\right)  } \label{eqnPR65}%
\end{equation}
we can present (\ref{eqnPR63}) in the form%

\begin{equation}
\omega_{k\left(  q0\right)  }=-\frac{1}{2ee^{0\left(  0\right)  }}I_{k\left(
q\right)  m\left(  p\right)  }\pi^{m\left(  p\right)  }-\frac{e^{0\left(
p\right)  }}{2e^{0\left(  0\right)  }}I_{k\left(  q\right)  m\left(  p\right)
}E^{m\left(  a\right)  n\left(  b\right)  }\omega_{n\left(  ab\right)  }%
+\frac{e^{0\left(  a\right)  }}{e^{0\left(  0\right)  }}\omega_{k\left(
aq\right)  }. \label{eqnPR66}%
\end{equation}

This is general solution valid in all dimensions and $3D$ form of it is just a
limited case. In $3D$ all spatial components have only two values, $1,2,$ for
both internal and external indices and, in addition, $I_{k\left(  q\right)
m\left(  p\right)  }$ becomes antisymmetric as $A,B,C$ and $E$. Taking this
into consideration, direct but simple calculations for all four possible
components $\omega_{1\left(  10\right)  }$, $\omega_{2\left(  20\right)  }$,
$\omega_{1\left(  20\right)  }$ and $\omega_{2\left(  10\right)  }$ lead to
cancellation of two last terms. So, (\ref{eqnPR66}) gives the same
$\omega_{k\left(  q0\right)  }$ for $D=3$ as was found in \cite{3D}. For the
Hamiltonian method a separate treatment of $3D$ case is not needed and all
results can be obtained from a general solution. We demonstrated this for
(\ref{eqnPR66}) but it is also valid for other results, as it will be clear
from our further calculations.

Now we consider \textbf{second group} of primary constraints $\phi^{k\left(
0\right)  }$ and $\Phi^{p\left(  km\right)  }$%

\begin{equation}
\phi^{k\left(  0\right)  }=\pi^{k\left(  0\right)  }-eB^{k\left(  0\right)
0\left(  p\right)  m\left(  q\right)  }\omega_{m\left(  pq\right)  }=0,\text{
\ \ \ \ } \label{eqnPR67}%
\end{equation}

\begin{equation}
\Phi^{p\left(  km\right)  }=\Pi^{p\left(  km\right)  }=0 \label{eqnPR68}%
\end{equation}
and their time development%

\begin{equation}
\dot{\phi}^{k\left(  0\right)  }=-\frac{\delta H_{c}}{\delta e_{k\left(
0\right)  }}-\Lambda_{m\left(  pq\right)  }eB^{k\left(  0\right)  0\left(
p\right)  m\left(  q\right)  }, \label{eqnPR69}%
\end{equation}

\begin{equation}
\dot{\Pi}^{p\left(  mn\right)  }=-\frac{\delta H_{c}}{\delta\omega_{p\left(
mn\right)  }}+\underset{=0\text{ (if }D=3\text{)}}{\underbrace{\lambda
_{k\left(  q\right)  }eB^{k\left(  q\right)  0\left(  m\right)  p\left(
n\right)  }}}+\lambda_{k\left(  0\right)  }eB^{k\left(  0\right)  0\left(
m\right)  p\left(  n\right)  }. \label{eqnPR70}%
\end{equation}

In general case neither a short cut (the system (\ref{eqnPR67}-\ref{eqnPR68}))
nor all equations including the corresponding multipliers (\ref{eqnPR67}%
-\ref{eqnPR70}) can be solved because the number of equations is smaller than
the number of unknowns. The only exception here is $3D$ case where the number
of equations and unknowns is the same because the number of independent
components for all $\omega_{m\left(  pq\right)  }$, $\Pi^{p\left(  km\right)
}$, $\lambda_{k\left(  0\right)  },$ and $\Lambda_{m\left(  pq\right)  }$ is
just two \cite{3D}. Note that in $3D$ case this group is completely decoupled
from the first one (\ref{eqnPR52}-\ref{eqnPR55}). In higher than three
dimensions we have next generation of constraints (secondary)%

\begin{equation}
\chi^{p\left(  mn\right)  }=-\frac{\delta H_{c}}{\delta\omega_{p\left(
mn\right)  }} \label{eqnPR71}%
\end{equation}
and only after calculation of its time development we have enough equations to
find all fields and associated with them multipliers. We will solve these
equations in Section VI and illustrate again that a separate treatment of $3D$
is not needed because $3D-$limit of a general solution gives the same result
as \cite{3D}, as was demonstrated for (\ref{eqnPR66}).

Finally we consider the \textbf{third group} of primary constraints
$\Pi^{0\left(  \alpha\beta\right)  }$ and $\pi^{0\left(  \rho\right)  }.$ Time
development of them is
\begin{equation}
\dot{\Pi}^{0\left(  \alpha\beta\right)  }=eB^{m\left(  \rho\right)  k\left(
\alpha\right)  0\left(  \beta\right)  }e_{m\left(  \rho\right)  ,k}%
-eA^{0\left(  \alpha\right)  k\left(  \beta^{\prime}\right)  }\omega
_{k~~\beta^{\prime})}^{~(\beta}+eA^{0\left(  \beta\right)  k\left(
\beta^{\prime}\right)  }\omega_{k~~\beta^{\prime})}^{~(\alpha}=\chi^{0\left(
\alpha\beta\right)  }, \label{eqnPR72}%
\end{equation}

\[
\dot{\pi}^{0\left(  \rho\right)  }=\frac{\delta}{\delta e_{0\left(
\rho\right)  }}\left(  \underset{=0\text{ (if }D=3\text{)}}{\underbrace
{eB^{n\left(  \nu\right)  k\left(  \alpha\right)  m\left(  \beta\right)  }}%
}\right)  e_{n\left(  \nu\right)  ,k}\omega_{m\left(  \alpha\beta\right)
}-\underset{=0\text{ (if }D=3\text{ and }\rho,\alpha,\beta\neq0\text{) }%
}{\underbrace{\left(  eB^{0\left(  \rho\right)  k\left(  \alpha\right)
m\left(  \beta\right)  }\omega_{m\left(  \alpha\beta\right)  }\right)  _{,k}}}%
\]

\begin{equation}
-\frac{\delta}{\delta e_{0\left(  \rho\right)  }}\underset{=0\text{ (if
}D=3\text{)}}{\underbrace{eA^{k\left(  p\right)  m\left(  q\right)  }%
\omega_{k\left(  pn\right)  }\omega_{m~~q)}^{~(n}}}-\frac{\delta}{\delta
e_{0\left(  \rho\right)  }}2eA^{k\left(  p\right)  m\left(  0\right)  }%
\omega_{k\left(  pn\right)  }\omega_{m~~0)}^{~(n} \label{eqnPR73}%
\end{equation}

\[
-\underset{=0\text{ (if }D=3\text{ and }\rho\neq0\text{)}}{\underbrace
{\frac{\delta}{\delta e_{0\left(  \rho\right)  }}eA^{k\left(  p\right)
m\left(  q\right)  }\omega_{k\left(  p0\right)  }\omega_{m~~q)}^{~(0}}}%
=\chi^{0\left(  \rho\right)  }.
\]

Note that both equations (\ref{eqnPR72}) and (\ref{eqnPR73}) do not have
multipliers (the consequence of zero PBs of $\Pi^{0\left(  \alpha\beta\right)
}$ and $\pi^{0\left(  \rho\right)  }$ with the rest of primary constraints).
So, time development of $\Pi^{0\left(  \alpha\beta\right)  }$ and
$\pi^{0\left(  \rho\right)  }$ leads to secondary constraints (we cannot find
the corresponding multipliers). Note also that the temporal connections
($\omega_{0\left(  \alpha\beta\right)  }$) are absent in both constraints, so
time development of the secondary constraints $\chi^{0\left(  \alpha
\beta\right)  }$ and $\chi^{0\left(  \rho\right)  }$cannot give us an equation
to find $\Lambda_{0\left(  \alpha\beta\right)  }$ (PBs of primary constraint
$\Pi^{0\left(  \alpha\beta\right)  }$ are zero with both secondary
constraints). PBs of both constraints with primary $\pi^{0\left(  \rho\right)
}$ are also zero (almost manifestly) and cannot give us an equation to find
$\lambda_{0\left(  \rho\right)  }$. This is the result based on $ABC$
properties. Second variation of each term in (\ref{eqnPR73}) leads to
antisymmetric combination with two external zeros, e.g.%

\begin{equation}
\frac{\delta}{\delta e_{0\left(  \sigma\right)  }}\left(  eB^{m\left(
\rho\right)  k\left(  \alpha\right)  0\left(  \beta\right)  }\right)
=eC^{0\left(  \sigma\right)  m\left(  \rho\right)  k\left(  \alpha\right)
0\left(  \beta\right)  }=0. \label{eqnPR74}%
\end{equation}

All these simple results are equivalent with three dimensional case \cite{3D}
and provide strong indication that in all dimensions PBs among primary and
secondary constraints are zero. We call this \textquotedblleft
indication\textquotedblright\ because after solving second class constraints
we have to substitute solutions for spatial connections into (\ref{eqnPR73})
that will lead to it modification. However, as we will demonstrate later, such
substitutions do not affect PBs and zero PBs among primary and secondary
constraints are unaltered. So far we have the result that looks the same in
all dimensions and equivalent with considered before $3D$ case \cite{3D}. What
is the difference in higher dimensions? The difference is obviously in the
secondary constraint $\chi^{0\left(  \rho\right)  }$ (we will call it
\textquotedblleft translational\textquotedblright,\ as in $3D$ case) where we
have many \textquotedblleft invisible\textquotedblright\ in three dimension
terms (see (\ref{eqnPR73})) that have to change drastically this constraint
and make it much richer compare with $3D$ case. Secondary rotational
constraint $\chi^{0\left(  \alpha\beta\right)  }$ is different compare to
$\chi^{0\left(  \rho\right)  }$, it does not have any three dimensional
peculiarities as $\chi^{0\left(  \rho\right)  }$ and looks absolutely the same
in three \cite{3D} and all higher dimensions.

\section{Some simple preliminary results}

If the Lagrangian with some variables is defined in all dimensions, then the
Hamiltonian analysis based on such variables should also be independent on a
particular dimension. Quite simple and straightforward first steps of the
Dirac procedure performed in previous Sections allow to make some conclusions
and perform quick calculations. It is clear that at least part of PB algebra
of first class constraints should be the same as in three dimensions, as well
as some parts of transformations found in $3D$.

Let us briefly review the results of three dimensional case. The total
Hamiltonian after elimination of second class constraints is \cite{3D}
(disregarding a total derivative)%

\begin{equation}
H_{T}=-e_{0\left(  \rho\right)  }\chi^{0\left(  \rho\right)  }-\omega
_{0\left(  \alpha\beta\right)  }\chi^{0\left(  \alpha\beta\right)  }+\dot
{e}_{0\left(  \rho\right)  }\pi^{0\left(  \rho\right)  }+\dot{\omega
}_{0\left(  \alpha\beta\right)  }\Pi^{0\left(  \alpha\beta\right)  }.
\label{eqnPR80}%
\end{equation}

Algebra of PB among constraints is the following: all PBs among primary and
secondary constraints are zero and secondary constraints have ordinary
(Poincar\'{e}) algebra as in ordinary field theories (no structure functions
or non-localities - derivatives of delta functions)%

\begin{equation}
\left\{  \chi^{0\left(  \rho\right)  },\chi^{0\left(  \gamma\right)
}\right\}  =0, \label{eqnPR81}%
\end{equation}

\begin{equation}
\left\{  \chi^{0\left(  \alpha\beta\right)  },\chi^{0\left(  \rho\right)
}\right\}  =\frac{1}{2}\eta^{\left(  \beta\right)  \left(  \rho\right)  }%
\chi^{0\left(  \alpha\right)  }-\frac{1}{2}\eta^{\left(  \alpha\right)
\left(  \rho\right)  }\chi^{0\left(  \beta\right)  }, \label{eqnPR82}%
\end{equation}

\begin{equation}
\left\{  \chi^{0\left(  \alpha\beta\right)  },\chi^{0\left(  \mu\nu\right)
}\right\}  =\frac{1}{2}\eta^{\left(  \beta\right)  \left(  \mu\right)  }%
\chi^{0\left(  \alpha\nu\right)  }-\frac{1}{2}\eta^{\left(  \alpha\right)
\left(  \mu\right)  }\chi^{0\left(  \beta\nu\right)  }+\frac{1}{2}%
\eta^{\left(  \beta\right)  \left(  \nu\right)  }\chi^{0\left(  \mu
\alpha\right)  }-\frac{1}{2}\eta^{\left(  \alpha\right)  \left(  \nu\right)
}\chi^{0\left(  \mu\beta\right)  }. \label{eqnPR83}%
\end{equation}

The simplicity of the Hamiltonian and algebra of constraints (all of them are
first class) makes derivation of generators straightforward and gauge
invariance of all independent fields immediately follows \cite{3D}. The gauge
transformations can be cast into a covariant form but we have to remember that
this is the result of calculations in three dimensional case where many terms
(see (\ref{eqnPR52}, \ref{eqnPR53}, \ref{eqnPR70}, \ref{eqnPR71},
\ref{eqnPR73})) were not taken into account. Transformations for the first
order formulation of N-bein gravity for $D>3$ should be modified.

Transformations for $3D$ are \cite{3D}%

\begin{equation}
\delta e_{\gamma\left(  \lambda\right)  }=-t_{\left(  \lambda\right)  ,\gamma
}-\omega_{\gamma(\lambda}^{~~~\rho)}t_{\left(  \rho\right)  }-\frac{1}%
{2}\left(  e_{\gamma}^{\left(  \alpha\right)  }\delta_{\left(  \lambda\right)
}^{\left(  \beta\right)  }-e_{\gamma}^{\left(  \beta\right)  }\delta_{\left(
\lambda\right)  }^{\left(  \alpha\right)  }\right)  r_{\left(  \alpha
\beta\right)  }, \label{eqnPR84}%
\end{equation}

\begin{equation}
\delta\omega_{\gamma\left(  \sigma\lambda\right)  }=-r_{\left(  \sigma
\lambda\right)  ,\gamma}-\left(  \omega_{\gamma~~\lambda)}^{~(\alpha}%
\delta_{\left(  \sigma\right)  }^{\left(  \beta\right)  }-\omega
_{\gamma~~\sigma)}^{~(\alpha}\delta_{\left(  \lambda\right)  }^{\left(
\beta\right)  }\right)  r_{\left(  \alpha\beta\right)  }. \label{eqnPR85}%
\end{equation}
Here $t_{\left(  \rho\right)  }$ and $r_{\left(  \alpha\beta\right)  }$ are
the translational and rotational gauge parameters, respectively.

Any conclusion about higher dimensions based on three dimensional case should
be made with a great care. It is well known fact that the first order N-bein
Lagrangian is not invariant under translational (proportional to $t_{\left(
\rho\right)  }$) part of (\ref{eqnPR84}-\ref{eqnPR85}) and only rotational
part of these transformations can be promoted from three to any dimension
higher than two. Translational part of three dimensional case is not gauge
invariance in higher dimensions which is the expected result after neglecting
so many terms in (\ref{eqnPR52}, \ref{eqnPR53}, \ref{eqnPR70}, \ref{eqnPR71},
\ref{eqnPR73}). However, such simple observation is not sufficient to make any
general conclusion about gauge invariance in higher dimensions and especially
to say, based only on three dimensional results and without calculations in
higher dimensions, that N-bein gravity is not a Poincar\'{e} gauge theory.
Such a conclusion is in contradiction with an ordinary logic. Much more
reasonable expectation based on first steps of the Dirac procedure should be
quite different: in higher dimensions a translational part of transformations
should be different but algebra of constraints should be unchanged despite
modifications of constraints themselves.

Gauge transformations (\ref{eqnPR84}, \ref{eqnPR85}) were obtained in
\cite{3D} using the Castellani procedure \cite{Cast} and are the result of the
PB algebra of constraints (\ref{eqnPR81}-\ref{eqnPR83}). We know that
rotational invariance is the same in all dimensions $D>2$ and at least part of
algebra (\ref{eqnPR81}-\ref{eqnPR83}) must to be the same to preserve a
corresponding part of generators. In \cite{3D} we built such generators using
only this algebra. According to the Castellani procedure they are given by%

\begin{equation}
G=G_{\left(  1\right)  }^{\left(  \rho\right)  }\dot{t}_{\left(  \rho\right)
}+G_{\left(  0\right)  }^{\left(  \rho\right)  }t_{\left(  \rho\right)
}+G_{\left(  1\right)  }^{\left(  \alpha\beta\right)  }\dot{r}_{\left(
\alpha\beta\right)  }+G_{\left(  0\right)  }^{\left(  \alpha\beta\right)
}r_{\left(  \alpha\beta\right)  }. \label{eqnPR86}%
\end{equation}
The functions $G_{\left(  1\right)  }$ in (\ref{eqnPR86}) are the primary constraints%

\begin{equation}
G_{\left(  1\right)  }^{\left(  \rho\right)  }=\pi^{0\left(  \rho\right)
},\text{ \ \ }G_{\left(  1\right)  }^{\left(  \alpha\beta\right)  }%
=\Pi^{0\left(  \alpha\beta\right)  } \label{eqnPR87}%
\end{equation}
and $G_{\left(  0\right)  }$ are defined using the following relations
\cite{Cast}%

\begin{equation}
G_{\left(  0\right)  }^{\left(  \rho\right)  }\left(  x\right)  =-\left\{
\pi^{0\left(  \rho\right)  }\left(  x\right)  ,H_{T}\right\}  +\int\left[
\tilde{\alpha}_{\gamma}^{\rho}\left(  x,y\right)  \pi^{0\left(  \gamma\right)
}\left(  y\right)  +\tilde{\alpha}_{\left(  \alpha\beta\right)  }^{\rho
}\left(  x,y\right)  \Pi^{0\left(  \alpha\beta\right)  }\left(  y\right)
\right]  d^{D-1}y, \label{eqnPR88}%
\end{equation}

\begin{equation}
G_{\left(  0\right)  }^{\left(  \alpha\beta\right)  }\left(  x\right)
=-\left\{  \Pi^{0\left(  \alpha\beta\right)  }\left(  x\right)  ,H_{T}%
\right\}  +\int\left[  \tilde{\alpha}_{\gamma}^{\left(  \alpha\beta\right)
}\left(  x,y\right)  \pi^{0\left(  \gamma\right)  }\left(  y\right)
+\tilde{\alpha}_{\left(  \nu\mu\right)  }^{\left(  \alpha\beta\right)
}\left(  x,y\right)  \Pi^{0\left(  \nu\mu\right)  }\left(  y\right)  \right]
d^{D-1}y, \label{eqnPR89}%
\end{equation}
where the functions $\tilde{\alpha}_{\left(  ..\right)  }^{\left(  ..\right)
}\left(  x,y\right)  $ have to be chosen in such a way that the chains end at
primary constraints%

\begin{equation}
\left\{  G_{\left(  0\right)  }^{\sigma},H_{T}\right\}  =primary.
\label{eqnPR90}%
\end{equation}

To construct the generator (\ref{eqnPR86}), we have to find $\tilde{\alpha
}_{\left(  ..\right)  }^{\left(  ..\right)  }\left(  x,y\right)  $ using
condition (\ref{eqnPR90}). This calculation, because of the simple PBs among
the constraints, is straightforward:%

\begin{equation}
\left\{  G_{\left(  0\right)  }^{\left(  \rho\right)  }\left(  x\right)
,H_{T}\right\}  =\left\{  -\chi^{0\left(  \rho\right)  }\left(  x\right)
+\int\left[  \tilde{\alpha}_{\gamma}^{\rho}\left(  x,y\right)  \pi^{0\left(
\gamma\right)  }\left(  y\right)  +\tilde{\alpha}_{\left(  \alpha\beta\right)
}^{\rho}\left(  x,y\right)  \Pi^{0\left(  \alpha\beta\right)  }\left(
y\right)  \right]  d^{D-1}y,H_{T}\right\}  =0, \label{eqnPR91}%
\end{equation}

\begin{equation}
\left\{  G_{\left(  0\right)  }^{\left(  \alpha\beta\right)  }\left(
x\right)  ,H_{T}\right\}  =\left\{  -\chi^{0\left(  \alpha\beta\right)
}\left(  x\right)  +\int\left[  \tilde{\alpha}_{\gamma}^{\left(  \alpha
\beta\right)  }\left(  x,y\right)  \pi^{0\left(  \gamma\right)  }\left(
y\right)  +\tilde{\alpha}_{\left(  \nu\mu\right)  }^{\left(  \alpha
\beta\right)  }\left(  x,y\right)  \Pi^{0\left(  \nu\mu\right)  }\left(
y\right)  \right]  d^{D-1}y,H_{T}\right\}  =0 \label{eqnPR92}%
\end{equation}
where $H_{T}$ can be replaced by $H_{c}=-e_{0\left(  \sigma\right)  }%
\chi^{0\left(  \sigma\right)  }-\omega_{0\left(  \sigma\lambda\right)  }%
\chi^{0\left(  \sigma\lambda\right)  }$, because PBs among primary constraints
themselves and among primary and secondary constraints are zero. (These
calculations are simpler compare with the Hamiltonian formulation of metric
gravity \cite{KKRV}.)

From (\ref{eqnPR91}) and (\ref{eqnPR92}) and the PBs among first class
constraints we found all the functions $\tilde{\alpha}_{\left(  ..\right)
}^{\left(  ..\right)  }\left(  x,y\right)  $ in (\ref{eqnPR88}, \ref{eqnPR89}):%

\begin{equation}
\tilde{\alpha}_{\left(  \alpha\beta\right)  }^{\left(  \rho\right)  }\left(
x,y\right)  =0, \label{eqnPR93}%
\end{equation}

\begin{equation}
\tilde{\alpha}_{\left(  \gamma\right)  }^{\left(  \rho\right)  }\left(
x,y\right)  =\omega_{0(\gamma}^{~~~\rho)}\delta\left(  x-y\right)  ,
\label{eqnPR94}%
\end{equation}

\begin{equation}
\tilde{\alpha}_{\left(  \gamma\right)  }^{\left(  \alpha\beta\right)  }\left(
x,y\right)  =\frac{1}{2}\left(  e_{0}^{\left(  \alpha\right)  }\delta_{\left(
\gamma\right)  }^{\left(  \beta\right)  }-e_{0}^{\left(  \beta\right)  }%
\delta_{\left(  \gamma\right)  }^{\left(  \alpha\right)  }\right)
\delta\left(  x-y\right)  , \label{eqnPR95}%
\end{equation}

\begin{equation}
\tilde{\alpha}_{\left(  \nu\mu\right)  }^{\left(  \alpha\beta\right)  }\left(
x,y\right)  =\left(  \omega_{0~~\mu)}^{~(\alpha}\delta_{\left(  \nu\right)
}^{\left(  \beta\right)  }-\omega_{0~~\nu)}^{~(\alpha}\delta_{\left(
\mu\right)  }^{\left(  \beta\right)  }\right)  \delta\left(  x-y\right)  .
\label{eqnPR96}%
\end{equation}

This completes the derivation of the generator (\ref{eqnPR86}) as now%

\begin{equation}
G_{\left(  0\right)  }^{\left(  \rho\right)  }=-\chi^{0\left(  \rho\right)
}+\omega_{0(\gamma}^{~~~\rho)}\pi^{0\left(  \gamma\right)  }, \label{eqnPR98}%
\end{equation}
and%

\begin{equation}
G_{\left(  0\right)  }^{\left(  \alpha\beta\right)  }=-\chi^{0\left(
\alpha\beta\right)  }+\frac{1}{2}\left(  e_{0}^{\left(  \alpha\right)  }%
\delta_{\left(  \gamma\right)  }^{\left(  \beta\right)  }-e_{0}^{\left(
\beta\right)  }\delta_{\left(  \gamma\right)  }^{\left(  \alpha\right)
}\right)  \pi^{0\left(  \gamma\right)  }+\omega_{0~~\mu)}^{~(\alpha}%
\Pi^{0\left(  \beta\mu\right)  }-\omega_{0~\text{~}\mu)}^{~(\beta}%
\Pi^{0\left(  \alpha\mu\right)  }. \label{eqnPR99}%
\end{equation}

Using%

\begin{equation}
\delta\left(  field\right)  =\left\{  G,field\right\}  =\left\{  G_{\left(
1\right)  }^{\left(  \rho\right)  }\dot{t}_{\left(  \rho\right)  }+G_{\left(
0\right)  }^{\left(  \rho\right)  }t_{\left(  \rho\right)  }+G_{\left(
1\right)  }^{\left(  \alpha\beta\right)  }\dot{r}_{\left(  \alpha\beta\right)
}+G_{\left(  0\right)  }^{\left(  \alpha\beta\right)  }r_{\left(  \alpha
\beta\right)  },field\right\}  \label{eqnPR100}%
\end{equation}
we can find the gauge transformations of fields.

Because rotational invariance is the same in all dimensions, the absence of
dimensional peculiarities in (\ref{eqnPR72}) is consistent with this fact.
Part of the generator (\ref{eqnPR86}) responsible for this transformation must
be the same in all dimensions ($D>2$), so as the corresponding $\tilde{\alpha
}_{\left(  ..\right)  }^{\left(  ..\right)  }\left(  x,y\right)  $
(\ref{eqnPR95}, \ref{eqnPR96}) that follow from (\ref{eqnPR92}). This makes
even small deviations from the Poincar\'{e} algebra almost impossible. In
addition, whatever modifications of a translational constraint we have in
higher dimensions they suppose to preserve the limit of three dimensional case
where, as we know, we have the Poincar\'{e} algebra.

Let us provide some simple calculations that are based on and supported by our
reasonings. The rotational constraint is the same in all dimensions,
consequently, the solution for $\omega_{m\left(  pq\right)  }$ (different in
higher dimension, see (\ref{eqnPR67}-\ref{eqnPR71})) should not affect
$\chi^{0\left(  \alpha\beta\right)  }$. We should be able to find this
constraint in any dimension without solving for $\omega_{m\left(  pq\right)
}$. In $3D$ we obtained \cite{3D}%

\begin{equation}
\chi^{0\left(  \alpha\beta\right)  }=\frac{1}{2}e_{k}^{\left(  \alpha\right)
}\pi^{k\left(  \beta\right)  }-\frac{1}{2}e_{k}^{\left(  \beta\right)  }%
\pi^{k\left(  \alpha\right)  }-eB^{\gamma\left(  \rho\right)  k\left(
\alpha\right)  0\left(  \beta\right)  }e_{\gamma\left(  \rho\right)  ,k}
\label{eqnPR110}%
\end{equation}
and demonstrated that such a constraint satisfies (\ref{eqnPR83}) without any
reference to a particular dimension.

Without solving for $\omega_{m\left(  pq\right)  }$ in higher dimensions using
the secondary constraint, we have the relation from the primary constraint
(\ref{eqnPR31})%

\begin{equation}
\pi^{k\left(  \rho\right)  }=eB^{k\left(  \rho\right)  0\left(  \alpha\right)
m\left(  \beta\right)  }\omega_{m\left(  \alpha\beta\right)  }.
\label{eqnPR111}%
\end{equation}
We substitute this into the first two terms of (\ref{eqnPR110}) and, after
short rearrangements and using properties of $B$ (\ref{eqnPR6}), we obtain%

\begin{equation}
\frac{1}{2}e_{k}^{\left(  \alpha\right)  }\pi^{k\left(  \beta\right)  }%
-\frac{1}{2}e_{k}^{\left(  \beta\right)  }\pi^{k\left(  \alpha\right)
}=-eA^{0\left(  \alpha\right)  k\left(  \gamma\right)  }\omega_{k~~\gamma
)}^{~(\beta}+eA^{0\left(  \beta\right)  k\left(  \gamma\right)  }%
\omega_{k~~\gamma)}^{~(\alpha} \label{eqnPR112}%
\end{equation}
which is exactly the expression that we have in a general case (\ref{eqnPR72}%
). We, almost at once, have the rotational constraint, moreover, as it was
demonstrated (see discussion after (\ref{eqnPR73})), before we used a solution
of the secondary constraint, its PBs with primary constraints $\Pi^{0\left(
\alpha\beta\right)  }$ and $\pi^{0\left(  \rho\right)  }$ are zero and after
substitution of (\ref{eqnPR111}) (expression of spatial connections in terms
of momenta) they remain to be zero%

\begin{equation}
\left\{  \Pi^{0\left(  \alpha\beta\right)  },\chi^{0\left(  \alpha
\beta\right)  }\right\}  =\left\{  \pi^{0\left(  \rho\right)  },\chi^{0\left(
\alpha\beta\right)  }\right\}  =0. \label{eqnPR114}%
\end{equation}

So, as it was calculated in three dimensional case \cite{3D}, PBs among
(\ref{eqnPR110}) are (\ref{eqnPR83}). We demonstrated that this part of PB
algebra is the same in all dimensions.

In a general expression of the translational constraint we also have one
contribution that can be almost immediately obtained without solution for
$\omega_{m\left(  pq\right)  }$, just by using primary second class
constraints (\ref{eqnPR111}). The second term of (\ref{eqnPR73}) consists of
exactly the same combination (\ref{eqnPR111}) and upon substitution gives a
simple contribution into the secondary translational constraint in all
dimensions and such a contribution supports three dimensional limit (see
\cite{3D})%

\begin{equation}
\chi^{0\left(  \rho\right)  }=\pi_{,k}^{k\left(  \rho\right)  }+...
\label{eqnPR115}%
\end{equation}

This part by itself (as it was before substitution) has zero PBs with both
primary constraints and supports all relations of the Poincar\'{e} algebra
found in three dimensions (\ref{eqnPR81}, \ref{eqnPR82}). To consider the rest
of contributions in (\ref{eqnPR73}) the biggest part of which is zero in three
dimensional case, we have to find the solution for $\omega_{m\left(
pq\right)  }$. So, go back to the second group of equations (\ref{eqnPR67}%
-\ref{eqnPR71}).

\section{Finding the solution for $\omega_{m\left(  pq\right)  }$}

Using the first group of equations (\ref{eqnPR52}-\ref{eqnPR55}), components
of a spin connection $\omega_{m\left(  p0\right)  }$ and the corresponding to
them momenta can be solved as in $3D$ \cite{3D}, of course, with some extra
contributions. Now to find $\omega_{m\left(  pq\right)  }$ we consider two
equations (\ref{eqnPR67}) and (\ref{eqnPR70}) from the second group in $D>3$:%

\begin{equation}
\pi^{k\left(  0\right)  }+2ee^{0\left(  0\right)  }\gamma^{k\left(  p\right)
}\gamma^{m\left(  q\right)  }\omega_{m\left(  pq\right)  }=0,\text{\ \ \ }
\label{eqnPR120}%
\end{equation}

\begin{equation}
-\frac{\delta H_{c}}{\delta\omega_{p\left(  mn\right)  }}+\lambda_{k\left(
q\right)  }eB^{k\left(  q\right)  0\left(  m\right)  p\left(  n\right)
}-\lambda_{k\left(  0\right)  }ee^{0\left(  0\right)  }E^{k\left(  m\right)
p\left(  n\right)  }=0. \label{eqnPR121}%
\end{equation}

First equation is slightly modified using (\ref{eqnPR59}, \ref{eqnPR60}) and
antisymmetry of connections. In (\ref{eqnPR121}) the solutions for
$\omega_{k\left(  p0\right)  }$ (\ref{eqnPR66}) and the multiplier
$\lambda_{k\left(  q\right)  }$ (\ref{eqnPR64}) have to be substituted. We
have two sets of equations to solve for two sets of variables $\omega
_{m\left(  pq\right)  }$ and $\lambda_{k\left(  0\right)  }$. The number of
unknowns and equations are the same in any dimension $D>2$. This system of
equations is a perfect illustration of the fact (which is often hidden in
simple theories) that $\lambda_{k\left(  0\right)  }$ are true Lagrange
multipliers: we have to find a solution for $\omega_{p\left(  mn\right)  }$
with extra condition (\ref{eqnPR120}) and the Dirac method automatically gives
correct settings for the problem of finding conditional extremum.

To perform substitutions of (\ref{eqnPR64}) and (\ref{eqnPR66}), the explicit
form of variation $-\frac{\delta H_{c}}{\delta\omega_{p\left(  mn\right)  }}$
is needed%

\[
-\frac{\delta H_{c}}{\delta\omega_{p\left(  mn\right)  }}=eB^{\gamma\left(
\rho\right)  k\left(  m\right)  p\left(  n\right)  }e_{\gamma\left(
\rho\right)  ,k}-eA^{p\left(  m\right)  k\left(  q\right)  }\omega
_{k~~q)}^{~(n}+eA^{p\left(  n\right)  k\left(  q\right)  }\omega_{k~~q)}^{~(m}%
\]

\begin{equation}
-eA^{p\left(  m\right)  k\left(  0\right)  }\omega_{k~~0)}^{~(n}+eA^{p\left(
n\right)  k\left(  0\right)  }\omega_{k~~0)}^{~(m}-eA^{p\left(  m\right)
0\left(  \alpha\right)  }\omega_{0~~\alpha)}^{~(n}+eA^{p\left(  n\right)
0\left(  \alpha\right)  }\omega_{0~~\alpha)}^{~(m}, \label{eqnPR122}%
\end{equation}
as well as $\frac{\delta H_{c}}{\delta\omega_{p\left(  m0\right)  }}:$%

\[
\frac{\delta H_{c}}{\delta\omega_{p\left(  m0\right)  }}=-eB^{\gamma\left(
q\right)  k\left(  m\right)  p\left(  0\right)  }e_{\gamma\left(  q\right)
,k}%
\]

\begin{equation}
+eA^{p\left(  m\right)  n\left(  q\right)  }\omega_{n~~q)}^{~(0}-eA^{p\left(
0\right)  n\left(  q\right)  }\omega_{n~~q)}^{~(m}+eA^{p\left(  m\right)
0\left(  q\right)  }\omega_{0~~q)}^{~(0}-eA^{p\left(  0\right)  0\left(
q\right)  }\omega_{0~~q)}^{~(m}. \label{eqnPR123}%
\end{equation}

Solutions of previous equations suggest transition of all expressions into
\textquotedblleft$\gamma-E$\textquotedblright-form that after short
rearrangements, using (\ref{eqnPR65}) and%

\begin{equation}
A^{p\left(  m\right)  k\left(  q\right)  }=E^{p\left(  m\right)  k\left(
q\right)  }+\frac{e^{0\left(  m\right)  }}{e^{0\left(  0\right)  }}A^{p\left(
0\right)  k\left(  q\right)  }+A^{p\left(  m\right)  k\left(  0\right)  }%
\frac{e^{0\left(  q\right)  }}{e^{0\left(  0\right)  }}, \label{eqnPR123a}%
\end{equation}
converts (\ref{eqnPR121}-\ref{eqnPR123}) into%

\[
-E^{p\left(  m\right)  k\left(  q\right)  }\omega_{k~~q)}^{~(n}+E^{p\left(
n\right)  k\left(  q\right)  }\omega_{k~~q)}^{~(m}+\frac{e^{0\left(  m\right)
}}{e^{0\left(  0\right)  }}E^{p\left(  n\right)  n^{\prime}\left(  q^{\prime
}\right)  }\frac{e^{0\left(  a\right)  }}{e_{\left(  0\right)  }^{0}}%
\omega_{n^{\prime}\left(  aq^{\prime}\right)  }-\frac{e^{0\left(  n\right)  }%
}{e^{0\left(  0\right)  }}E^{p\left(  m\right)  n^{\prime}\left(  q^{\prime
}\right)  }\frac{e^{0\left(  a\right)  }}{e_{\left(  0\right)  }^{0}}%
\omega_{n^{\prime}\left(  aq^{\prime}\right)  }%
\]

\[
-\frac{e^{0\left(  q\right)  }}{e^{0\left(  0\right)  }}E^{k\left(  m\right)
p\left(  n\right)  }I_{k\left(  q\right)  p^{\prime}\left(  m^{\prime}\right)
}A^{p^{\prime}\left(  0\right)  n\left(  q^{\prime}\right)  }\omega
_{n~~q^{\prime})}^{~(m^{\prime}}-\frac{e^{0\left(  q\right)  }}{e^{0\left(
0\right)  }}E^{k\left(  m\right)  p\left(  n\right)  }I_{k\left(  q\right)
p^{\prime}\left(  m^{\prime}\right)  }\frac{e^{0\left(  m^{\prime}\right)  }%
}{e^{0\left(  0\right)  }}A^{p^{\prime}\left(  0\right)  n\left(  q^{\prime
}\right)  }\frac{e^{0\left(  a\right)  }}{e_{\left(  0\right)  }^{0}}%
\omega_{n\left(  aq^{\prime}\right)  }%
\]

\[
+D^{p\left(  mn\right)  }\left(  e_{\mu\left(  \nu\right)  ,k}\right)
+D^{p\left(  mn\right)  }\left(  \omega_{0\left(  \alpha\beta\right)
}\right)  +D^{p\left(  mn\right)  }\left(  \pi^{k\left(  m\right)  }\right)
+D^{p\left(  mn\right)  }\left(  \pi^{k\left(  0\right)  }\right)
\]

\begin{equation}
-\lambda_{k\left(  0\right)  }e^{0\left(  0\right)  }E^{k\left(  m\right)
p\left(  n\right)  }=0 \label{eqnPR124}%
\end{equation}
where%

\begin{equation}
D^{p\left(  mn\right)  }\left(  e_{\mu\left(  \nu\right)  ,k}\right)
=B^{\gamma\left(  \rho\right)  k\left(  m\right)  p\left(  n\right)
}e_{\gamma\left(  \rho\right)  ,k}+B^{k\left(  q\right)  0\left(  m\right)
p\left(  n\right)  }I_{k\left(  q\right)  p^{\prime}\left(  m^{\prime}\right)
}\frac{1}{e^{0\left(  0\right)  }}B^{\gamma\left(  q^{\prime}\right)
k^{\prime}\left(  m^{\prime}\right)  p^{\prime}\left(  0\right)  }%
e_{\gamma\left(  q^{\prime}\right)  ,k^{\prime}}; \label{eqnPR125}%
\end{equation}

\[
D^{p\left(  mn\right)  }\left(  \omega_{0\left(  \alpha\beta\right)  }\right)
=-A^{p\left(  m\right)  0\left(  \alpha\right)  }\omega_{0~~\alpha)}%
^{~(n}+A^{p\left(  n\right)  0\left(  \alpha\right)  }\omega_{0~~\alpha
)}^{~(m}%
\]

\begin{equation}
-B^{k\left(  q\right)  0\left(  m\right)  p\left(  n\right)  }I_{k\left(
q\right)  p^{\prime}\left(  m^{\prime}\right)  }\frac{1}{e^{0\left(  0\right)
}}\left(  A^{p^{\prime}\left(  m^{\prime}\right)  0\left(  q^{\prime}\right)
}\omega_{0~~q^{\prime})}^{~(0}-A^{p^{\prime}\left(  0\right)  0\left(
q^{\prime}\right)  }\omega_{0~~q^{\prime})}^{~(m^{\prime}}\right)  ;
\label{eqnPR126}%
\end{equation}

\[
D^{p\left(  mn\right)  }\left(  \pi^{k\left(  m\right)  }\right)
=+A^{p\left(  m\right)  k\left(  0\right)  }\frac{1}{2ee^{0\left(  0\right)
}}I_{k\left(  n\uparrow\right)  m^{\prime}\left(  p^{\prime}\right)  }%
\pi^{m^{\prime}\left(  p^{\prime}\right)  }-A^{p\left(  n\right)  k\left(
0\right)  }\frac{1}{2ee^{0\left(  0\right)  }}I_{k\left(  m\uparrow\right)
m^{\prime}\left(  p^{\prime}\right)  }\pi^{m^{\prime}\left(  p^{\prime
}\right)  }%
\]

\[
+\frac{e^{0\left(  q\right)  }}{e^{0\left(  0\right)  }}E^{k\left(  m\right)
p\left(  n\right)  }I_{k\left(  q\right)  p^{\prime}\left(  m^{\prime}\right)
}A^{p^{\prime}\left(  m^{\prime}\right)  n\left(  q^{\prime}\right)  }\frac
{1}{2ee_{\left(  0\right)  }^{0}}I_{n\left(  q^{\prime}\right)  a\left(
b\right)  }\pi^{a\left(  b\right)  }%
\]

\begin{equation}
-\frac{e^{0\left(  m\right)  }}{e^{0\left(  0\right)  }}A^{p\left(  n\right)
n^{\prime}\left(  q^{\prime}\right)  }\frac{1}{2ee_{\left(  0\right)  }^{0}%
}I_{n^{\prime}\left(  q^{\prime}\right)  a\left(  b\right)  }\pi^{a\left(
b\right)  }+\frac{e^{0\left(  n\right)  }}{e^{0\left(  0\right)  }}A^{p\left(
m\right)  n^{\prime}\left(  q^{\prime}\right)  }\frac{1}{2ee_{\left(
0\right)  }^{0}}I_{n^{\prime}\left(  q^{\prime}\right)  a\left(  b\right)
}\pi^{a\left(  b\right)  }; \label{eqnPR127}%
\end{equation}

\[
D^{p\left(  mn\right)  }\left(  \pi^{k\left(  0\right)  }\right)  =A^{p\left(
m\right)  k\left(  0\right)  }\frac{e^{0\left(  p^{\prime}\right)  }%
}{2e^{0\left(  0\right)  }}I_{k\left(  n\uparrow\right)  m^{\prime}\left(
p^{\prime}\right)  }E^{m^{\prime}\left(  a\right)  n\left(  b\right)  }%
\omega_{n\left(  ab\right)  }%
\]

\[
-A^{p\left(  n\right)  k\left(  0\right)  }\frac{e^{0\left(  p^{\prime
}\right)  }}{2e^{0\left(  0\right)  }}I_{k\left(  m\uparrow\right)  m^{\prime
}\left(  p^{\prime}\right)  }E^{m^{\prime}\left(  a\right)  n\left(  b\right)
}\omega_{n\left(  ab\right)  }%
\]

\begin{equation}
+\frac{e^{0\left(  q\right)  }}{e^{0\left(  0\right)  }}E^{k\left(  m\right)
p\left(  n\right)  }I_{k\left(  q\right)  p^{\prime}\left(  m^{\prime}\right)
}A^{p^{\prime}\left(  m^{\prime}\right)  n\left(  q^{\prime}\right)  }%
\frac{e^{0\left(  c\right)  }}{2e_{\left(  0\right)  }^{0}}I_{n\left(
q^{\prime}\right)  d\left(  c\right)  }E^{d\left(  a\right)  n\left(
b\right)  }\omega_{n\left(  ab\right)  } \label{eqnPR128}%
\end{equation}

\[
-\frac{e^{0\left(  m\right)  }}{e^{0\left(  0\right)  }}A^{p\left(  n\right)
n^{\prime}\left(  q^{\prime}\right)  }\frac{e^{0\left(  c\right)  }%
}{2e_{\left(  0\right)  }^{0}}I_{n^{\prime}\left(  q^{\prime}\right)  d\left(
c\right)  }E^{d\left(  a\right)  n\left(  b\right)  }\omega_{n\left(
ab\right)  }+\frac{e^{0\left(  n\right)  }}{e^{0\left(  0\right)  }%
}A^{p\left(  m\right)  n^{\prime}\left(  q^{\prime}\right)  }\frac{e^{0\left(
c\right)  }}{2e_{\left(  0\right)  }^{0}}I_{n^{\prime}\left(  q^{\prime
}\right)  d\left(  c\right)  }E^{d\left(  a\right)  n\left(  b\right)  }%
\omega_{n\left(  ab\right)  };
\]

\begin{equation}
D^{p\left(  mn\right)  }\left(  \omega_{0\left(  \alpha\beta\right)  }\right)
=-A^{p\left(  m\right)  0\left(  \alpha\right)  }\omega_{0~~\alpha)}%
^{~(n}+A^{p\left(  n\right)  0\left(  \alpha\right)  }\omega_{0~~\alpha
)}^{~(m} \label{eqnPR128b}%
\end{equation}

\[
-B^{k\left(  q\right)  0\left(  m\right)  p\left(  n\right)  }I_{k\left(
q\right)  p^{\prime}\left(  m^{\prime}\right)  }\frac{1}{e^{0\left(  0\right)
}}\left(  A^{p^{\prime}\left(  m^{\prime}\right)  0\left(  q^{\prime}\right)
}\omega_{0~~q^{\prime})}^{~(0}-A^{p^{\prime}\left(  0\right)  0\left(
q^{\prime}\right)  }\omega_{0~~q^{\prime})}^{~(m^{\prime}}\right)  .
\]
All terms in equation (\ref{eqnPR124}) with a particular contraction
$E^{d\left(  a\right)  n\left(  b\right)  }\omega_{n\left(  ab\right)  }$ are
in $D^{p\left(  mn\right)  }\left(  \pi^{k\left(  0\right)  }\right)  $
because they are expressible in terms of momenta%

\begin{equation}
-\frac{1}{ee^{0\left(  0\right)  }}\pi^{k\left(  0\right)  }=E^{k\left(
p\right)  m\left(  q\right)  }\omega_{m\left(  pq\right)  }=2\gamma^{k\left(
p\right)  }\gamma^{m\left(  q\right)  }\omega_{m\left(  pq\right)  }.
\label{eqnPR128a}%
\end{equation}

As in a covariant case considered in Introduction, to solve equation
(\ref{eqnPR124}) (similar with (\ref{eqnPR6a})) we have to have combinations
with internal indices only, which in this case are%

\begin{equation}
\gamma^{k\left(  m\right)  }\omega_{k~~q)}^{~(n}\equiv\tilde{\omega
}_{.....~~q)}^{m(n}, \label{eqnPR129}%
\end{equation}
and contract (\ref{eqnPR124}) with $e_{p}^{\left(  s\right)  }$ (we just mimic
a solution in a covariant case, see (\ref{eqnPR19})). Note also that, compare
with a covariant case (\ref{eqnPR7}), in (\ref{eqnPR128a}) we have contraction
in spatial indices only.

The resulting equation is (all free indices are internal)%

\[
\tilde{\omega}^{m(ns)}-\tilde{\omega}^{n(ms)}-\frac{e^{0\left(  m\right)  }%
}{e^{0\left(  0\right)  }}\frac{e_{\left(  a\right)  }^{0}}{e_{\left(
0\right)  }^{0}}\tilde{\omega}^{n\left(  as\right)  }+\frac{e^{0\left(
n\right)  }}{e^{0\left(  0\right)  }}\frac{e_{\left(  a\right)  }^{0}%
}{e_{\left(  0\right)  }^{0}}\tilde{\omega}^{m\left(  as\right)  }%
\]

\[
-\tilde{\eta}^{sm}\tilde{\omega}_{..~~q)}^{~q(n}+\tilde{\eta}^{sn}%
\tilde{\omega}_{....~~q)}^{~q(m}+\tilde{\eta}^{sn}\frac{e^{0\left(  m\right)
}}{e^{0\left(  0\right)  }}\frac{e^{0\left(  a\right)  }}{e_{\left(  0\right)
}^{0}}\tilde{\omega}_{..\left(  aq^{\prime}\right)  }^{q^{\prime}}-\tilde
{\eta}^{sm}\frac{e^{0\left(  n\right)  }}{e^{0\left(  0\right)  }}%
\frac{e^{0\left(  a\right)  }}{e_{\left(  0\right)  }^{0}}\tilde{\omega
}_{..\left(  aq^{\prime}\right)  }^{q^{\prime}}%
\]

\[
-\frac{e^{0\left(  q\right)  }}{e^{0\left(  0\right)  }}\left(  \gamma
^{k\left(  m\right)  }\tilde{\eta}^{sn}-\gamma^{k\left(  n\right)  }%
\tilde{\eta}^{sm}\right)  I_{k\left(  q\right)  p^{\prime}\left(  m^{\prime
}\right)  }A^{p^{\prime}\left(  0\right)  n\left(  q^{\prime}\right)  }%
\omega_{n~~q^{\prime})}^{~(m^{\prime}}%
\]

\[
-\frac{e^{0\left(  q\right)  }}{e^{0\left(  0\right)  }}\left(  \gamma
^{k\left(  m\right)  }\tilde{\eta}^{sn}-\gamma^{k\left(  n\right)  }%
\tilde{\eta}^{sm}\right)  I_{k\left(  q\right)  p^{\prime}\left(  m^{\prime
}\right)  }\frac{e^{0\left(  m^{\prime}\right)  }}{e^{0\left(  0\right)  }%
}A^{p^{\prime}\left(  0\right)  n\left(  q^{\prime}\right)  }\frac{e^{0\left(
a\right)  }}{e_{\left(  0\right)  }^{0}}\omega_{n\left(  aq^{\prime}\right)  }%
\]

\[
+\tilde{D}^{s\left(  mn\right)  }\left(  \pi^{k\left(  m\right)  }\right)
+\tilde{D}^{s\left(  mn\right)  }\left(  \pi^{k\left(  0\right)  }\right)
+\tilde{D}^{s\left(  mn\right)  }\left(  e_{\mu\left(  \nu\right)  ,k}\right)
+\tilde{D}^{s\left(  mn\right)  }\left(  \omega_{0\left(  \alpha\beta\right)
}\right)
\]

\begin{equation}
-e^{0\left(  0\right)  }\left(  \tilde{\lambda}^{m}\tilde{\eta}^{sn}%
-\tilde{\lambda}^{n}\tilde{\eta}^{sm}\right)  =0 \label{eqnPR130}%
\end{equation}
where%

\begin{equation}
\tilde{D}^{s\left(  mn\right)  }\equiv e_{p}^{\left(  s\right)  }D^{p\left(
mn\right)  } \label{eqnPR130a}%
\end{equation}
and%

\begin{equation}
\gamma^{k\left(  n\right)  }\lambda_{k\left(  0\right)  }\equiv\tilde{\lambda
}^{n}. \label{eqnPR131}%
\end{equation}

The solution for free part of (\ref{eqnPR130}), two first terms, is known, and
we have to eliminate the contractions (as we did this in (\ref{eqnPR12})). We
have once contracted terms with two free indices in the first line of
(\ref{eqnPR130}), $e_{\left(  a\right)  }^{0}\tilde{\omega}^{m\left(
as\right)  }$, and with one free index (third line), such as $\tilde{\eta
}^{sn}\tilde{X}^{m}$ (the explicit form of $\tilde{X}^{m}$ can be read off
from (\ref{eqnPR130})). First, as elimination of \textquotedblleft
trace\textquotedblright\ in covariant case, we consider contraction of
(\ref{eqnPR130}) with $\tilde{\eta}_{ms}$ that gives%

\begin{equation}
\tilde{\omega}_{s}^{.(ns)}+\frac{e^{0\left(  n\right)  }}{e^{0\left(
0\right)  }}\frac{e_{\left(  a\right)  }^{0}}{e_{\left(  0\right)  }^{0}%
}\tilde{\omega}_{s}^{..\left(  as\right)  } \label{eqnPR132}%
\end{equation}

\[
-\left(  D-2\right)  \tilde{\omega}_{..~~q)}^{~q(n}-\left(  D-2\right)
\frac{e^{0\left(  n\right)  }}{e^{0\left(  0\right)  }}\frac{e^{0\left(
a\right)  }}{e_{\left(  0\right)  }^{0}}\tilde{\omega}_{..\left(  aq^{\prime
}\right)  }^{q^{\prime}}%
\]

\[
+\frac{e^{0\left(  q\right)  }}{e^{0\left(  0\right)  }}\gamma^{k\left(
n\right)  }\left(  D-2\right)  \tilde{\eta}^{sm}I_{k\left(  q\right)
p^{\prime}\left(  m^{\prime}\right)  }A^{p^{\prime}\left(  0\right)  n\left(
q^{\prime}\right)  }\omega_{n~~q^{\prime})}^{~(m^{\prime}}%
\]

\[
+\frac{e^{0\left(  q\right)  }}{e^{0\left(  0\right)  }}\gamma^{k\left(
n\right)  }\left(  D-2\right)  I_{k\left(  q\right)  p^{\prime}\left(
m^{\prime}\right)  }\frac{e^{0\left(  m^{\prime}\right)  }}{e^{0\left(
0\right)  }}A^{p^{\prime}\left(  0\right)  n\left(  q^{\prime}\right)  }%
\frac{e^{0\left(  a\right)  }}{e_{\left(  0\right)  }^{0}}\omega_{n\left(
aq^{\prime}\right)  }%
\]

\[
+\tilde{D}_{m}^{..\left(  mn\right)  }\left(  \pi^{k\left(  m\right)
}\right)  +\tilde{D}_{m}^{..\left(  mn\right)  }\left(  \pi^{k\left(
0\right)  }\right)  +\tilde{D}_{m}^{..\left(  mn\right)  }\left(
e_{\mu\left(  \nu\right)  ,k}\right)  +\tilde{D}_{m}^{..\left(  mn\right)
}\left(  \omega_{0\left(  \alpha\beta\right)  }\right)  +\tilde{\lambda}%
^{n}e^{0\left(  0\right)  }\left(  D-2\right)  =0.
\]

As before, $\tilde{\omega}_{s}^{.(ns)}$ is known and expressible in terms of
$\pi^{k\left(  0\right)  }$ using (\ref{eqnPR128a}) and we can solve this
equation for the multiplier $\tilde{\lambda}^{n}$%

\begin{equation}
\tilde{\lambda}^{n}=-\frac{D-3}{D-2}\frac{1}{e^{0\left(  0\right)  }}\left(
\tilde{\omega}_{s}^{.(ns)}+\frac{e^{0\left(  n\right)  }}{e^{0\left(
0\right)  }}\frac{e_{\left(  a\right)  }^{0}}{e_{\left(  0\right)  }^{0}%
}\tilde{\omega}_{s}^{..\left(  as\right)  }\right)  \label{eqnPR132a}%
\end{equation}

\[
-\frac{1}{e^{0\left(  0\right)  }}\frac{e^{0\left(  q\right)  }}{e^{0\left(
0\right)  }}\gamma^{k\left(  n\right)  }\tilde{\eta}^{sm}I_{k\left(  q\right)
p^{\prime}\left(  m^{\prime}\right)  }A^{p^{\prime}\left(  0\right)  n\left(
q^{\prime}\right)  }\omega_{n~~q^{\prime})}^{~(m^{\prime}}%
\]

\[
-\frac{1}{e^{0\left(  0\right)  }}\frac{e^{0\left(  q\right)  }}{e^{0\left(
0\right)  }}\gamma^{k\left(  n\right)  }I_{k\left(  q\right)  p^{\prime
}\left(  m^{\prime}\right)  }\frac{e^{0\left(  m^{\prime}\right)  }%
}{e^{0\left(  0\right)  }}A^{p^{\prime}\left(  0\right)  n\left(  q^{\prime
}\right)  }\frac{e^{0\left(  a\right)  }}{e_{\left(  0\right)  }^{0}}%
\omega_{n\left(  aq^{\prime}\right)  }%
\]

\[
-\frac{1}{D-2}\frac{1}{e^{0\left(  0\right)  }}\left(  \tilde{D}%
_{s}^{..\left(  sn\right)  }\left(  \pi^{k\left(  m\right)  }\right)
+\tilde{D}_{s}^{..\left(  sn\right)  }\left(  \pi^{k\left(  0\right)
}\right)  +\tilde{D}_{s}^{..\left(  sn\right)  }\left(  e_{\mu\left(
\nu\right)  ,k}\right)  +\tilde{D}_{s}^{..\left(  sn\right)  }\left(
\omega_{0\left(  \alpha\beta\right)  }\right)  \right)
\]
and substitute its solution back into (\ref{eqnPR130}) (as elimination of
\textquotedblleft trace\textquotedblright\ in covariant case, see
(\ref{eqnPR14}-\ref{eqnPR15}))%

\[
\tilde{\omega}^{m(ns)}-\tilde{\omega}^{n(ms)}-\frac{e^{0\left(  m\right)  }%
}{e^{0\left(  0\right)  }}\frac{e_{\left(  a\right)  }^{0}}{e_{\left(
0\right)  }^{0}}\tilde{\omega}^{n\left(  as\right)  }+\frac{e^{0\left(
n\right)  }}{e^{0\left(  0\right)  }}\frac{e_{\left(  a\right)  }^{0}%
}{e_{\left(  0\right)  }^{0}}\tilde{\omega}^{m\left(  as\right)  }%
\]

\begin{equation}
+\tilde{D}^{\prime s\left(  mn\right)  }\left(  \pi^{k\left(  m\right)
}\right)  +\tilde{D}^{\prime s\left(  mn\right)  }\left(  \pi^{k\left(
0\right)  }\right)  +\tilde{D}^{\prime s\left(  mn\right)  }\left(
e_{\mu\left(  \nu\right)  ,k}\right)  +\tilde{D}^{\prime s\left(  mn\right)
}\left(  \omega_{0\left(  \alpha\beta\right)  }\right)  =0 \label{eqnPR133}%
\end{equation}
where%

\[
\tilde{D}^{\prime s\left(  mn\right)  }\left(  \pi^{k\left(  0\right)
}\right)  \equiv\tilde{D}^{s\left(  mn\right)  }\left(  \pi^{k\left(
0\right)  }\right)  +\frac{\tilde{\eta}^{sn}}{\left(  D-2\right)  }\left(
\tilde{\omega}_{s}^{.(ms)}+\frac{e^{0\left(  m\right)  }}{e^{0\left(
0\right)  }}\frac{e_{\left(  a\right)  }^{0}}{e_{\left(  0\right)  }^{0}%
}\tilde{\omega}_{s}^{..\left(  as\right)  }+\tilde{D}_{a}^{..\left(
am\right)  }\left(  \pi^{k\left(  0\right)  }\right)  \right)
\]

\begin{equation}
-\frac{\tilde{\eta}^{sm}}{\left(  D-2\right)  }\left(  \tilde{\omega}%
_{s}^{.(ns)}+\frac{e^{0\left(  n\right)  }}{e^{0\left(  0\right)  }}%
\frac{e_{\left(  a\right)  }^{0}}{e_{\left(  0\right)  }^{0}}\tilde{\omega
}_{s}^{..\left(  as\right)  }+\tilde{D}_{a}^{..\left(  an\right)  }\left(
\pi^{k\left(  0\right)  }\right)  \right)  , \label{eqnPR134}%
\end{equation}

\begin{equation}
\tilde{D}^{\prime s\left(  mn\right)  }\left(  \pi^{k\left(  m\right)
}\right)  =\tilde{D}^{s\left(  mn\right)  }\left(  \pi^{k\left(  m\right)
}\right)  +\frac{\tilde{\eta}^{sn}}{D-2}\tilde{D}_{a}^{..\left(  am\right)
}\left(  \pi^{k\left(  m\right)  }\right)  -\frac{\tilde{\eta}^{sm}}%
{D-2}\tilde{D}_{a}^{..\left(  an\right)  }\left(  \pi^{k\left(  m\right)
}\right)  \label{eqnPR135}%
\end{equation}
and similar expressions can be written for $\tilde{D}^{\prime s\left(
mn\right)  }\left(  e_{\mu\left(  \nu\right)  ,k}\right)  $, $\tilde
{D}^{\prime s\left(  mn\right)  }\left(  \omega_{0\left(  \alpha\beta\right)
}\right)  $. Note that the second line in (\ref{eqnPR130}) vanishes after
substituting to (\ref{eqnPR130}) the multiplier $\tilde{\lambda}^{n}$ from
(\ref{eqn132a}). Note that the same dimensional coefficient, $\frac{1}{D-2}$,
appears, as in a covariant case, reflecting the same fact: first order
formulation (Lagrangian or Hamiltonian) is not valid in two dimensions. This
coefficient was already uncounted in the definition of $I_{k\left(  m\right)
n\left(  p\right)  }$ (\ref{eqnPR61}).

What is important that the direct calculations (using (\ref{eqnPR135}),
(\ref{eqnPR130a}), and (\ref{eqnPR128b}) ) shows that%

\begin{equation}
\tilde{D}^{\prime s\left(  mn\right)  }\left(  \omega_{0\left(  \alpha
\beta\right)  }\right)  =0. \label{eqnPR136}%
\end{equation}

This result is very significant: there are no temporal connections in a
solution for spatial connections and so in the secondary translational
constraint. We do not have reappearance of temporal connections that were
absent in (\ref{eqnPR73}). This immediately allows to make a conclusion that
the secondary translational constraint has zero PB with primary rotational one
(again, as in $3D$ case \cite{3D}).

We present equation (\ref{eqnPR133}) using short notation%

\[
\tilde{\omega}^{m(ns)}-\tilde{\omega}^{n(ms)}-\frac{e^{0\left(  m\right)  }%
}{e^{0\left(  0\right)  }}\frac{e_{\left(  a\right)  }^{0}}{e_{\left(
0\right)  }^{0}}\tilde{\omega}^{n\left(  as\right)  }+\frac{e^{0\left(
n\right)  }}{e^{0\left(  0\right)  }}\frac{e_{\left(  a\right)  }^{0}%
}{e_{\left(  0\right)  }^{0}}\tilde{\omega}^{m\left(  as\right)  }=
\]

\begin{equation}
\tilde{D}^{\prime\prime s\left(  mn\right)  }\left(  \pi^{k\left(  m\right)
},\pi^{k\left(  0\right)  },e_{\mu\left(  \nu\right)  ,k}\right)
\label{eqnPR137}%
\end{equation}
where%

\begin{equation}
\tilde{D}^{\prime\prime s\left(  mn\right)  }\left(  \pi^{k\left(  m\right)
},\pi^{k\left(  0\right)  },e_{\mu\left(  \nu\right)  ,k}\right)
\equiv-\tilde{D}^{\prime s\left(  mn\right)  }\left(  \pi^{k\left(  m\right)
}\right)  -\tilde{D}^{\prime s\left(  mn\right)  }\left(  \pi^{k\left(
0\right)  }\right)  -\tilde{D}^{\prime s\left(  mn\right)  }\left(
e_{\mu\left(  \nu\right)  ,k}\right)  \label{eqnPR138}%
\end{equation}
and the part $\tilde{D}^{\prime s\left(  mn\right)  }\left(  \omega_{0\left(
\alpha\beta\right)  }\right)  $ is not here by virtue of (\ref{eqnPR136}). The
solution of (\ref{eqnPR137}) is a little bit more involved compare with a
covariant case (\ref{eqnPR14}). Part of terms with contractions were
eliminated by solving for multipliers but we still have additional contraction
in one index (third and fourth terms of (\ref{eqnPR137})). We were not able to
eliminate these terms by further contractions, i.e. contraction with
$e_{\left(  s\right)  }^{0}$ leads only to a relation (that we will use), not
to an elimination of it,%

\begin{equation}
e_{\left(  s\right)  }^{0}\tilde{\omega}^{~m(ns)}-e_{\left(  s\right)  }%
^{0}\tilde{\omega}^{~n(ms)}=e_{\left(  s\right)  }^{0}\tilde{D}^{\prime\prime
s\left(  mn\right)  }. \label{eqnPR139}%
\end{equation}

So, we have to perform Einstein's permutation first (as in (\ref{eqnPR16})):
$\left(  mns\right)  +\left(  smn\right)  -\left(  nsm\right)  $, and try to
eliminate a contraction after that. After permutation of (\ref{eqnPR137}) we obtain%

\[
2\tilde{\omega}^{~m(ns)}+\frac{e^{0\left(  m\right)  }}{e^{0\left(  0\right)
}e_{\left(  0\right)  }^{0}}\left(  e_{\left(  a\right)  }^{0}\omega^{s\left(
an\right)  }-e_{\left(  a\right)  }^{0}\tilde{\omega}^{z\left(  an\right)
}\right)
\]

\begin{equation}
-\frac{e^{0\left(  s\right)  }}{e^{0\left(  0\right)  }e_{\left(  0\right)
}^{0}}\left(  e_{\left(  a\right)  }^{0}\omega^{n\left(  am\right)
}+e_{\left(  a\right)  }^{0}\tilde{\omega}^{m\left(  an\right)  }\right)
+\frac{e^{0\left(  n\right)  }}{e^{0\left(  0\right)  }e_{\left(  0\right)
}^{0}}\left(  e_{\left(  a\right)  }^{0}\omega^{m\left(  as\right)
}+e_{\left(  a\right)  }^{0}\tilde{\omega}^{s\left(  am\right)  }\right)
\label{eqnPR140}%
\end{equation}

\[
=\tilde{D}^{\prime\prime s\left(  mn\right)  }+\tilde{D}^{\prime\prime
n\left(  sm\right)  }-\tilde{D}^{\prime\prime m\left(  ns\right)  }.
\]
The first bracket can be eliminated using (\ref{eqnPR139}) and for the last
two we can eliminate half of contributions that gives%

\begin{equation}
2\tilde{\omega}^{~m(ns)}-2\frac{e^{0\left(  s\right)  }}{e^{0\left(  0\right)
}e_{\left(  0\right)  }^{0}}e_{\left(  a\right)  }^{0}\tilde{\omega}%
^{~m(an)}+2\frac{e^{0\left(  n\right)  }}{e^{0\left(  0\right)  }e_{\left(
0\right)  }^{0}}e_{a}^{0}\omega^{m\left(  as\right)  } \label{eqnPR141}%
\end{equation}

\[
=\tilde{D}^{\prime\prime s\left(  mn\right)  }+\tilde{D}^{\prime\prime
n\left(  sm\right)  }-\tilde{D}^{\prime\prime m\left(  ns\right)  }%
-\frac{e^{0\left(  m\right)  }}{e^{0\left(  0\right)  }e_{\left(  0\right)
}^{0}}e_{\left(  a\right)  }^{0}\tilde{D}^{\prime\prime a\left(  ns\right)
}+\frac{e^{0\left(  s\right)  }}{e^{0\left(  0\right)  }e_{\left(  0\right)
}^{0}}e_{\left(  a\right)  }^{0}\tilde{D}^{\prime\prime a\left(  mn\right)
}-\frac{e^{0\left(  n\right)  }}{e^{0\left(  0\right)  }e_{\left(  0\right)
}^{0}}e_{\left(  a\right)  }^{0}\tilde{D}^{\prime\prime a\left(  ms\right)
}.
\]
In this form terms $e_{\left(  a\right)  }^{0}\tilde{\omega}^{~m(an)}$ will
can be found by contraction with $e_{\left(  n\right)  }^{0}$%

\[
2e_{\left(  n\right)  }^{0}\tilde{\omega}^{~m(ns)}\frac{g^{00}}{e^{0\left(
0\right)  }e_{\left(  0\right)  }^{0}}=e_{\left(  n\right)  }^{0}\tilde
{D}^{\prime\prime s\left(  mn\right)  }+e_{\left(  n\right)  }^{0}\tilde
{D}^{\prime\prime n\left(  sm\right)  }-e_{\left(  n\right)  }^{0}\tilde
{D}^{\prime\prime m\left(  ns\right)  }%
\]

\begin{equation}
-\frac{e^{0\left(  m\right)  }}{e^{0\left(  0\right)  }e_{\left(  0\right)
}^{0}}e_{\left(  n\right)  }^{0}e_{\left(  a\right)  }^{0}\tilde{D}%
^{\prime\prime a\left(  ns\right)  }+\frac{e^{0\left(  s\right)  }%
}{e^{0\left(  0\right)  }e_{\left(  0\right)  }^{0}}e_{\left(  n\right)  }%
^{0}e_{\left(  a\right)  }^{0}\tilde{D}^{\prime\prime a\left(  mn\right)
}-\frac{e_{\left(  n\right)  }^{0}e^{0\left(  n\right)  }}{e^{0\left(
0\right)  }e_{\left(  0\right)  }^{0}}e_{\left(  a\right)  }^{0}\tilde
{D}^{\prime\prime a\left(  ms\right)  } \label{eqnPR142}%
\end{equation}
where $g^{00}$ is just a short notation for $e^{0\left(  \mu\right)
}e_{\left(  \mu\right)  }^{0}$.

Equation (\ref{eqnPR141}) after expressing $e_{\left(  a\right)  }^{0}%
\tilde{\omega}^{~m(an)}$ using (\ref{eqnPR142}) gives us the final solution:%

\begin{equation}
2\tilde{\omega}^{~m(ns)}=\tilde{D}^{\prime\prime s\left(  mn\right)  }%
+\tilde{D}^{\prime\prime n\left(  sm\right)  }-\tilde{D}^{\prime\prime
m\left(  ns\right)  } \label{eqnPR143}%
\end{equation}

\[
-\frac{e^{0\left(  m\right)  }}{e^{0\left(  0\right)  }e_{\left(  0\right)
}^{0}}e_{\left(  a\right)  }^{0}\tilde{D}^{\prime\prime a\left(  ns\right)
}+\frac{e^{0\left(  s\right)  }}{e^{0\left(  0\right)  }e_{\left(  0\right)
}^{0}}e_{\left(  a\right)  }^{0}\tilde{D}^{\prime\prime a\left(  mn\right)
}-\frac{e^{0\left(  n\right)  }}{e^{0\left(  0\right)  }e_{\left(  0\right)
}^{0}}e_{\left(  a\right)  }^{0}\tilde{D}^{\prime\prime a\left(  ms\right)  }%
\]

\[
+\frac{e^{0\left(  s\right)  }}{g^{00}}\left[  e_{\left(  a\right)  }%
^{0}\tilde{D}^{\prime\prime n\left(  ma\right)  }+e_{\left(  a\right)  }%
^{0}\tilde{D}^{\prime\prime a\left(  nm\right)  }-e_{\left(  a\right)  }%
^{0}\tilde{D}^{\prime\prime m\left(  an\right)  }\right]
\]

\[
+\frac{e^{0\left(  s\right)  }}{g^{00}}\left[  -\frac{e^{0\left(  m\right)  }%
}{e^{0\left(  0\right)  }e_{\left(  0\right)  }^{0}}e_{\left(  b\right)  }%
^{0}e_{\left(  a\right)  }^{0}\tilde{D}^{\prime\prime a\left(  bn\right)
}-\frac{e_{\left(  b\right)  }^{0}e^{0\left(  b\right)  }}{e^{0\left(
0\right)  }e_{\left(  0\right)  }^{0}}e_{\left(  a\right)  }^{0}\tilde
{D}^{\prime\prime a\left(  mn\right)  }\right]
\]

\[
-\frac{e^{0\left(  n\right)  }}{g^{00}}\left[  e_{\left(  a\right)  }%
^{0}\tilde{D}^{\prime\prime s\left(  ma\right)  }+e_{\left(  a\right)  }%
^{0}\tilde{D}^{\prime\prime a\left(  sm\right)  }-e_{\left(  a\right)  }%
^{0}\tilde{D}^{\prime\prime m\left(  as\right)  }\right]
\]

\[
-\frac{e^{0\left(  n\right)  }}{g^{00}}\left[  -\frac{e^{0\left(  m\right)  }%
}{e^{0\left(  0\right)  }e_{\left(  0\right)  }^{0}}e_{\left(  b\right)  }%
^{0}e_{\left(  a\right)  }^{0}\tilde{D}^{\prime\prime a\left(  bs\right)
}-\frac{e_{\left(  b\right)  }^{0}e^{0\left(  b\right)  }}{e^{0\left(
0\right)  }e_{\left(  0\right)  }^{0}}e_{\left(  a\right)  }^{0}\tilde
{D}^{\prime\prime a\left(  ms\right)  }\right]  .
\]

Note that the solution given by (\ref{eqnPR143}) is manifestly antisymmetric,
as it should be. In addition, because of linearity of $\tilde{D}^{\prime\prime
a\left(  ms\right)  },$ which is, in turn linear in contribution with
dependence on $\pi^{k\left(  0\right)  }$, $\pi^{k\left(  m\right)  },$ and
$e_{\mu\left(  \nu\right)  ,k}$ (see (\ref{eqnPR138})), we can calculate these
contributions separately. Let us, as an example, consider the result for
$\tilde{\omega}^{~m(ns)}\left(  \pi^{k\left(  0\right)  }\right)  $.

Using explicit form of $D^{p\left(  ns\right)  }$ (\ref{eqnPR128}), after
performing contractions and going to \textquotedblleft tilde\textquotedblright%
\ notation (\ref{eqnPR130a}), we obtain%

\begin{equation}
\tilde{D}^{s\left(  mn\right)  }\left(  \pi^{k\left(  0\right)  }\right)
=-e_{0}^{\left(  s\right)  }\left(  e^{0\left(  m\right)  }\tilde{V}%
^{n}-e^{0\left(  n\right)  }\tilde{V}^{m}\right)  \label{eqnPR144}%
\end{equation}
where we introduce a short notation (solution of (\ref{eqnPR128a}))%

\begin{equation}
\tilde{V}^{s}\equiv\tilde{\omega}_{q}^{..\left(  sq\right)  }=-\frac
{1}{2ee^{0\left(  0\right)  }}e_{k}^{\left(  s\right)  }\pi^{k\left(
0\right)  }. \label{eqnPR145}%
\end{equation}
According to the definition (\ref{eqnPR134}), we have for $\tilde{D}^{\prime
s\left(  mn\right)  }\left(  \pi^{k\left(  0\right)  }\right)  $%

\begin{equation}
\tilde{D}^{\prime s\left(  mn\right)  }\left(  \pi^{k\left(  0\right)
}\right)  =-e_{0}^{\left(  s\right)  }\left(  e^{0\left(  m\right)  }\tilde
{V}^{n}-e^{0\left(  n\right)  }\tilde{V}^{m}\right)  \label{eqnPR146}%
\end{equation}

\[
+\frac{\tilde{\eta}^{sn}}{D-2}\left(  \tilde{V}^{m}+\frac{e^{0\left(
m\right)  }}{e^{0\left(  0\right)  }}\frac{e_{\left(  a\right)  }^{0}%
}{e_{\left(  0\right)  }^{0}}\tilde{V}^{a}-\tilde{V}^{m}+e_{0\left(  0\right)
}e^{0\left(  0\right)  }\tilde{V}^{m}+e^{0\left(  m\right)  }e_{0\left(
a\right)  }\tilde{V}^{a}\right)
\]

\[
-\frac{\tilde{\eta}^{sm}}{D-2}\left(  \tilde{V}^{n}+\frac{e^{0\left(
n\right)  }}{e^{0\left(  0\right)  }}\frac{e_{\left(  a\right)  }^{0}%
}{e_{\left(  0\right)  }^{0}}\tilde{V}^{a}-\tilde{V}^{n}+e_{0\left(  0\right)
}e^{0\left(  0\right)  }\tilde{V}^{n}+e^{0\left(  n\right)  }e_{0\left(
a\right)  }\tilde{V}^{a}\right)  .
\]
Substitution it into (\ref{eqnPR143}) after some simplifications gives%

\begin{equation}
2\tilde{\omega}^{~m(ns)}\left(  \pi^{k\left(  0\right)  }\right)  =\frac
{1}{D-2}\left(  \tilde{\eta}^{sm}\tilde{V}^{n}-\tilde{\eta}^{nm}\tilde{V}%
^{s}\right)  +e^{0\left(  m\right)  }\left(  e_{0}^{\left(  s\right)  }%
\tilde{V}^{n}-e_{0}^{\left(  n\right)  }\tilde{V}^{s}\right)  \label{eqnPR147}%
\end{equation}

\[
+\left(  \tilde{V}^{n}e^{0\left(  s\right)  }-\tilde{V}^{s}e^{0\left(
n\right)  }\right)  \frac{e_{\left(  a\right)  }^{0}e_{0}^{\left(  a\right)
}}{g^{00}}e^{0\left(  m\right)  }-e_{0}^{\left(  m\right)  }\left(
e^{0\left(  n\right)  }\tilde{V}^{s}-e^{0\left(  s\right)  }\tilde{V}%
^{n}\right)  \frac{e_{\left(  0\right)  }^{0}e^{0\left(  0\right)  }}{g^{00}}%
\]

\[
+\left(  e_{0}^{\left(  n\right)  }e^{0\left(  s\right)  }-e_{0}^{\left(
s\right)  }e^{0\left(  n\right)  }\right)  \tilde{V}^{m}\frac{e_{\left(
0\right)  }^{0}e^{0\left(  0\right)  }}{g^{00}}+\frac{e^{0\left(  m\right)  }%
}{g^{00}}\left(  e_{0}^{\left(  n\right)  }e^{0\left(  s\right)  }-e^{0\left(
n\right)  }e_{0}^{\left(  s\right)  }\right)  e_{\left(  a\right)  }^{0}%
\tilde{V}^{a}%
\]

\[
+\frac{2\tilde{\eta}^{sm}}{D-2}\left[  \left(  -e_{0\left(  c\right)
}e^{0\left(  c\right)  }\tilde{V}^{n}+e^{0\left(  n\right)  }e_{0\left(
c\right)  }\tilde{V}^{c}\right)  -\frac{e^{0\left(  n\right)  }}{g^{00}%
}e_{\left(  a\right)  }^{0}\left(  -e_{0\left(  c\right)  }e^{0\left(
c\right)  }\tilde{V}^{a}+e^{0\left(  a\right)  }e_{0\left(  c\right)  }%
\tilde{V}^{c}\right)  \right]
\]

\[
-\frac{2\tilde{\eta}^{nm}}{D-2}\left[  \left(  -e_{0\left(  c\right)
}e^{0\left(  c\right)  }\tilde{V}^{s}+e^{0\left(  s\right)  }e_{0\left(
c\right)  }\tilde{V}^{c}\right)  -\frac{e^{0\left(  s\right)  }}{g^{00}%
}e_{\left(  a\right)  }^{0}\left(  -e_{0\left(  c\right)  }e^{0\left(
c\right)  }\tilde{V}^{a}+e^{0\left(  a\right)  }e_{0\left(  c\right)  }%
\tilde{V}^{c}\right)  \right]  .
\]

The solution for spatial connections (only dependence on $\pi^{k\left(
0\right)  }$) is quite big and one possible way to check it is to consider a
\textquotedblleft trace\textquotedblright\ of it by contraction with
$\tilde{\eta}_{ms}$. After not long calculations it gives the correct result%

\begin{equation}
2\tilde{\omega}_{c}^{.(nc)}\left(  \pi^{k\left(  0\right)  }\right)
=-\frac{1}{ee^{0\left(  0\right)  }}e_{k}^{\left(  n\right)  }\pi^{k\left(
0\right)  }. \label{eqnPR148}%
\end{equation}

Second consistency check is to consider three dimensional case which is also
an illustration that the Dirac approach is valid in all dimensions. There are
only two possible independent connections in three dimensional case,
$\tilde{\omega}^{~1(12)}$ and $\tilde{\omega}^{~2(12)}$, and by direct
calculation for both combinations we obtain%

\[
2\tilde{\omega}_{1}^{~(21)}\left(  \pi^{k\left(  0\right)  }\right)  =\frac
{1}{ee^{0\left(  0\right)  }}e_{k}^{\left(  2\right)  }\pi^{k\left(  0\right)
},\qquad2\tilde{\omega}_{2}^{~(12)}\left(  \pi^{k\left(  0\right)  }\right)
=\frac{1}{ee^{0\left(  0\right)  }}e_{k}^{\left(  1\right)  }\pi^{k\left(
0\right)  }.
\]

Actually, direct calculations to find $\omega_{p~~q)}^{~(n}$ can be avoided,
as it follows from the general \textquotedblleft trace\textquotedblright%
\ relation (\ref{eqnPR148}), because in three dimensions (due to antisymmetry)%

\[
\tilde{\omega}_{c}^{.(1c)}=\tilde{\omega}_{1}^{.(11)}+\tilde{\omega}%
_{2}^{.(12)}=\tilde{\omega}_{2}^{.(12)}%
\]
and using the inverse to (\ref{eqnPR129})%

\begin{equation}
\omega_{p~~q)}^{~(n}=e_{\left(  m\right)  p}\tilde{\omega}_{.....~~q)}^{m(n}
\label{eqnPR149}%
\end{equation}
we can find three dimensional expressions \cite{3D}, e.g.%

\begin{equation}
\omega_{1~~2)}^{~(1}=e_{\left(  m\right)  1}\tilde{\omega}_{.....~~2)}%
^{m(1}=e_{\left(  1\right)  1}\tilde{\omega}_{.....~~2)}^{1(1}+e_{\left(
2\right)  1}\tilde{\omega}_{.....~~2)}^{2(1}. \label{eqnPR150}%
\end{equation}

With the solution for $\tilde{\omega}^{~m(ns)}$ (\ref{eqnPR134}) (or
$\omega_{m}^{..(ns)}$, using (\ref{eqnPR149})) and found before for
$\omega_{m\left(  p0\right)  }$ (\ref{eqnPR66}), all spatial connections are
eliminated, as in $3D$ case, and we can find the reduced Hamiltonian with
fewer variables (without all pairs of canonical variables: spatial spin
connections and their momenta). The reduced Hamiltonian is the subject of next Section.

\section{ The reduced Hamiltonian}

Now we have the solution for all spatial connections, (\ref{eqnPR134}) and
(\ref{eqnPR66}), and can substitute them into the original Hamiltonian
(\ref{eqnPR31}) to obtain the reduced Hamiltonian, $\hat{H}_{T}$, with fewer
number of variables%

\begin{equation}
\hat{H}_{T}=\hat{H}_{c}+\pi^{0\left(  \rho\right)  }\dot{e}_{0\left(
\rho\right)  }+\Pi^{0\left(  \alpha\beta\right)  }\dot{\omega}_{0\left(
\alpha\beta\right)  }, \label{eqnPR160}%
\end{equation}

\begin{equation}
\hat{H}_{c}\left(  \pi^{k\left(  \rho\right)  },e_{\mu\left(  \alpha\right)
},\omega_{0\left(  \alpha\beta\right)  }\right)  =-eB^{\gamma\left(
\rho\right)  k\left(  \alpha\right)  \nu\left(  \beta\right)  }e_{\gamma
\left(  \rho\right)  ,k}\omega_{\nu\left(  \alpha\beta\right)  }%
+eA^{\mu\left(  \alpha\right)  \nu\left(  \beta\right)  }\omega_{\mu\left(
\alpha\gamma\right)  }\omega_{\nu~~\beta)}^{~(\gamma}. \label{eqnPR161}%
\end{equation}

Only two primary constraints survive the reduction and spatial connections
$\omega_{k\left(  \alpha\beta\right)  }$ in (\ref{eqnPR161}) are just short
notation; their expressions are given by (\ref{eqnPR134}) and (\ref{eqnPR66}).
Let us analyze the canonical part of $\hat{H}_{T}$. There is one, simple part
of $\hat{H}_{c}$, with contributions linear in temporal connections that we
combine together keeping the rest of terms $\hat{H}_{c}^{\prime}$ separately
(see Section V)%

\begin{equation}
\hat{H}_{c}=\left(  \frac{1}{2}e_{k}^{\left(  \alpha\right)  }\pi^{k\left(
\beta\right)  }-\frac{1}{2}e_{k}^{\left(  \beta\right)  }\pi^{k\left(
\alpha\right)  }-eB^{\gamma\left(  \rho\right)  k\left(  \alpha\right)
0\left(  \beta\right)  }e_{\gamma\left(  \rho\right)  ,k}\right)
\omega_{0\left(  \alpha\beta\right)  }+\hat{H}_{c}^{\prime}, \label{eqnPR162}%
\end{equation}

\begin{equation}
\hat{H}_{c}^{\prime}=-eB^{\gamma\left(  \rho\right)  k\left(  \alpha\right)
m\left(  \beta\right)  }e_{\gamma\left(  \rho\right)  ,k}\omega_{m\left(
\alpha\beta\right)  }+eA^{k\left(  \alpha\right)  m\left(  \beta\right)
}\omega_{k\left(  \alpha\gamma\right)  }\omega_{m~~\beta)}^{~(\gamma}.
\label{eqnPR163}%
\end{equation}

The expression in brackets is the rotational constraint $\chi^{0\left(
\alpha\beta\right)  }$ and, as expected, in all dimensions it is the same as
in three dimensions. So, this part of the reduced Hamiltonian can be written as%

\begin{equation}
\hat{H}_{c}=-\omega_{0\left(  \alpha\beta\right)  }\chi^{0\left(  \alpha
\beta\right)  }+\hat{H}_{c}^{\prime}. \label{eqnPR164}%
\end{equation}

Moreover (it is not difficult to check and it was demonstrated in our
\cite{3D}), this secondary rotational constraint has zero PBs with both
primary constraints and PB for two rotational constraints gives the
Poincar\'{e} relation (see (\ref{eqnPR83})) in all dimensions. Note that this
is the only part of the Hamiltonian with temporal connections (solutions of
second class constraints for spatial connections are independent on temporal
one, as we demonstrated in previous Section).

Let us continue with (\ref{eqnPR163}). Here we have also one simple
contribution which becomes transparent after separation of first term of
(\ref{eqnPR163})%

\begin{equation}
-eB^{\gamma\left(  \rho\right)  k\left(  \alpha\right)  m\left(  \beta\right)
}e_{\gamma\left(  \rho\right)  ,k}\omega_{m\left(  \alpha\beta\right)
}=-eB^{0\left(  \rho\right)  k\left(  \alpha\right)  m\left(  \beta\right)
}e_{0\left(  \rho\right)  ,k}\omega_{m\left(  \alpha\beta\right)
}-eB^{n\left(  \rho\right)  k\left(  \alpha\right)  m\left(  \beta\right)
}e_{n\left(  \rho\right)  ,k}\omega_{m\left(  \alpha\beta\right)  }.
\label{eqnPR170}%
\end{equation}

First part, non-zero in three dimensions, was also discussed and is quite
simple (see Section V)%

\begin{equation}
-eB^{0\left(  \rho\right)  k\left(  \alpha\right)  m\left(  \beta\right)
}e_{0\left(  \rho\right)  ,k}\omega_{m\left(  \alpha\beta\right)  }%
=-\pi^{k\left(  \rho\right)  }e_{0\left(  \rho\right)  ,k}. \label{eqnPR171}%
\end{equation}
It gives simple (first) contribution $\chi_{1}^{0\left(  \rho\right)  }$ into
the secondary translational constraint in all dimensions, the same that was
found in three dimensions \cite{3D},%

\begin{equation}
\chi_{1}^{0\left(  \rho\right)  }=\pi_{,k}^{k\left(  \rho\right)  }.
\label{eqnPR172}%
\end{equation}
This contribution has zero PBs with all primary constraints and by itself
gives all relations of the Poincar\'{e} algebra (\ref{eqnPR81}-\ref{eqnPR82})
and, after integrations by part, allows to write the corresponding part of the
Hamiltonian as%

\begin{equation}
\hat{H}_{c1}^{\prime}=-e_{0\left(  \rho\right)  }\chi_{1}^{0\left(
\rho\right)  }. \label{eqnPR173}%
\end{equation}

The second term of (\ref{eqnPR170}) is manifestly zero in three dimensions
(there are no three distinct values for spatial components of external indices
in three spacetime dimensions - only two spatial components are available) and
the same happens with many terms quadratic in spatial connections (see
(\ref{eqnPR73})).

Let us discuss the effect of these (\textquotedblleft
invisible\textquotedblright\ in three dimensions) terms on the translational
constraint and on PBs among constraints found in three dimensions \cite{3D}.

First, we substitute our solution for $\omega_{m\left(  p0\right)  }$
(\ref{eqnPR66}) into $\hat{H}_{c}^{\prime}$ (\ref{eqnPR163}) (remember that
one small contribution is already found and only the second term of
(\ref{eqnPR170}) is left), where we explicitly separate two kinds of spatial
connections ($\omega_{m\left(  p0\right)  }$ and $\omega_{m\left(  pq\right)
}$)%

\begin{equation}
\hat{H}_{c}^{\prime}=-eB^{n\left(  \rho\right)  k\left(  p\right)  m\left(
q\right)  }e_{n\left(  \rho\right)  ,k}\omega_{m\left(  pq\right)
}-2eB^{n\left(  q\right)  k\left(  p\right)  m\left(  0\right)  }e_{n\left(
q\right)  ,k}\omega_{m\left(  p0\right)  } \label{eqnPR174}%
\end{equation}

\[
+eA^{k\left(  p\right)  m\left(  q\right)  }\omega_{k\left(  pn\right)
}\omega_{m~~q)}^{~(n}+eA^{k\left(  p\right)  m\left(  q\right)  }%
\omega_{k\left(  p0\right)  }\omega_{m~~q)}^{~(0}+2eA^{k\left(  0\right)
m\left(  q\right)  }\omega_{k\left(  0p\right)  }\omega_{m~~q)}^{~(p}.
\]
The result of such a substitution converts $\hat{H}_{c}^{\prime}$ into the
expression where we have three groups of terms%

\begin{equation}
\hat{H}_{c}^{\prime}=\hat{H}_{c}^{\prime}\left(  0\right)  +\hat{H}%
_{c}^{\prime}\left(  1\right)  +\hat{H}_{c}^{\prime}\left(  2\right)
\label{eqnPR175}%
\end{equation}
classified by order of spatial connections $\omega_{m\left(  pq\right)  }$
($\omega_{m\left(  p0\right)  }$ is substituted). For the first part, $\hat
{H}_{c}^{\prime}\left(  0\right)  $, we have%

\[
\hat{H}_{c}^{\prime}\left(  0\right)  =eB^{n\left(  q\right)  k\left(
p\right)  m\left(  0\right)  }e_{n\left(  q\right)  ,k}\left[  \frac
{1}{ee^{0\left(  0\right)  }}I_{m\left(  p\right)  a\left(  b\right)  }%
\pi^{a\left(  b\right)  }\right]
\]

\begin{equation}
-\frac{1}{4}eA^{k\left(  p\right)  m\left(  q\right)  }\left[  \frac
{1}{ee^{0\left(  0\right)  }}I_{k\left(  p\right)  a\left(  b\right)  }%
\pi^{a\left(  b\right)  }\right]  \left[  \frac{1}{ee_{\left(  0\right)  }%
^{0}}I_{m\left(  q\right)  a\left(  b\right)  }\pi^{a\left(  b\right)
}\right]  . \label{eqnPR176}%
\end{equation}

Let us analyze contributions into the secondary constraint that are created
but this part (we repeat, the first term of (\ref{eqnPR176}) is manifestly
zero in three dimensions)%

\begin{equation}
\dot{\pi}^{0\left(  \sigma\right)  }=\left\{  \pi^{0\left(  \sigma\right)
},\hat{H}_{c}^{\prime}\left(  0\right)  \right\}  =-\frac{\delta\hat{H}%
_{c}^{\prime}\left(  0\right)  }{\delta e_{0\left(  \sigma\right)  }}.
\label{eqnPR177}%
\end{equation}
We have%

\begin{equation}
-\frac{\delta\hat{H}_{c}^{\prime}\left(  0\right)  }{\delta e_{0\left(
\sigma\right)  }}=-\frac{\delta}{\delta e_{0\left(  \sigma\right)  }}\left(
eB^{n\left(  q\right)  k\left(  p\right)  m\left(  0\right)  }\right)  \left[
e_{n\left(  q\right)  ,k}\right]  \left[  \frac{1}{ee^{0\left(  0\right)  }%
}I_{m\left(  p\right)  a\left(  b\right)  }\pi^{a\left(  b\right)  }\right]
\label{eqnPR178}%
\end{equation}

\[
+\frac{1}{4}\frac{\delta}{\delta e_{0\left(  \sigma\right)  }}\left(
eA^{k\left(  p\right)  m\left(  q\right)  }\right)  \left[  \frac
{1}{ee^{0\left(  0\right)  }}I_{k\left(  p\right)  a\left(  b\right)  }%
\pi^{a\left(  b\right)  }\right]  \left[  \frac{1}{ee_{\left(  0\right)  }%
^{0}}I_{m\left(  q\right)  a\left(  b\right)  }\pi^{a\left(  b\right)
}\right]  =\chi^{0\left(  \sigma\right)  }\left(  0\right)  .
\]

Note that variation of expressions in square brackets is zero. This part of
the secondary constraint has obviously zero PBs with primary rotational and
also with primary translational constraints just because of antisymmetry of
$ABC$ functions - the only part which is affected by a variation. For example,%

\begin{equation}
\frac{\delta}{\delta e_{0\left(  \tau\right)  }}\frac{\delta}{\delta
e_{0\left(  \sigma\right)  }}\left(  eA^{k\left(  p\right)  m\left(  q\right)
}\right)  =\frac{\delta}{\delta e_{0\left(  \tau\right)  }}\left(
eB^{0\left(  \sigma\right)  k\left(  p\right)  m\left(  q\right)  }\right)
=eC^{0\left(  \tau\right)  0\left(  \sigma\right)  k\left(  p\right)  m\left(
q\right)  }=0. \label{eqnPR179}%
\end{equation}

So, at least for this part, despite appearance of additional terms in higher
dimensions, properties which were found in three dimensions \cite{3D} survive.
Let us check possibility to present this part of the Hamiltonian as a linear
combination of components of a translational constraint. Again, $ABC$
properties make calculations simple. Let us illustrate this. In the second
line we have%

\begin{equation}
\frac{\delta}{\delta e_{0\left(  \sigma\right)  }}\left(  eA^{k\left(
p\right)  m\left(  q\right)  }\right)  =eB^{0\left(  \sigma\right)  k\left(
p\right)  m\left(  q\right)  } \label{eqnPR180}%
\end{equation}
that we have to contract with $e_{0\left(  \sigma\right)  }$ with a hope to
have $\hat{H}_{c}^{\prime}\left(  0\right)  =-e_{0\left(  \sigma\right)  }%
\chi^{0\left(  \sigma\right)  }\left(  0\right)  $. We used properties of $B$
(see (\ref{eqnPR6})) which is a kind of expansion in an external index but
similar relation exists for the internal one%

\begin{equation}
B^{\gamma\left(  \rho\right)  \mu\left(  \alpha\right)  \nu\left(
\beta\right)  }=e^{\gamma\left(  \rho\right)  }A^{\mu\left(  \alpha\right)
\nu\left(  \beta\right)  }+e^{\mu\left(  \rho\right)  }A^{\nu\left(
\alpha\right)  \gamma\left(  \beta\right)  }+e^{\nu\left(  \rho\right)
}A^{\gamma\left(  \alpha\right)  \mu\left(  \beta\right)  } \label{eqnPR181}%
\end{equation}
In our case it gives%

\begin{equation}
\frac{\delta}{\delta e_{0\left(  \sigma\right)  }}\left(  eA^{k\left(
p\right)  m\left(  q\right)  }\right)  =e\left(  e^{0\left(  \sigma\right)
}A^{k\left(  p\right)  m\left(  q\right)  }+e^{k\left(  \sigma\right)
}A^{m\left(  p\right)  0\left(  q\right)  }+e^{m\left(  \sigma\right)
}A^{0\left(  p\right)  k\left(  q\right)  }\right)  \label{eqnPR182}%
\end{equation}
and after contraction we have%

\begin{equation}
e_{0\left(  \sigma\right)  }\frac{\delta}{\delta e_{0\left(  \sigma\right)  }%
}\left(  eA^{k\left(  p\right)  m\left(  q\right)  }\right)  =eA^{k\left(
p\right)  m\left(  q\right)  }. \label{eqnPR183}%
\end{equation}

A similar, external, expansion exists for $C$%

\[
C^{\sigma\left(  \tau\right)  \gamma\left(  \rho\right)  \mu\left(
\alpha\right)  \nu\left(  \beta\right)  }=
\]

\begin{equation}
e^{\sigma\left(  \tau\right)  }B^{\gamma\left(  \rho\right)  \mu\left(
\alpha\right)  \nu\left(  \beta\right)  }-e^{\gamma\left(  \tau\right)
}B^{\mu\left(  \rho\right)  \nu\left(  \alpha\right)  \sigma\left(
\beta\right)  }+e^{\mu\left(  \tau\right)  }B^{\nu\left(  \rho\right)
\sigma\left(  \alpha\right)  \gamma\left(  \beta\right)  }-e^{\nu\left(
\tau\right)  }B^{\sigma\left(  \rho\right)  \gamma\left(  \alpha\right)
\mu\left(  \beta\right)  }. \label{eqnPR184}%
\end{equation}

In short, the translational constraint has additional contributions in higher
dimensions compare with $3D$ but properties of a constraint and possibility to
present the Hamiltonian as a linear combination of it survives. For this part
we demonstrated that%

\begin{equation}
\hat{H}_{c}^{\prime}\left(  0\right)  =-e_{0\left(  \sigma\right)  }%
\chi^{0\left(  \sigma\right)  }\left(  0\right)  \label{eqnPR185}%
\end{equation}
which is the same relation as in $3D$ case (\ref{eqnPR80}). Of course, to make
the final conclusion, all terms have to be considered, so we are looking now
for contributions linear and quadratic in $\omega_{m\left(  pq\right)  }$
which turned to be better to consider together ($\hat{H}_{c}^{\prime}\left(
1\right)  +\hat{H}_{c}^{\prime}\left(  2\right)  $).

Here we collect what is left after separating $\hat{H}_{c}^{\prime}\left(
0\right)  $ (we advise not to simplify expressions)%

\begin{equation}
\hat{H}_{c}^{\prime}\left(  1\right)  +\hat{H}_{c}^{\prime}\left(  2\right)
=\hat{H}_{c}^{\prime}\left(  1+2\right)  = \label{eqnPR186}%
\end{equation}

\[
-eB^{n\left(  \rho\right)  k\left(  p\right)  m\left(  q\right)  }\left[
e_{n\left(  \rho\right)  ,k}\right]  \omega_{m\left(  pq\right)
}+2eB^{n\left(  q\right)  k\left(  p\right)  m\left(  0\right)  }\left[
e_{n\left(  q\right)  ,k}\right]  \left[  \frac{e^{0\left(  d\right)  }%
}{2e^{0\left(  0\right)  }}I_{m\left(  p\right)  c\left(  d\right)
}E^{c\left(  a\right)  f\left(  b\right)  }\omega_{f\left(  ab\right)  }%
-\frac{e^{0\left(  a\right)  }}{e^{0\left(  0\right)  }}\omega_{m\left(
ap\right)  }\right]
\]

\[
+eA^{k\left(  p\right)  m\left(  q\right)  }\omega_{k\left(  pn\right)
}\omega_{m~~q)}^{~(n}%
\]

\[
+2eA^{k\left(  0\right)  m\left(  q\right)  }\omega_{m~~q)}^{~(p}\left[
\frac{1}{2ee^{0\left(  0\right)  }}I_{k\left(  p\right)  a\left(  b\right)
}\pi^{a\left(  b\right)  }+\frac{e^{0\left(  d\right)  }}{2e^{0\left(
0\right)  }}I_{k\left(  p\right)  c\left(  d\right)  }E^{c\left(  a\right)
f\left(  b\right)  }\omega_{f\left(  ab\right)  }-\frac{e^{0\left(  a\right)
}}{e^{0\left(  0\right)  }}\omega_{k\left(  ap\right)  }\right]
\]

\[
-eA^{k\left(  p\right)  m\left(  q\right)  }\left[  \frac{1}{ee^{0\left(
0\right)  }}I_{k\left(  p\right)  a\left(  b\right)  }\pi^{a\left(  b\right)
}\right]  \left[  \frac{e^{0\left(  d\right)  }}{2e_{\left(  0\right)  }^{0}%
}I_{m\left(  q\right)  c\left(  d\right)  }E^{c\left(  a\right)  f\left(
b\right)  }\omega_{f\left(  ab\right)  }-\frac{e^{0\left(  a\right)  }%
}{e_{\left(  0\right)  }^{0}}\omega_{m\left(  aq\right)  }\right]
\]

\[
-eA^{k\left(  p\right)  m\left(  q\right)  }\left[  \frac{e^{0\left(
d\right)  }}{2e^{0\left(  0\right)  }}I_{k\left(  p\right)  c\left(  d\right)
}E^{c\left(  a\right)  f\left(  b\right)  }\omega_{f\left(  ab\right)  }%
-\frac{e^{0\left(  a\right)  }}{e^{0\left(  0\right)  }}\omega_{k\left(
ap\right)  }\right]  \left[  \frac{e^{0\left(  d\right)  }}{2e_{\left(
0\right)  }^{0}}I_{m\left(  q\right)  c\left(  d\right)  }E^{c\left(
a\right)  f\left(  b\right)  }\omega_{f\left(  ab\right)  }-\frac{e^{0\left(
a\right)  }}{e_{\left(  0\right)  }^{0}}\omega_{m\left(  aq\right)  }\right]
.
\]

To find a contribution into the secondary constraint from this part we, as
with part $\hat{H}_{c}^{\prime}\left(  0\right)  $, have to perform variation
of this expression with respect to $e_{0\left(  \sigma\right)  }$. Instead of
explicit substitution of the solution for $\omega_{m\left(  pq\right)  }$
(which would drug us into extremely cumbersome calculations), we use the following%

\begin{equation}
\frac{\delta\hat{H}_{c}^{\prime}\left(  1+2\right)  }{\delta e_{0\left(
\sigma\right)  }}=\frac{\partial\hat{H}_{c}^{\prime}\left(  1+2\right)
}{\partial e_{0\left(  \sigma\right)  }}+\frac{\partial\hat{H}_{c}^{\prime
}\left(  1+2\right)  }{\partial\omega_{x\left(  yz\right)  }}\frac
{\partial\omega_{x\left(  yz\right)  }}{\partial e_{0\left(  \sigma\right)  }%
}. \label{eqnPR187}%
\end{equation}

First variation is extremely simple and, actually, is exactly the same as what
we did in $\hat{H}_{c}^{\prime}\left(  0\right)  $-part. Variation of all
expressions in square brackets are zero because of%

\begin{equation}
\frac{\partial}{\partial e_{0\left(  \sigma\right)  }}\left(  \frac
{1}{ee^{0\left(  0\right)  }}\right)  =\frac{\partial}{\partial e_{0\left(
\sigma\right)  }}\left(  \frac{e^{0\left(  a\right)  }}{e^{0\left(  0\right)
}}\right)  =\frac{\partial}{\partial e_{0\left(  \sigma\right)  }}I_{k\left(
p\right)  a\left(  b\right)  }=\frac{\partial}{\partial e_{0\left(
\sigma\right)  }}\gamma^{k\left(  m\right)  }=\frac{\partial}{\partial
e_{0\left(  \sigma\right)  }}E^{c\left(  a\right)  f\left(  b\right)  }=0.
\label{eqnPR188}%
\end{equation}
As in previous case, we have to consider only variations of $eA$ and $eB$
which gives the same, as for $\hat{H}_{c}^{\prime}\left(  0\right)  ,$ result:%

\begin{equation}
\hat{H}_{c}^{\prime}\left(  1+2\right)  =-e_{0\left(  \sigma\right)  }%
\chi^{0\left(  \sigma\right)  }\left(  1+2\right)  . \label{eqnPR189}%
\end{equation}

This is the final answer and the Hamiltonian is the linear combination of
secondary constraints (up to a total derivative) as in three dimensional case
if, of course, the second part of variation in (\ref{eqnPR187}) gives zero.
Let us prove this. It is obvious from the solution for $\omega_{x\left(
yz\right)  }$ (see (\ref{eqnPR143})) that there are terms with non-zero
variation in it but contraction $\frac{\partial\hat{H}_{c}^{\prime}\left(
1+2\right)  }{\partial\omega_{x\left(  yz\right)  }}\frac{\partial
\omega_{x\left(  yz\right)  }}{\partial e_{0\left(  \sigma\right)  }}$ in
(\ref{eqnPR187}) still can be zero. Again, the direct substitution here is too
long and we can try to relate $\frac{\partial\hat{H}_{c}^{\prime}\left(
1+2\right)  }{\partial\omega_{x\left(  yz\right)  }}$ with the known equation
for a connection (of course, with one where the multipliers were already eliminated).

So, as we did in obtaining (\ref{eqnPR124}), we perform variation of
$\frac{\partial\hat{H}_{c}^{\prime}\left(  1+2\right)  }{\partial
\omega_{x\left(  yz\right)  }}$ and present the result using \textquotedblleft%
$\gamma-E$\textquotedblright\ notation%

\begin{equation}
\frac{\partial\hat{H}_{c}^{\prime}\left(  1+2\right)  }{\partial
\omega_{x\left(  yz\right)  }}= \label{eqnPR190}%
\end{equation}

\[
+eE^{x\left(  y\right)  m\left(  q\right)  }\omega_{m~~q)}^{~(z}-eE^{x\left(
z\right)  m\left(  q\right)  }\omega_{m~~q)}^{~(y}+eE^{k\left(  p\right)
x\left(  z\right)  }\frac{e^{0\left(  y\right)  }}{e_{\left(  0\right)  }^{0}%
}\frac{e^{0\left(  a\right)  }}{e^{0\left(  0\right)  }}\omega_{k\left(
ap\right)  }-eE^{k\left(  p\right)  x\left(  y\right)  }\frac{e^{0\left(
a\right)  }}{e^{0\left(  0\right)  }}\frac{e^{0\left(  z\right)  }}{e_{\left(
0\right)  }^{0}}\omega_{k\left(  ap\right)  }%
\]

\[
+eA^{k\left(  0\right)  m\left(  q\right)  }\omega_{m~~q)}^{~(p}%
\frac{e^{0\left(  d\right)  }}{e^{0\left(  0\right)  }}I_{k\left(  p\right)
c\left(  d\right)  }E^{c\left(  y\right)  x\left(  z\right)  }+eA^{k\left(
p\right)  m\left(  0\right)  }\frac{e^{0\left(  q\right)  }}{e^{0\left(
0\right)  }}\frac{e^{0\left(  a\right)  }}{e^{0\left(  0\right)  }}%
\omega_{k\left(  ap\right)  }\frac{e^{0\left(  d\right)  }}{e_{\left(
0\right)  }^{0}}I_{m\left(  q\right)  c\left(  d\right)  }E^{c\left(
y\right)  x\left(  z\right)  }%
\]

\[
+\hat{D}^{x\left(  yz\right)  }\left(  e_{\mu\left(  \nu\right)  ,k}\right)
+\hat{D}^{x\left(  yz\right)  }\left(  \pi^{k\left(  m\right)  }\right)
+\hat{D}^{x\left(  yz\right)  }\left(  \pi^{k\left(  0\right)  }\right)
\]
where, as in (\ref{eqnPR124}), we introduce $\hat{D}^{x\left(  yz\right)  }$
for similar terms which are%

\[
\hat{D}^{x\left(  yz\right)  }\left(  e_{\mu\left(  \nu\right)  ,k}\right)
=-eB^{n\left(  \rho\right)  k\left(  y\right)  x\left(  z\right)  }e_{n\left(
\rho\right)  ,k}+eB^{n\left(  q\right)  k\left(  p\right)  m\left(  0\right)
}e_{n\left(  q\right)  ,k}\frac{e^{0\left(  d\right)  }}{e^{0\left(  0\right)
}}I_{m\left(  p\right)  c\left(  d\right)  }E^{c\left(  y\right)  x\left(
z\right)  }%
\]

\begin{equation}
-eB^{n\left(  q\right)  k\left(  z\right)  x\left(  0\right)  }e_{n\left(
q\right)  ,k}\frac{e^{0\left(  y\right)  }}{e^{0\left(  0\right)  }%
}+eB^{n\left(  q\right)  k\left(  y\right)  x\left(  0\right)  }e_{n\left(
q\right)  ,k}\frac{e^{0\left(  z\right)  }}{e^{0\left(  0\right)  }};
\label{eqnPR191}%
\end{equation}

\[
\hat{D}^{x\left(  yz\right)  }\left(  \pi^{k\left(  m\right)  }\right)
=+eA^{k\left(  0\right)  x\left(  z\right)  }\left[  \frac{1}{2ee^{0\left(
0\right)  }}I_{k\left(  y\uparrow\right)  a\left(  b\right)  }\pi^{a\left(
b\right)  }\right]  -eA^{k\left(  0\right)  x\left(  y\right)  }\left[
\frac{1}{2ee^{0\left(  0\right)  }}I_{k\left(  z\uparrow\right)  a\left(
b\right)  }\pi^{a\left(  b\right)  }\right]
\]

\[
-eA^{k\left(  p\right)  m\left(  q\right)  }\left[  \frac{1}{2ee^{0\left(
0\right)  }}I_{k\left(  p\right)  a\left(  b\right)  }\pi^{a\left(  b\right)
}\right]  \left[  \frac{e^{0\left(  d\right)  }}{e_{\left(  0\right)  }^{0}%
}I_{m\left(  q\right)  c\left(  d\right)  }E^{c\left(  y\right)  x\left(
z\right)  }\right]
\]

\begin{equation}
+eA^{k\left(  p\right)  x\left(  z\right)  }\left[  \frac{1}{2ee^{0\left(
0\right)  }}I_{k\left(  p\right)  a\left(  b\right)  }\pi^{a\left(  b\right)
}\right]  \frac{e^{0\left(  y\right)  }}{e_{\left(  0\right)  }^{0}%
}-eA^{k\left(  p\right)  x\left(  y\right)  }\left[  \frac{1}{2ee^{0\left(
0\right)  }}I_{k\left(  p\right)  a\left(  b\right)  }\pi^{a\left(  b\right)
}\right]  \frac{e^{0\left(  z\right)  }}{e_{\left(  0\right)  }^{0}};
\label{eqnPR192}%
\end{equation}

\[
\hat{D}^{x\left(  yz\right)  }\left(  \pi^{k\left(  0\right)  }\right)
=+eA^{k\left(  0\right)  x\left(  z\right)  }\frac{e^{0\left(  d\right)  }%
}{2e^{0\left(  0\right)  }}I_{k\left(  y\uparrow\right)  c\left(  d\right)
}E^{c\left(  a\right)  f\left(  b\right)  }\omega_{f\left(  ab\right)
}-eA^{k\left(  0\right)  x\left(  y\right)  }\frac{e^{0\left(  d\right)  }%
}{2e^{0\left(  0\right)  }}I_{k\left(  z\uparrow\right)  c\left(  d\right)
}E^{c\left(  a\right)  f\left(  b\right)  }\omega_{f\left(  ab\right)  }%
\]

\[
-e\left(  \frac{e^{0\left(  p\right)  }}{e^{0\left(  0\right)  }}A^{k\left(
0\right)  m\left(  q\right)  }+A^{k\left(  p\right)  m\left(  0\right)  }%
\frac{e^{0\left(  q\right)  }}{e^{0\left(  0\right)  }}\right)  \frac
{e^{0\left(  d\right)  }}{2e^{0\left(  0\right)  }}I_{k\left(  p\right)
c\left(  d\right)  }E^{c\left(  a\right)  f\left(  b\right)  }\omega_{f\left(
ab\right)  }\frac{e^{0\left(  d\right)  }}{e_{\left(  0\right)  }^{0}%
}I_{m\left(  q\right)  c\left(  d\right)  }E^{c\left(  y\right)  x\left(
z\right)  }%
\]

\begin{equation}
-e\frac{e^{0\left(  q\right)  }}{2e^{0\left(  0\right)  }}E^{m\left(
a\right)  f\left(  b\right)  }\omega_{f\left(  ab\right)  }\frac{e^{0\left(
d\right)  }}{e_{\left(  0\right)  }^{0}}I_{m\left(  q\right)  c\left(
d\right)  }E^{c\left(  y\right)  x\left(  z\right)  } \label{eqnPR193}%
\end{equation}

\[
+e\frac{e^{0\left(  p\right)  }}{e^{0\left(  0\right)  }}A^{k\left(  0\right)
x\left(  z\right)  }\frac{e^{0\left(  d\right)  }}{2e^{0\left(  0\right)  }%
}I_{k\left(  p\right)  c\left(  d\right)  }E^{c\left(  a\right)  f\left(
b\right)  }\omega_{f\left(  ab\right)  }\frac{e^{0\left(  y\right)  }%
}{e_{\left(  0\right)  }^{0}}-e\frac{e^{0\left(  p\right)  }}{e^{0\left(
0\right)  }}A^{k\left(  0\right)  x\left(  y\right)  }\frac{e^{0\left(
d\right)  }}{2e^{0\left(  0\right)  }}I_{k\left(  p\right)  c\left(  d\right)
}E^{c\left(  a\right)  f\left(  b\right)  }\omega_{f\left(  ab\right)  }%
\frac{e^{0\left(  z\right)  }}{e_{\left(  0\right)  }^{0}}.
\]

Equation (\ref{eqnPR190}), or rather the result of variation $\frac
{\partial\hat{H}_{c}^{\prime}\left(  1+2\right)  }{\partial\omega_{x\left(
yz\right)  }}$ (it is not an equation of motion), we compare with the similar
equation (\ref{eqnPR130}), but not in \textquotedblleft
tilde\textquotedblright\ notation which is the result of further contractions
that were needed for use of the Einstein permutation, and not one
(\ref{eqnPR124}) which has the form of (\ref{eqnPR190}), but with the
multiplier. We substitute the solution for the multiplier (\ref{eqnPR132a})
into (\ref{eqnPR124}) and obtain the equation with which we compare
(\ref{eqnPR190}). We use the same letters ($x\left(  yz\right)  $) for free
indices to make a comparison more transparent. Of course, terms with temporal
connections are cancelled out, as it was found before, as well as some
additional cancellation occurs, and the final result is%

\[
-E^{x\left(  y\right)  k\left(  q\right)  }\omega_{k~~q)}^{~(z}+E^{x\left(
z\right)  k\left(  q\right)  }\omega_{k~~q)}^{~(y}+\frac{e^{0\left(  y\right)
}}{e^{0\left(  0\right)  }}E^{x\left(  z\right)  n^{\prime}\left(  q^{\prime
}\right)  }\frac{e^{0\left(  a\right)  }}{e_{\left(  0\right)  }^{0}}%
\omega_{n^{\prime}\left(  aq^{\prime}\right)  }-\frac{e^{0\left(  z\right)  }%
}{e^{0\left(  0\right)  }}E^{x\left(  y\right)  n^{\prime}\left(  q^{\prime
}\right)  }\frac{e^{0\left(  a\right)  }}{e_{\left(  0\right)  }^{0}}%
\omega_{n^{\prime}\left(  aq^{\prime}\right)  }%
\]

\begin{equation}
+D^{x\left(  yz\right)  }\left(  \pi^{k\left(  m\right)  }\right)
+\gamma^{x\left(  z\right)  }\frac{1}{\left(  D-2\right)  }\left(
\tilde{\omega}_{s}^{.(ys)}+\frac{e^{0\left(  y\right)  }}{e^{0\left(
0\right)  }}\frac{e_{\left(  a\right)  }^{0}}{e_{\left(  0\right)  }^{0}%
}\tilde{\omega}_{s}^{..\left(  as\right)  }+\tilde{D}_{a}^{..\left(
ay\right)  }\left(  \pi^{k\left(  m\right)  }\right)  \right)
\label{eqnPR194}%
\end{equation}

\[
-\gamma^{x\left(  y\right)  }\frac{1}{\left(  D-2\right)  }\left(
\tilde{\omega}_{s}^{.(zs)}+\frac{e^{0\left(  z\right)  }}{e^{0\left(
0\right)  }}\frac{e_{\left(  a\right)  }^{0}}{e_{\left(  0\right)  }^{0}%
}\tilde{\omega}_{s}^{..\left(  as\right)  }+\tilde{D}_{m}^{..\left(
mz\right)  }\left(  \pi^{k\left(  m\right)  }\right)  \right)
\]

\[
+D^{x\left(  yz\right)  }\left(  \pi^{k\left(  0\right)  }\right)
+\gamma^{x\left(  z\right)  }\frac{1}{D-2}\tilde{D}_{a}^{..\left(  ay\right)
}\left(  \pi^{k\left(  0\right)  }\right)  -\gamma^{x\left(  y\right)  }%
\frac{1}{D-2}\tilde{D}_{m}^{..\left(  mz\right)  }\left(  \pi^{k\left(
0\right)  }\right)
\]

\[
+D^{x\left(  yz\right)  }\left(  e_{\mu\left(  \nu\right)  ,k}\right)
+\gamma^{x\left(  z\right)  }\frac{1}{D-2}\tilde{D}_{a}^{..\left(  ay\right)
}\left(  e_{\mu\left(  \nu\right)  ,k}\right)  -\gamma^{x\left(  y\right)
}\frac{1}{D-2}\tilde{D}_{m}^{..\left(  mz\right)  }\left(  e_{\mu\left(
\nu\right)  ,k}\right)  =0.
\]

Now, to simplify calculations, we can multiply this equality by $e$ and add to
RHS of (\ref{eqnPR190}). The most difficult for calculation part (first line
of (\ref{eqnPR190})) will disappear. The second line of (\ref{eqnPR190})
survives but the rest is already expressed in terms of momenta and N-beins,
and we avoided a cumbersome substitution. Moreover, let us look at the second
line of (\ref{eqnPR190}). Both contributions have $x\left(  yz\right)  $
indices in the following form: $E^{c\left(  y\right)  x\left(  z\right)  }.$
We almost immediately have zero for the corresponding contributions%

\begin{equation}
\frac{\partial\hat{H}_{c}^{\prime}\left(  1+2\right)  }{\partial
\omega_{x\left(  yz\right)  }}\frac{\partial\omega_{x\left(  yz\right)  }%
}{\partial e_{0\left(  \sigma\right)  }}=X_{c}E^{c\left(  y\right)  x\left(
z\right)  }\frac{\partial\omega_{x\left(  yz\right)  }}{\partial e_{0\left(
\sigma\right)  }}=X_{c}\frac{\partial E^{c\left(  y\right)  x\left(  z\right)
}\omega_{x\left(  yz\right)  }}{\partial e_{0\left(  \sigma\right)  }}=0.
\label{eqnPR195}%
\end{equation}

Here, because variation of $E^{c\left(  y\right)  x\left(  z\right)  }$ is
zero (\ref{eqnPR188}), we can interchange order of contraction and variation
and obtain the combination%

\begin{equation}
E^{c\left(  y\right)  x\left(  z\right)  }\omega_{x\left(  yz\right)
}=2\gamma^{c\left(  y\right)  }\gamma^{x\left(  z\right)  }\omega_{x\left(
yz\right)  }=-\frac{1}{ee^{0\left(  0\right)  }}\pi^{c\left(  0\right)  },
\label{eqnPR196}%
\end{equation}
variation of which with respect to $e_{0\left(  \sigma\right)  }$ is zero
(\ref{eqnPR188}). Similarly, in the rest of expression we consider separately
three different contributions which are proportional to $\pi^{k\left(
0\right)  }$, $\pi^{k\left(  m\right)  }$, and terms without momenta (with
derivatives of tetrads) and check this relation. Note that in (\ref{eqnPR194})
all terms proportional to $\frac{1}{D-2}$ have a factor $\gamma^{x\left(
z\right)  }$ or $\gamma^{x\left(  y\right)  }$. These factors, as in
(\ref{eqnPR195}), can be moved under variation and contracted with
$\omega_{x\left(  yz\right)  }$. This gives similar to (\ref{eqnPR196}) result%

\begin{equation}
\gamma^{x\left(  z\right)  }\omega_{x\left(  yz\right)  }=-\frac
{1}{ee^{0\left(  0\right)  }\left(  D-2\right)  }e_{m\left(  y\right)  }%
\pi^{m\left(  0\right)  } \label{eqnPR197}%
\end{equation}
which has zero variation. The rest of terms have the following structure%

\begin{equation}
\hat{D}^{x\left(  yz\right)  }\left(  \pi^{k\left(  m\right)  }\right)
+eD^{x\left(  yz\right)  }\left(  \pi^{k\left(  m\right)  }\right)
\label{eqnPR198}%
\end{equation}
and there are similar expressions for $\pi^{k\left(  0\right)  }$ and
$e_{\mu\left(  \nu\right)  ,k}$. By performing the same as above operations,
all of them give zero. So, the only non-zero contributions come from the first
term in (\ref{eqnPR187}), and (\ref{eqnPR189}) is the final (complete) result.
As in $3D$, the canonical Hamiltonian is a linear combination of secondary
translational and rotational constraints.

To prove that translational secondary and primary constraints have zero PBs
(as in $3D$), we have to find second variation of the secondary translational
constraint, i.e. we have to calculate%

\begin{equation}
\frac{\delta}{\delta e_{0\left(  \tau\right)  }}\frac{\delta\hat{H}%
_{c}^{\prime}\left(  1+2\right)  }{\delta e_{0\left(  \sigma\right)  }}%
=\frac{\partial}{\partial e_{0\left(  \tau\right)  }}\frac{\partial\hat{H}%
_{c}^{\prime}\left(  1+2\right)  }{\partial e_{0\left(  \sigma\right)  }%
}+\frac{\partial}{\partial\omega_{x\left(  yz\right)  }}\frac{\partial\hat
{H}_{c}^{\prime}\left(  1+2\right)  }{\partial e_{0\left(  \sigma\right)  }%
}\cdot\frac{\partial\omega_{x\left(  yz\right)  }}{\partial e_{0\left(
\tau\right)  }}. \label{eqnPR199}%
\end{equation}
The first term is zero, just based on $ABC$ properties, as we have already
discussed. The second term needs consideration. We demonstrated above that in
all terms in $\frac{\partial\hat{H}_{c}^{\prime}\left(  1+2\right)  }%
{\partial\omega_{x\left(  yz\right)  }}$ we have parts (as (\ref{eqnPR196}) or
(\ref{eqnPR197})) that are unaffected by variation with respect to components
$e_{0\left(  \sigma\right)  }$, so we can freely move them trough both
variations, e.g.%

\[
\frac{\partial}{\partial e_{0\left(  \sigma\right)  }}\frac{\partial
X_{c}E^{c\left(  y\right)  x\left(  z\right)  }}{\partial\omega_{x\left(
yz\right)  }}\cdot\frac{\partial\omega_{x\left(  yz\right)  }}{\partial
e_{0\left(  \tau\right)  }}=E^{c\left(  y\right)  x\left(  z\right)  }%
\frac{\partial}{\partial e_{0\left(  \sigma\right)  }}\frac{\partial X_{c}%
}{\partial\omega_{x\left(  yz\right)  }}\cdot\frac{\partial\omega_{x\left(
yz\right)  }}{\partial e_{0\left(  \tau\right)  }}=
\]

\begin{equation}
\frac{\partial}{\partial e_{0\left(  \sigma\right)  }}\frac{\partial X_{c}%
}{\partial\omega_{x\left(  yz\right)  }}\cdot E^{c\left(  y\right)  x\left(
z\right)  }\frac{\partial\omega_{x\left(  yz\right)  }}{\partial e_{0\left(
\tau\right)  }}=\frac{\partial}{\partial e_{0\left(  \sigma\right)  }}%
\frac{\partial X_{c}}{\partial\omega_{x\left(  yz\right)  }}\cdot
\frac{\partial E^{c\left(  y\right)  x\left(  z\right)  }\omega_{x\left(
yz\right)  }}{\partial e_{0\left(  \tau\right)  }}=0. \label{eqnPR199a}%
\end{equation}

We almost at once obtained the expected result: in all dimensions PBs among
primary and secondary translational constraints are zero. In this Section we
demonstrated that, despite having so many additional terms that are present in
higher than three dimensions and much richer structure of the secondary
translational constraint, the reduced Hamiltonian has the same structure is in
three dimensions: it is a linear combination of secondary constraints and all
secondary constraints have zero PBs with all primary. Next step of the Dirac
procedure is to prove its closure and consider PBs of secondary constraints
with the Hamiltonian. Because the Hamiltonian is a linear combination of
constraints, these calculations are equivalent to calculations of PBs among
all constraints. Complete calculations of PBs among secondary constraints are
in progress and only few contributions are checked but they so far support the
Poincar\'{e} algebra. Possible consequences of our calculations is the subject
of next Section.

\section{Discussion}

We start our discussion by briefly summarizing the obtained so far results for
the Hamiltonian formulation of N-bein gravity.

a) In \textit{any dimension} after elimination of second class constraints and
corresponding to them variables (spatial connections and conjugate to them
momenta) the total Hamiltonian of N-bein gravity is%

\[
H_{T}\left(  e_{\mu\left(  \rho\right)  },\pi^{\mu\left(  \rho\right)
},\omega_{0\left(  \alpha\beta\right)  },\Pi^{0\left(  \alpha\beta\right)
}\right)  =
\]

\begin{equation}
\underset{H_{c}}{\underbrace{-e_{0\left(  \rho\right)  }\chi^{0\left(
\rho\right)  }-\omega_{0\left(  \alpha\beta\right)  }\chi^{0\left(
\alpha\beta\right)  }}}+\dot{e}_{0\left(  \rho\right)  }\pi^{0\left(
\rho\right)  }+\dot{\omega}_{0\left(  \alpha\beta\right)  }\Pi^{0\left(
\alpha\beta\right)  }. \label{eqnPR200}%
\end{equation}

b) In \textit{any dimension }the\textit{ }canonical\textit{ }part of the
Hamiltonian ( $H_{c}$) is a linear combination of secondary constraints
($\chi^{0\left(  \rho\right)  }$, $\chi^{0\left(  \alpha\beta\right)  }$)
which are time development of two primary constraints ($\pi^{0\left(
\rho\right)  }$, $\Pi^{0\left(  \alpha\beta\right)  }$)%

\begin{equation}
\dot{\pi}^{0\left(  \rho\right)  }=\left\{  \pi^{0\left(  \rho\right)  }%
,H_{T}\right\}  =\chi^{0\left(  \rho\right)  }\text{, \ \ \ \ \ }\Pi^{0\left(
\alpha\beta\right)  }=\left\{  \Pi^{0\left(  \alpha\beta\right)  }%
,H_{T}\right\}  =\chi^{0\left(  \alpha\beta\right)  }\text{\ }
\label{eqnPR200a}%
\end{equation}
and all PBs among primary and secondary constraints are zero:%

\begin{equation}
\left\{  \pi^{0\left(  \rho\right)  },\chi^{0\left(  \sigma\right)  }\right\}
=\left\{  \pi^{0\left(  \rho\right)  },\chi^{0\left(  \alpha\beta\right)
}\right\}  =\left\{  \Pi^{0\left(  \alpha\beta\right)  },\chi^{0\left(
\sigma\right)  }\right\}  =\left\{  \Pi^{0\left(  \alpha\beta\right)  }%
,\chi^{0\left(  \mu\nu\right)  }\right\}  =0. \label{eqnPR200b}%
\end{equation}

c) In \textit{any dimension}%

\begin{equation}
\left\{  \chi^{0\left(  \alpha\beta\right)  },\chi^{0\left(  \mu\nu\right)
}\right\}  =\frac{1}{2}\tilde{\eta}^{\beta\mu}\chi^{0\left(  \alpha\nu\right)
}-\frac{1}{2}\tilde{\eta}^{\alpha\mu}\chi^{0\left(  \beta\nu\right)  }%
+\frac{1}{2}\tilde{\eta}^{\beta\nu}\chi^{0\left(  \mu\alpha\right)  }-\frac
{1}{2}\tilde{\eta}^{\alpha\nu}\chi^{0\left(  \mu\beta\right)  }
\label{eqnPR200c}%
\end{equation}
and in three dimensional case also%

\begin{equation}
\left\{  \chi^{0\left(  \alpha\beta\right)  },\chi^{0\left(  \rho\right)
}\right\}  =\frac{1}{2}\tilde{\eta}^{\beta\rho}\chi^{0\left(  \alpha\right)
}-\frac{1}{2}\tilde{\eta}^{\alpha\rho}\chi^{0\left(  \beta\right)  }
\label{eqnPR200d}%
\end{equation}
that must be true in \textit{any} dimension to preserve a rotational
invariance and%

\begin{equation}
\left\{  \chi^{0\left(  \rho\right)  },\chi^{0\left(  \gamma\right)
}\right\}  =0. \label{eqnPR200e}%
\end{equation}

The proof of (\ref{eqnPR200d}) and (\ref{eqnPR200e}) for any dimension is a
quite involved calculation taking into account the complexity of
$\chi^{0\left(  \rho\right)  }$ in higher than three dimensions. We checked a
few terms for the general $\chi^{0\left(  \rho\right)  }$ and did not find
contradictions with (\ref{eqnPR200d}-\ref{eqnPR200e}). In addition, because
rotational invariance is the same in all dimensions, PB (\ref{eqnPR200e}) must
be the same in all dimensions, which follows from the Castellani procedure.
Moreover, general expressions for constraints and all calculated so far
properties, in particular the secondary translational constraint
$\chi^{0\left(  \rho\right)  }$, satisfy $3D$ limit, i.e. equivalent with
found before \cite{3D} where all calculations were performed using
simplifications of $3D$ case right from the beginning.

The above results and observations seems to us sufficient to make the
following conjecture: \textit{in any dimension the algebra of secondary
constraints is Poincar\'{e} (N-bein gravity is the Poincar\'{e} gauge theory),
and consequently, N-bein gravity has rotational and translational gauge
invariance}.

For someone our conjecture can sound very reasonable, for others, maybe, even
not reasonable at all. However, contrary to many well-known conjectures that
no one knows how to prove or disprove, our conjecture is accompanied by the
mathematically well-defined procedure of proving or disproving it: calculate
PB among two translational constraints (\ref{eqnPR200e}). Moreover, because it
is always easier to disprove something: one counter-example is enough. We even
can suggest, seems to us, a relatively simple calculation: consider PB,
$\left\{  \chi^{0\left(  0\right)  },\chi^{0\left(  k\right)  }\right\}  $,
keeping only quadratic in $\pi^{k\left(  0\right)  }$ contributions (note that
there are no such contributions in $3D$ \cite{3D}). The result will be the
third order in momenta $\pi^{k\left(  0\right)  }$. If the result is not zero,
our conjecture is wrong and N-bein gravity either has algebra of first class
constraints different from Poincar\'{e} or, at least, third generation of
constraints will appear that must be second class. In such a case the only
gauge invariance would be rotational. There would be no translational or any
other invariances.

Let us discuss consequences of our conjecture. With the same algebra of
constraints in all dimensions calculation of generators is independent on a
dimension and we can just use results for the generator that was obtained for
the Poincar\'{e} algebra in $3D$ case and recalculate transformations in parts
where the full translational constraint is present. This simplicity is the
reflection of the fact that \textit{PB algebra} of first class constraints
defines a generator and explicit form of constraints is irrelevant, especially
when the Hamiltonian is a linear combination of constraints as (\ref{eqnPR200}).

We immediately have a gauge generator (using three dimensional result
\cite{3D})%

\begin{equation}
G=G_{t}+G_{r} \label{eqnPR201}%
\end{equation}
where translational and rotational parts are%

\begin{equation}
G=\pi^{0\left(  \rho\right)  }\dot{t}_{\left(  \rho\right)  }+\left(
-\chi^{0\left(  \rho\right)  }+\omega_{0(\gamma}^{~~~\rho)}\pi^{0\left(
\gamma\right)  }\right)  t_{\left(  \rho\right)  }, \label{eqnPR202}%
\end{equation}

\begin{equation}
G_{r}=\Pi^{0\left(  \alpha\beta\right)  }\dot{r}_{\left(  \alpha\beta\right)
}+\left(  -\chi^{0\left(  \alpha\beta\right)  }+\frac{1}{2}\left(
e_{0}^{\left(  \alpha\right)  }\pi^{0\left(  \beta\right)  }-e_{0}^{\left(
\beta\right)  }\pi^{0\left(  \alpha\right)  }\right)  +\omega_{0~~\mu
)}^{~(\alpha}\Pi^{0\left(  \beta\mu\right)  }-\omega_{0~\text{~}\mu)}%
^{~(\beta}\Pi^{0\left(  \alpha\mu\right)  }\right)  r_{\left(  \alpha
\beta\right)  }. \label{eqnPR203}%
\end{equation}

The only difference with $3D$ case is in translational part where
$\chi^{0\left(  \rho\right)  }$ is much richer. The knowledge of a generator
allows us to find transformations of all fields and any combinations of them
(in particular, transformations of secondary constraints)%

\begin{equation}
\delta(...)=\left\{  G,\left(  ...\right)  \right\}  . \label{eqnPR204}%
\end{equation}

The total Hamiltonian (\ref{eqnPR200}) is the result of the Hamiltonian
reduction \cite{AOP}: elimination of all spatial connections. This simple
Hamiltonian can be converted by the inverse Legendre transformations into
equivalent to it Lagrangian%

\begin{equation}
L\left(  \pi^{k\left(  \rho\right)  },e_{\mu\left(  \nu\right)  }%
,\omega_{0\left(  \alpha\beta\right)  }\right)  =\pi^{k\left(  \rho\right)
}\dot{e}_{k\left(  \rho\right)  }+e_{0\left(  \rho\right)  }\chi^{0\left(
\rho\right)  }+\omega_{0\left(  \alpha\beta\right)  }\chi^{0\left(
\alpha\beta\right)  }. \label{eqnPR205}%
\end{equation}

This is a different first order formulation of the original Einstein-Cartan
action obtained from the reduced Hamiltonian. As in the case of usual first
order formulation discussed in Introduction, the correctness of
(\ref{eqnPR205}) can be proven by elimination of auxiliary fields using a
variational method. The result: the same equation of motion for N-beins, the
only independent variables in the second order formulation.

The Hamiltonian and Lagrangian reductions must be equivalent \cite{AOP} and
the Lagrangian (\ref{eqnPR205}) obtained from the reduced Hamiltonian has to
be derivable also in the Lagrangian approach. Here the method of Lagrange
multipliers, used by Ostrogradsky \cite{Ostr} for the Hamiltonian formulation
of higher derivatives actions, should be implemented. We apply this method to
the standard first order formulation (\ref{eqnPR1}) but with different purpose
(it is already linear in derivatives) - to simplify elimination of spatial
connections. So, we define a coefficient in front of time derivatives of
N-bein as $\pi$ $^{k\left(  \rho\right)  }=eB^{k\left(  \rho\right)  0\left(
\alpha\right)  \nu\left(  \beta\right)  }\omega_{\nu\left(  \alpha
\beta\right)  }$ ($\pi$ $^{k\left(  \rho\right)  }$ is just an auxiliary
field) and, to keep equivalence, we add this redefinition using the Lagrange
multiplier $\Lambda_{k\left(  \rho\right)  }$ (this guaranties equivalence
with the original Lagrangian \cite{Comm})%

\[
L\left(  \pi^{k\left(  \rho\right)  },e_{\mu\left(  \nu\right)  }%
,\omega_{0\left(  \alpha\beta\right)  },\omega_{k\left(  \alpha\beta\right)
},\Lambda_{k\left(  \rho\right)  }\right)  =\pi^{k\left(  \rho\right)  }%
\dot{e}_{k\left(  \rho\right)  }+\Lambda_{k\left(  \rho\right)  }\left(
\pi^{k\left(  \rho\right)  }-eB^{k\left(  \rho\right)  0\left(  \alpha\right)
\nu\left(  \beta\right)  }\omega_{\nu\left(  \alpha\beta\right)  }\right)
\]

\begin{equation}
+eB^{\gamma\left(  \rho\right)  k\left(  \alpha\right)  \nu\left(
\beta\right)  }e_{\gamma\left(  \rho\right)  ,k}\omega_{\nu\left(  \alpha
\beta\right)  }-eA^{\mu\left(  \alpha\right)  \nu\left(  \beta\right)  }%
\omega_{\mu\left(  \alpha\gamma\right)  }\omega_{\nu~~\beta)}^{~(\gamma}.
\label{eqnPR206}%
\end{equation}

Performing variation with respect to $\Lambda_{k\left(  \rho\right)  }$ and
$\omega_{m\left(  \alpha\gamma\right)  }$, we obtain exactly the same
equations as in the Hamiltonian approach (\ref{eqnPR67}) and (\ref{eqnPR70})
(or (\ref{eqnPR120}) and (\ref{eqnPR121})) and after elimination of these
fields we have as a result (\ref{eqnPR205}). This is exactly what we did in
the Hamiltonian approach: they are equivalent, as it should be, and there is
no advantage in amount of calculations. However, even performance of such an
operation cannot be motivated by pure Lagrangian methods: why should one
introduce even more fields to already a first order Lagrangian? Contrary, in
the Hamiltonian approach the reason for such modifications is to simplify
solutions of second class constraints and to have a formulation with only
first class constraints that allows to find gauge transformations.

Let us discuss invariance of the Lagrangian (\ref{eqnPR205}) using the gauge
generators (\ref{eqnPR202}), (\ref{eqnPR203}) which are built in the
corresponding Hamiltonian formulation (\ref{eqnPR200}). First, we start from
the rotational invariance of (\ref{eqnPR205})%

\begin{equation}
\delta_{r}L\left(  \pi^{k\left(  \rho\right)  },e_{\mu\left(  \nu\right)
},\omega_{0\left(  \alpha\beta\right)  }\right)  = \label{eqnPR207}%
\end{equation}

\[
\delta_{r}\pi^{k\left(  \rho\right)  }\dot{e}_{k\left(  \rho\right)  }%
+\pi^{k\left(  \rho\right)  }\delta_{r}\dot{e}_{k\left(  \rho\right)  }%
+\delta_{r}e_{0\left(  \rho\right)  }\chi^{0\left(  \rho\right)  }+e_{0\left(
\rho\right)  }\delta_{r}\chi^{0\left(  \rho\right)  }+\delta_{r}%
\omega_{0\left(  \alpha\beta\right)  }\chi^{0\left(  \alpha\beta\right)
}+\omega_{0\left(  \alpha\beta\right)  }\delta_{r}\chi^{0\left(  \alpha
\beta\right)  }.
\]
All variations which are needed here are easy to find using the generator
(\ref{eqnPR203}) and (\ref{eqnPR204})%

\begin{equation}
\delta_{r}e_{0\left(  \rho\right)  }=-\frac{1}{2}\left(  e_{0}^{\left(
\alpha\right)  }r_{\left(  \rho\beta\right)  }-e_{0}^{\left(  \beta\right)
}r_{\left(  \rho\beta\right)  }\right)  , \label{eqnPR208}%
\end{equation}

\begin{equation}
\delta_{r}\omega_{0\left(  \alpha\beta\right)  }=-\dot{r}_{\left(  \alpha
\beta\right)  }-\left(  \omega_{0~~\beta)}^{~(\gamma}r_{\left(  \gamma
\alpha\right)  }-\omega_{0~~\alpha)}^{~(\gamma}r_{\left(  \gamma\beta\right)
}\right)  , \label{eqnPR209}%
\end{equation}

\begin{equation}
\delta_{r}e_{k\left(  \rho\right)  }=r_{\left(  \alpha\beta\right)  }%
\frac{\delta}{\delta\pi^{k\left(  \rho\right)  }}\chi^{0\left(  \alpha
\beta\right)  }=r_{\left(  \alpha\rho\right)  }e_{k}^{\left(  \alpha\right)
}, \label{eqnPR210}%
\end{equation}

\begin{equation}
\delta_{r}\pi^{k\left(  \rho\right)  }=-r_{\left(  \alpha\beta\right)  }%
\frac{\delta}{\delta e_{k\left(  \rho\right)  }}\chi^{0\left(  \alpha
\beta\right)  }=-r_{\left(  \rho\beta\right)  }\pi^{k\left(  \beta\right)
}-eB^{k\left(  \rho\right)  m\left(  \alpha\right)  0\left(  \beta\right)
}\left(  r_{\left(  \alpha\beta\right)  }\right)  _{,m}, \label{eqnPR211}%
\end{equation}

\begin{equation}
\delta_{r}\chi^{0\left(  \rho\right)  }=-r_{(\alpha}^{...\rho)}\chi^{0\left(
\alpha\right)  }, \label{eqnPR212}%
\end{equation}

\begin{equation}
\delta_{r}\chi^{0\left(  \alpha\beta\right)  }=-r_{(\gamma}^{...\alpha)}%
\chi^{0\left(  \gamma\beta\right)  }+r_{(\gamma}^{...\beta)}\chi^{0\left(
\gamma\alpha\right)  }. \label{eqnPR213}%
\end{equation}

Note that only explicit form of the rotational constraint (\ref{eqnPR110}) was
used (see (\ref{eqnPR210}), (\ref{eqnPR211})) and this again explains why
found in $3D$ rotational invariance remains to be the same in all dimensions
despite modification of the translational constraint. Calculation of
(\ref{eqnPR212}) and (\ref{eqnPR213}) are based on algebra of secondary
constraints but not on explicit form of constraints. Note that our assumption
$\left\{  \chi^{0\left(  \alpha\right)  },\chi^{0\left(  \beta\right)
}\right\}  =0$ (\ref{eqnPR200e}) was not applied here but (\ref{eqnPR200d})
was essential for (\ref{eqnPR212}) and (\ref{eqnPR213}). This provides
explicit illustration of the argument to support our conjecture that known
rotational invariance of N-bein gravity imposes severe restrictions on
possible modification of the Poincar\'{e} algebra in higher dimensions:
(\ref{eqnPR200d}) must be correct in any dimension.

Substitution of (\ref{eqnPR208}-\ref{eqnPR213}) into (\ref{eqnPR207}) gives%

\begin{equation}
\delta_{r}L\left(  \pi^{k\left(  \rho\right)  },e_{\mu\left(  \nu\right)
},\omega_{0\left(  \alpha\beta\right)  }\right)  = \label{eqnPR214}%
\end{equation}

\[
\left(  -r_{...\beta)}^{(\rho}\pi^{k\left(  \beta\right)  }-eB^{k\left(
\rho\right)  m\left(  \alpha\right)  0\left(  \beta\right)  }\left(
r_{\left(  \alpha\beta\right)  }\right)  _{,m}\right)  \dot{e}_{k\left(
\rho\right)  }+\pi^{k\left(  \rho\right)  }\left(  r_{\left(  \alpha
\rho\right)  }e_{k}^{\left(  \alpha\right)  }\right)  _{,0}-\dot{r}_{\left(
\alpha\beta\right)  }\chi^{0\left(  \alpha\beta\right)  }.
\]

The contributions with the translational constraint automatically disappear
and the only property that was used is its PB with the rotational one.
Substitution of explicit form of the rotational constraint (\ref{eqnPR110}) gives%

\begin{equation}
\delta_{r}L\left(  \pi^{k\left(  \rho\right)  },e_{\mu\left(  \nu\right)
},\omega_{0\left(  \alpha\beta\right)  }\right)  =\left(  eA^{n\left(
\alpha\right)  0\left(  \beta\right)  }r_{\left(  \alpha\beta\right)
,n}\right)  _{,0}-\left(  eA^{n\left(  \alpha\right)  0\left(  \beta\right)
}r_{\left(  \alpha\beta\right)  ,0}\right)  _{,n} \label{eqnPR215}%
\end{equation}
that complete the proof of invariance. We substituted the explicit form of the
rotational constraint in (\ref{eqnPR210}), (\ref{eqnPR211}) and
(\ref{eqnPR214}) but even this is not necessary for proof of invariance.
Equation (\ref{eqnPR214}) can be written (see (\ref{eqnPR210}),(\ref{eqnPR211}%
)) as%

\[
\delta_{r}L\left(  \pi^{k\left(  \rho\right)  },e_{\mu\left(  \nu\right)
},\omega_{0\left(  \alpha\beta\right)  }\right)  =
\]

\begin{equation}
-r_{\left(  \alpha\beta\right)  }\frac{\delta\chi^{0\left(  \alpha
\beta\right)  }}{\delta e_{k\left(  \rho\right)  }}\dot{e}_{k\left(
\rho\right)  }+\pi^{k\left(  \rho\right)  }\left(  r_{\left(  \alpha
\beta\right)  }\frac{\delta\chi^{0\left(  \alpha\beta\right)  }}{\delta
\pi^{k\left(  \rho\right)  }}\right)  _{,0}-\dot{r}_{\left(  \alpha
\beta\right)  }\chi^{0\left(  \alpha\beta\right)  }. \label{eqnPR216}%
\end{equation}
After integrations by parts in the second and third terms we obtain%

\[
\delta_{r}L\left(  \pi^{k\left(  \rho\right)  },e_{\mu\left(  \nu\right)
},\omega_{0\left(  \alpha\beta\right)  }\right)  =r_{\left(  \alpha
\beta\right)  }\left(  \dot{\chi}^{0\left(  \alpha\beta\right)  }-\frac
{\delta\chi^{0\left(  \alpha\beta\right)  }}{\delta e_{k\left(  \rho\right)
}}\dot{e}_{k\left(  \rho\right)  }-\dot{\pi}^{k\left(  \rho\right)  }%
\frac{\delta\chi^{0\left(  \alpha\beta\right)  }}{\delta\pi^{k\left(
\rho\right)  }}\right)
\]

\begin{equation}
-\left(  r_{\left(  \alpha\beta\right)  }\left(  \chi^{0\left(  \alpha
\beta\right)  }-\pi^{k\left(  \rho\right)  }\frac{\delta\chi^{0\left(
\alpha\beta\right)  }}{\delta\pi^{k\left(  \rho\right)  }}\right)  \right)
_{,0} \label{eqnPR217}%
\end{equation}
where the first bracket is identically zero and the second term is the total
temporal derivative. The explicit form of the rotational constraint is not
needed to prove rotational invariance of the Lagrangian. Zero value for the
first bracket just follows from the definition of derivatives%

\begin{equation}
\dot{\chi}^{0\left(  \alpha\beta\right)  }=\frac{\delta\chi^{0\left(
\alpha\beta\right)  }}{\delta e_{k\left(  \rho\right)  }}\dot{e}_{k\left(
\rho\right)  }+\frac{\delta\chi^{0\left(  \alpha\beta\right)  }}{\delta
\pi^{k\left(  \rho\right)  }}\dot{\pi}^{k\left(  \rho\right)  }+\frac
{\delta\chi^{0\left(  \alpha\beta\right)  }}{\delta e_{0\left(  \rho\right)
}}\dot{e}_{0\left(  \rho\right)  } \label{eqnPR218}%
\end{equation}
where the last term, as we demonstrated (see (\ref{eqnPR200b}), $\left\{
\chi^{0\left(  \alpha\beta\right)  },\pi^{0\left(  \rho\right)  }\right\}
=0=\frac{\delta\chi^{0\left(  \alpha\beta\right)  }}{\delta e_{0\left(
\rho\right)  }}$), is zero.

Of course, to find explicit form of the transformation in (\ref{eqnPR210},
\ref{eqnPR211}) or total derivative (\ref{eqnPR215}), we need the exact
expression for the rotational constraint. We will not give further detail here
because they are the same in all dimensions and were discussed in \cite{3D}.

Now we consider translational invariance of (\ref{eqnPR205})%

\begin{equation}
\delta_{t}L=\delta_{t}L\left(  \pi^{k\left(  \rho\right)  },e_{\mu\left(
\nu\right)  },\omega_{0\left(  \alpha\beta\right)  }\right)  \label{eqnPR219}%
\end{equation}

\[
\delta_{t}\pi^{k\left(  \rho\right)  }\dot{e}_{k\left(  \rho\right)  }%
+\pi^{k\left(  \rho\right)  }\delta_{t}\dot{e}_{k\left(  \rho\right)  }%
+\delta_{t}e_{0\left(  \rho\right)  }\chi^{0\left(  \rho\right)  }+e_{0\left(
\rho\right)  }\delta_{t}\chi^{0\left(  \rho\right)  }+\delta_{t}%
\omega_{0\left(  \alpha\beta\right)  }\chi^{0\left(  \alpha\beta\right)
}+\omega_{0\left(  \alpha\beta\right)  }\delta_{t}\chi^{0\left(  \alpha
\beta\right)  }%
\]
where again we can find transformations of all presented fields and constraints%

\begin{equation}
\delta_{t}e_{0\left(  \rho\right)  }=-\dot{t}_{\left(  \rho\right)  }%
-\omega_{0(\rho}^{~~~..\alpha)}t_{\left(  \alpha\right)  }, \label{eqnPR220}%
\end{equation}

\begin{equation}
\delta_{t}\omega_{0\left(  \alpha\beta\right)  }=0, \label{eqnPR221}%
\end{equation}

\begin{equation}
\delta_{t}e_{k\left(  \rho\right)  }=t_{\left(  \alpha\right)  }\frac
{\delta\chi^{0\left(  \alpha\right)  }}{\delta\pi^{k\left(  \rho\right)  }},
\label{eqnPR222}%
\end{equation}

\begin{equation}
\delta_{t}\pi^{k\left(  \rho\right)  }=-t_{\left(  \alpha\right)  }%
\frac{\delta\chi^{0\left(  \alpha\right)  }}{\delta e_{k\left(  \rho\right)
}}, \label{eqnPR223}%
\end{equation}

\begin{equation}
\delta_{t}\chi^{0\left(  \rho\right)  }=0, \label{eqnPR224}%
\end{equation}

\begin{equation}
\delta_{t}\chi^{0\left(  \alpha\beta\right)  }=\frac{1}{2}t^{\beta}%
\chi^{0\left(  \alpha\right)  }-\frac{1}{2}t^{\left(  \alpha\right)  }%
\chi^{0\left(  \beta\right)  } \label{eqnPR225}%
\end{equation}
which are even simpler than for rotational invariance (of course, if explicit
calculations of variations in (\ref{eqnPR222}) and (\ref{eqnPR223}) are not
needed). Substitution of (\ref{eqnPR220}-\ref{eqnPR225}) into (\ref{eqnPR219}) gives%

\begin{equation}
\delta_{t}L=-t_{\left(  \alpha\right)  }\frac{\delta\chi^{0\left(
\alpha\right)  }}{\delta e_{k\left(  \rho\right)  }}\dot{e}_{k\left(
\rho\right)  }+\pi^{k\left(  \rho\right)  }\left(  t_{\left(  \alpha\right)
}\frac{\delta\chi^{0\left(  \alpha\right)  }}{\delta\pi^{k\left(  \rho\right)
}}\right)  _{,0}-\dot{t}_{\left(  \rho\right)  }\chi^{0\left(  \rho\right)  }.
\label{eqnPR226}%
\end{equation}

As in the case of rotational invariance (\ref{eqnPR216}) (where the
translational constraint automatically dropped out), here terms proportional
to the rotational constraint cancel out without using its explicit form. Note
that our assumption $\left\{  \chi^{0\left(  \alpha\right)  },\chi^{0\left(
\beta\right)  }\right\}  =0$ was imposed here and triviality of
(\ref{eqnPR224}) is the consequence of it.

The invariance of the Lagrangian (\ref{eqnPR205}) for $3D$ case can be easily
verified because the form of constraints is quite simple due to absence of
many terms presented in higher dimensions. We return to the expression for the
translational constraint in any dimension and use it to illustrate one more
time that there is nothing special about three dimensional case if the general
Dirac procedure is used, all $3D$ results follow from general expressions for constraints.

The translational constraint was obtained in Section VII%

\[
\chi^{0\left(  \rho\right)  }=\pi_{,k}^{k\left(  \rho\right)  }-\frac{\delta
}{\delta e_{0\left(  \rho\right)  }}\left(  eB^{n\left(  \rho\right)  k\left(
\alpha\right)  m\left(  \beta\right)  }\right)  e_{n\left(  \rho\right)
,k}\omega_{m\left(  \alpha\beta\right)  }-\frac{\delta}{\delta e_{0\left(
\rho\right)  }}\left(  eA^{k\left(  p\right)  m\left(  q\right)  }\right)
\omega_{k\left(  pn\right)  }\omega_{m~~q)}^{~(n}%
\]

\begin{equation}
-\frac{\delta}{\delta e_{0\left(  \rho\right)  }}\left(  eA^{k\left(
p\right)  m\left(  q\right)  }\right)  \omega_{k\left(  p0\right)  }%
\omega_{m~~q)}^{~(0}-2\frac{\delta}{\delta e_{0\left(  \rho\right)  }}\left(
eA^{k\left(  0\right)  m\left(  q\right)  }\right)  \omega_{k\left(
0p\right)  }\omega_{m~~q)}^{~(p} \label{eqnPR227}%
\end{equation}
where spatial connections are not independent fields but only a short notation
for solutions given by (\ref{eqnPR143}) and (\ref{eqnPR66}).

In $3D$ case the term with $B$ is manifestly zero (there are no three distinct
external spatial indices to support its antisymmetry for all pairs of
permutation $nkm$), as well as the last term in the first line (if $n=1\left(
2\right)  $ then $p=q=2\left(  1\right)  $ and $A$ is zero). The solutions for
spatial connections that we have to substitute into (\ref{eqnPR227}) are also
considerably simplified in three dimensions.

The general solution for $\omega_{k\left(  pn\right)  }$ (\ref{eqnPR143}) in
$3D$ limit gives%

\[
\lim_{D\longmapsto3}\omega_{k\left(  pq\right)  }=-\frac{1}{2ee^{0\left(
0\right)  }}I_{k\left(  p\right)  m\left(  q\right)  }\pi^{m\left(  0\right)
}.
\]
The solution for $\omega_{k\left(  p0\right)  }$ (\ref{eqnPR66}) in $3D$ case is%

\[
\lim_{D\longmapsto3}\omega_{k\left(  q0\right)  }=-\frac{1}{2ee^{0\left(
0\right)  }}I_{k\left(  q\right)  m\left(  p\right)  }\pi^{m\left(  p\right)
}.
\]

Performing variation in the second line of (\ref{eqnPR227}), we obtain $B$
with one external zero which is expressible in terms of $E$ that allows to
perform contraction with one of $I$ and, as a result, we obtain the secondary
translational constrain in $3D$ case. Note that all the above $3D$ results are
limits of general solutions and they are equivalent with \cite{3D}.

As it was shown in \cite{3D} with the explicit form of constraints, we can
derive transformations of $\pi^{k\left(  \rho\right)  }$ and $e_{k\left(
\rho\right)  }$. To find transformations of the original first order
Lagrangian which has different set of variables, we have to find
transformations of spatial connections first. Solutions to them (second class
constraints) have to be used (for $3D$ see \cite{3D}). Finally, if we are
interested in invariance of the original second order Lagrangian, we have to
find a transformation of N-beins in terms of N-beins (the only independent
variable) using the definition of a connection (\ref{eqnPR21}). Alternatively,
we can obtain this transformation from equivalence of (\ref{eqnPR205}) with
the original Lagrangian using its equations of motion and avoid intermediate
calculations of transformatons for spatial connections.

Now, in higher dimensions, we just demonstrate the complexity of
transformations due to many additional contributions and essential
modifications of the $3D$ result.

First, let us, as an example, consider only one (second term in
(\ref{eqnPR227})) which is zero in three dimensions%

\begin{equation}
\chi_{2}^{0\left(  \rho\right)  }=-\frac{\delta}{\delta e_{0\left(
\rho\right)  }}\left(  eB^{n\left(  \rho\right)  k\left(  \alpha\right)
m\left(  \beta\right)  }\right)  e_{n\left(  \rho\right)  ,k}\omega_{m\left(
\alpha\beta\right)  } \label{eqnPR235}%
\end{equation}
and its contribution into the constraint $\chi^{0\left(  0\right)  }$.
Performing variation and using $ABC$ properties, we obtain%

\begin{equation}
\chi_{2}^{0\left(  0\right)  }=ee^{0\left(  0\right)  }\left(  \tilde{\omega
}_{~\left(  pq\right)  }^{q}E^{n\left(  b\right)  k\left(  p\right)  }%
+\tilde{\omega}_{~\left(  pq\right)  }^{b}E^{n\left(  p\right)  k\left(
q\right)  }+\tilde{\omega}_{~\left(  pq\right)  }^{p}E^{n\left(  q\right)
k\left(  b\right)  }\right)  e_{n\left(  b\right)  ,k} \label{eqnPR236}%
\end{equation}
and even further restriction to a particular variation $\frac{\delta
\chi^{0\left(  0\right)  }}{\delta\pi^{k\left(  0\right)  }}$ (this choice is
dictated by a solution for $\omega_{m\left(  pq\right)  }\left(  \pi^{k\left(
0\right)  }\right)  $ \ given in Section VII). To find this single variation
$\frac{\delta\chi^{0\left(  0\right)  }}{\delta\pi^{k\left(  0\right)  }}$, we
have to perform variation of (\ref{eqnPR236}) where there are two terms with
\textquotedblleft traces\textquotedblright\ $\tilde{\omega}_{~\left(
pq\right)  }^{q}$ which are simple, but in the second term we have three
different indices and full solution is needed (see (\ref{eqnPR143})). Of
course, $3D$ limit is preserved on all stages of calculations, in particular,
for (\ref{eqnPR236}). To prove this we have to consider all possible
combinations of spatial indices, as $3D$ limit is not manifest in such a form.
For $n=k=1(2)$ it is zero because of antisymmetry of $E$, for non-equal
indices $n=1(2)$ $k=2(1)$ we have to consider also particular values, $\left(
1\right)  $ and $\left(  2\right)  $, and find all contributions, sum of which
is zero.

The appearance of additional contributions in the translational constraint and
a comparison of them with derivation of transformations in $3D$ case allows to
describe expected modifications (qualitatively) in transformations of
different fields%

\begin{equation}
\delta_{t}e_{\mu\left(  \alpha\right)  }\propto\text{ \ }e_{\nu\left(
\beta\right)  ,\gamma}\text{ and }\omega_{\nu\left(  \beta\sigma\right)  },
\label{eqnPR237}%
\end{equation}

\begin{equation}
\delta_{t}\omega_{\mu\left(  \alpha\beta\right)  }\propto\text{\ }\omega
_{\nu\left(  \tau\sigma\right)  ,\lambda}\text{ and }\omega_{\nu\left(
\tau\sigma\right)  }\times\omega_{\gamma\left(  \rho\varepsilon\right)  }.
\label{eqnPR238}%
\end{equation}

Of course, the completion of all, only partially described, calculations is
needed to have the explicit form of transformations but to prove invariance of
the Lagrangian (\ref{eqnPR205}) under translation in any dimension, we can
repeat simple steps as it was done for rotational invariance (\ref{eqnPR216}%
)-(\ref{eqnPR218}), because the algebra of first class constraints (not an
explicit form of constraints) defines invariance. Performing integration by
parts in (\ref{eqnPR226}) we obtain%

\begin{equation}
\delta_{t}L=t_{\left(  \alpha\right)  }\left(  \dot{\chi}^{0\left(
\alpha\right)  }-\frac{\delta\chi^{0\left(  \alpha\right)  }}{\delta
e_{k\left(  \rho\right)  }}\dot{e}_{k\left(  \rho\right)  }-\dot{\pi
}^{k\left(  \rho\right)  }\frac{\delta\chi^{0\left(  \alpha\right)  }}%
{\delta\pi^{k\left(  \rho\right)  }}\right)  + \label{eqnPR240}%
\end{equation}

\[
-\left(  t_{\left(  \alpha\right)  }\left(  \chi^{0\left(  \alpha\right)
}-\pi^{k\left(  \rho\right)  }\frac{\delta\chi^{0\left(  \alpha\right)  }%
}{\delta\pi^{k\left(  \rho\right)  }}\right)  \right)  _{,0}%
\]
where the first bracket is zero because%

\begin{equation}
\dot{\chi}^{0\left(  \alpha\right)  }=\frac{\delta\chi^{0\left(
\alpha\right)  }}{\delta e_{k\left(  \rho\right)  }}\dot{e}_{k\left(
\rho\right)  }+\frac{\delta\chi^{0\left(  \alpha\right)  }}{\delta
\pi^{k\left(  \rho\right)  }}\dot{\pi}^{k\left(  \rho\right)  }+\frac
{\delta\chi^{0\left(  \alpha\right)  }}{\delta e_{0\left(  \rho\right)  }}%
\dot{e}_{0\left(  \rho\right)  } \label{eqnPR241}%
\end{equation}
with the last term also zero which is again the consequence of the algebra of
constraints ($\left\{  \chi^{0\left(  \alpha\right)  },\pi^{0\left(
\rho\right)  }\right\}  =0=\frac{\delta\chi^{0\left(  \alpha\right)  }}{\delta
e_{0\left(  \rho\right)  }}$). In all dimensions the Lagrangian is invariant
under a translation, only algebra of constraints is needed to prove this, not
the explicit form of constraints. Of course, long calculations have to be
performed to find the explicit form of transformations (\ref{eqnPR222}%
-\ref{eqnPR223}) and the total derivative in (\ref{eqnPR240}). All qualitative
changes that one can expect in translational invariance in higher dimensions
were described in (\ref{eqnPR237}-\ref{eqnPR238}). We repeat that, based on
(\ref{eqnPR217}-\ref{eqnPR218}) and (\ref{eqnPR240}-\ref{eqnPR241}), the
invariance of the reduced Lagrangian in any dimension, as well as all
equivalent to it (second or first order) formulations, is the consequence of
(\ref{eqnPR200}-\ref{eqnPR200e}). So, only our conjecture has to be proven.
The complexity of general expression for the translational constraint makes
this task very difficult if direct substitution is used. We are trying to find
some short cuts using $ABC$ properties and/or something similar with what was
found for calculations of PB between primary and secondary translational
constraints in Section VII. Whatever result of calculation for PB among
translational constraints will produce we can make some conclusions.

\section{Conclusion}

The Hamiltonian formulation of constraint systems developed by Dirac is
indispensable in studies of complicated theories with unknown \textit{a
priori} gauge invariance. When original, not specialized to a particular
dimension, variables are used it allows to consider all dimensions at once and
see possible peculiarities and origin of them in particular dimensions.

The approach based on \textit{a priori} assumptions about gauge invariance,
attempts to build a gauge theory according to standard rules (as Yang-Mills)
and comparison with known theories with a hope to obtain equivalence, is not
productive. The transformations found by Witten \cite{Witten} recognizing
relation of the Einstein-Cartan action in three dimensions to the Chern-Simons
action is an interesting observation but not a mathematical method to find out
whether any field theoretical model in any dimension has a gauge symmetry or
not and to obtain gauge transformations of fields. His conclusion:
\textquotedblleft we cannot hope that four-dimensional gravity would be a
gauge theory \textit{in that sense }(Italic is our)\textquotedblright\ (as we
understood, it means Einstein-Cartan is not Chern-Simons in four dimensions)
is \textit{trivial} but, of course, \textit{correct}. According to Dirac
\cite{Diracbook}, a gauge theory is a theory that has first class constraints
which define the gauge invariance. And \textit{in this sense,} using the
well-defined and general procedure, we can always answer the question whether
we have a gauge theory or not. However, Witten's followers trying to apply
gauge transformations found in $3D$ case to the Eistein-Cartan action in
higher dimensions (obviously without success, after neglecting so many
contributions as it is clear from the Hamiltonian approach presented here)
made a \textit{non-trivial} conclusion that N-bein gravity is not Poincar\'{e}
gauge theory or that translational invariance is only property of $3D$ which,
as we claim, is \textit{incorrect}. Gauge invariance is defined by algebra of
PBs among first class constraints and, as we illustrated in Discussion, the
translational secondary constraint is much richer in higher dimensions but its
PBs, e.g. with primary constraints, remains the same. Despite of different
expression of the translational constraint, the algebra of first class
constraints might be the same in all dimensions $D>2$ although the form of
transformations of fields might be different in higher dimensions. It is clear
now that, for example, the gauge transformation of $\omega_{\nu\left(
\alpha\beta\right)  }$ in dimensions higher than three should also have a
dependence on a translational parameter.

The final answer on the question what gauge symmetry N-bein gravity has in
dimensions higher than three depends on correctness of our conjecture. If
$\left\{  \chi^{0\left(  \alpha\right)  },\chi^{0\left(  \beta\right)
}\right\}  =0$ then N-bein gravity is Poincar\'{e} gauge theory in all
dimensions and $3D$ is not special at all. If $\left\{  \chi^{0\left(
\alpha\right)  },\chi^{0\left(  \beta\right)  }\right\}  \neq0$ but
proportional to secondary constraints, i.e. $\left\{  \chi^{0\left(
\alpha\right)  },\chi^{0\left(  \beta\right)  }\right\}  =f^{0\left(
\beta\right)  }\chi^{0\left(  \alpha\right)  }-f^{0\left(  \alpha\right)
}\chi^{0\left(  \beta\right)  }$ (structure functions instead of structure
constraints are unavoidable but in $3D$ this bracket must be zero) we still
have closure of the Dirac procedure, all constraints are first class and new
generators can be found easily. In this case N-bein gravity is the gauge
theory but it is Poincar\'{e} only in $3D$ case. If $\left\{  \chi^{0\left(
\alpha\right)  },\chi^{0\left(  \beta\right)  }\right\}  \neq0$ and not
proportional to secondary constraints, then we have next generation of
constraints because multipliers would not be found on this stage (primary and
secondary constraints have zero PBs). In this case, at least tertiary
constraints would appear, all of them could not be first class, otherwise we
would have a negative number of degrees of freedom. We reject such a
nonphysical result. If tertiary, quarterly, etc. constraints appear for
consistency they should be second class, neither translational invariance nor
diffeomorphism would be gauge symmetries and only rotational gauge invariance
would survive.

The direct calculation of $\left\{  \chi^{0\left(  \alpha\right)  }%
,\chi^{0\left(  \beta\right)  }\right\}  $ is laborious because of complexity
of constraints in higher dimensions and we are trying to find a way of dealing
with such calculations. However, some results already allow to make the
following conclusion, respectively what PBs among translational constraints
are. It is clear that if all constraints ($\pi^{0\left(  \rho\right)  }$,
$\pi^{0\left(  \alpha\beta\right)  }$, $\chi^{0\left(  \rho\right)  }$,
$\chi^{0\left(  \alpha\beta\right)  }$) are first class we have two gauge
parameters, $t_{\left(  \rho\right)  }$ and $r_{\left(  \alpha\beta\right)  }%
$, which correspond to two primary constraints, $\pi^{0\left(  \rho\right)  }$
and $\pi^{0\left(  \alpha\beta\right)  }$. Both parameters have internal
indices, so there is no place for \textquotedblleft diffeomorphism
constraint\textquotedblright\ (spatial or full) and diffeomorphism is not a
\textit{gauge invariance} of N-bein gravity. All formulations that claim to
have the \textquotedblleft spatial diffeomorphism constraint\textquotedblright%
\ for tetrad gravity are the product of non-canonical change of variables. The
similar loss of full diffeomorphism invariance in the metric gravity was
discussed in \cite{KKRV}, \cite{FKK}. Actually, this non-canonical change of
variables for tetrad gravity has the same origin as in the metric gravity
\cite{Myths}. Of course, after a non-canonical change of variables is
performed, any connection with an original theory is lost. Loosely speaking
(as a mathematical result cannot be more correct or more incorrect), a
\textquotedblleft deviation\textquotedblright\ from a correct formulation in
case of tetrad gravity is more severe: the gauge parameter of diffeomorphism,
$\xi_{\mu}$, has an external index, whereas the gauge parameter of
translation, $t_{\left(  \rho\right)  }$, has an internal index. The only
possibility to reconcile translational invariance with diffeomorphism (of
course, full, not spatial) in the Hamiltonian formulation, where these two
symmetries cannot be present simultaneously as gauge symmetries (too many
primary constraints are needed), is to find a canonical transformation that
converts one into another. However, such a possibility seems to us quite
bleak, in particular, because the nature of these two parameters is so
different: translational invariance arises from the primary constraint
$\phi^{0\left(  \rho\right)  },$ whereas for diffeomorphism we need $\phi
^{\mu\left(  0\right)  }$.

Finally, if our conjecture is correct, the algebra of constraints is the
Poincar\'{e} which is an ordinary Lie algebra: no structure functions, no
derivatives of delta functions (non-locality). In contrast, formulations based
on non-canonical changes of variables leading to the \textquotedblleft spatial
diffeomorphism\textquotedblright\ constraint for tetrad gravity have non-local
algebra of constraints with structure functions. This algebra for a long time
is the source of many troubles and numerous speculations.

\textbf{Acknowledgements}

We would like to thank D.G.C. McKeon for many helpful discussions and for
invitation to give a talk on Theoretical Physics section at CAIMS. We are also
thankful to A.M. Frolov for numerous discussions during preparation of our
Report. The partial support of the Huron University College Faculty of Arts
and Social Science Research Grant Fund is greatly acknowledged.

\end{document}